\documentclass[12pt,draft]{article}
\usepackage{amssymb,axodraw}
\usepackage{epsfig}
\usepackage{equation}
\newcommand{\imag}{\Im {\rm m}}
\newcommand{\real}{\Re {\rm e}}

\def\lsim{\:\raisebox{-0.5ex}{$\stackrel{\textstyle<}{\sim}$}\:}

\def\slep{{\tilde{l}}}
\def\cb{\cos\beta}
\def\sb{\sin\beta}
\def\tb{\tan\beta}
\def\ctb{\cot\beta}
\def\cbz{\cos2\beta}
\def\sbz{\sin 2\beta}
\def\epem{$e^+e^-$}
\def\sw{\sin\theta_{\rm W}}
\def\swq{\sin^2\theta_{\rm W}}
\def\cw{\cos\theta_{\rm W}}
\def\cwq{\cos^2\theta_{\rm W}}

\def\tw{\tan\theta_{\rm W}}

\def\mw{M_W}
\def\mz{M_Z}
\def\cL{{\cal L}}
\def\cM{{\cal M}}

\def\cS{{\cal S}}

\def\pslash{p\hspace{-0.45em} /}
\def\kslash{k\hspace{-0.42em} /}
\def\beq{\begin{equation}}
\def\eeq{\end{equation}}
\def\beqa{\begin{eqnarray}}
\def\eeqa{\end{eqnarray}}
\newcommand{\ben}{\begin{subequations}}
\newcommand{\een}{\end{subequations}}
\newcommand{\beqq}{\begin{eqalignno}}
\newcommand{\eeqq}{\end{eqalignno}}

\begin{document}

\renewcommand{\thefootnote}{\fnsymbol{footnote}}

\begin{flushright}
TUM--HEP--490/02\\
hep--ph/0403054\\
 \today\\~\newline
\end{flushright}
\begin{center}
  {\Large\bf Systematic study of the impact of CP--violating phases of
    the MSSM on leptonic high--energy observables}  \\[8mm]

{\large Seong Youl Choi$^1$, Manuel Drees$^2$ and Benedikt
Gaissmaier$^2$} \\[4mm] 
{}$^1${\it  Department of Physics, Chonbuk National University, Chonju
561-756, Korea} \\
{}$^2${\it Physik Dept., TU M\"unchen, James Franck Str., D--85748
Garching, Germany}
\end{center}

\bigskip
\bigskip
\bigskip
\begin{abstract}

Low-energy results from measurements of leptonic dipole moments are
used to derive constraints on the CP--violating phases of the
dimensionful parameters of the minimal supersymmetric extension of the
standard model (MSSM). We use these (known) bounds to investigate the
impact of these phases on CP--even cross sections at high--energy
\epem\ and $e^-e^-$ colliders. To that end we define two measures of
the significance with which the existence of non--vanishing phases
could be deduced from the measurements of these cross sections. We
find that highly significant evidence for deviations from the
CP--conserving MSSM could be obtained at the next \epem\ collider even
if the electric dipole moment of the electron is very small or zero.
We also analyze a CP--odd final state polarization, which can be large
when two different charginos or neutralinos are produced. Finally, we
study correlations between the phase--sensitive observables.

\noindent
\end{abstract}
%
%PACS number(s): 11.30.Pb, 11.30.Er ?????????
\newpage
%%%%%%%%%%%%%%%%%%%%%%%%%%%%%%%%%%%%%%%%%%%%%%%%%%%%%%%%%%%%%%%%%
\section{Introduction}
\label{sec:introduction}
%%%%%%%%%%%%%%%%%%%%%%%%%%%%%%
\setcounter{footnote}{0}
\setcounter{equation}{0}
\renewcommand{\theequation} {\thesection.\arabic{equation}}
%%%%%%%%%%%%%%%%%%%%%%%%%%%%%%%%%%%%%%%%%%%%%%%%%%%%%%%%%%%%%%%%%

CP violation was observed first in the neutral kaon system
\cite{Kaon}, and has recently been found in $B-$meson decays
\cite{B-meson}. In addition, CP violation constitutes one of the
conditions for a dynamical generation of the cosmological baryon
asymmetry \cite{sakharov}. In the Standard Model (SM), which contains
only one physical neutral Higgs boson and assumes neutrinos to be
massless, the only source of CP violation is the complex phase of the
quark mixing matrix \cite{km}.\footnote{The observation \cite{nuosc}
of neutrino flavor oscillations opens the possibility that the
neutrino mass matrix contains non--trivial CP--violating phases, but
this has not yet been confirmed experimentally. In principle CP could
also be violated in the SM by the QCD $\theta-$term, but bounds on the
electric dipole moment of the neutron force $|\theta_{\rm QCD}|$ to be
$\lsim 10^{-10}$.}

Supersymmetry (SUSY) is now widely regarded to be the most plausible
extension of the SM; among other things, it stabilizes the gauge
hierarchy \cite{6} and allows the grand unification of all known gauge
interactions \cite{7}. Of course, supersymmetry most be (softly)
broken to be phenomenologically viable. In general this introduces a
large number of unknown parameters, many of which can be complex
\cite{DS}. In the most general minimal supersymmetric standard model
(MSSM) 44 phases cannot be removed by suitable redefinitions of fields
and remain as ``physical'' phases in the model. For example, they have
a direct impact on the mass spectra as they enter most mass matrices
in the Lagrangian. Of course, one can use more specific assumptions on
the soft breaking terms and/or an underlying GUT theory to get
simpler versions of the MSSM with a smaller number of parameters,
but the price for doing so is the loss of generality.

CP-violating phases associated with sfermions of the first and, to a
lesser extent, second generation, and with the chargino/neutralino
sector, are severely constrained by bounds on the electric dipole
moments of the electron, neutron and muon.  However, as emphasized in
\cite{9,nath,kane,bartl} cancellations between different diagrams
allow some combinations of these phases to be rather large even for a
sfermion mass spectrum accessible at the expected center--of--mass
energy of a possible next linear \epem\ collider (LC). Even in models
with universal boundary conditions for soft breaking mass at some very
high energy scale, the relative phase between the supersymmetric
Higgsino mass parameter $\mu$ and the universal trilinear soft
breaking parameter $A_0$ can be ${\cal O}(1)$ \cite{10}. If
universality is not assumed, the relative phase between the $U(1)_Y$
and $SU(2)$ gaugino masses may also be large.

In the past few years a lot of effort has been devoted to analyses of
the physics output that can be expected from experiments at the LC,
including a possible $e^-e^-$ option \cite{lcrev}. Work towards the
design of such a device has also made great progress. Today it is
assumed that it will (initially) have a center--of--mass energy
$\sqrt{s}$ in the range between 500 GeV and 1 TeV, an integrated
luminosity of at least several hundred ${\rm fb}^{-1}$, and adjustable
polarization for both beams. Detailed analyses \cite{lcrev,oldana}
have established that sparticles with mass $\lsim \sqrt{s}/2$ can
easily be discovered at an LC. Moreover, many of their properties
(masses, spins, some couplings) can be measured precisely.

Unfortunately most of these analyses \cite{oldana} show the dangerous
tendency to neglect phases, which are actually free parameters of the
model and are not necessarily negligibly small. Note that both masses
and couplings depend on these phases, which will hence have a direct
impact on sparticle production cross sections and decays.  Neglecting
non-vanishing phases when determining real parameters from
experimental data could thus lead to wrong inputs for attempts to
reconstruct the underlying theory at the unification scale.  On the
other hand, the construction of sizable and experimentally accessible
CP--violating observables is rather difficult in most of the
production channels at $e^+e^-$ colliders, as at least one secondary
decay has to be included in the analysis. At the tree level nonzero
CP--odd asymmetries can only result if the decaying particle has
nonzero spin, which should be at least partly reconstructed from its
decay products.

We therefore first perform a rather general analysis of the impact of
non--vanishing CP--odd phases on CP--even cross sections. We work in
the framework of the MSSM with non--vanishing CP phases. We assume
flavor universality for soft breaking terms associated with sfermions
of the first and second generation, but we do not assume any specific
model for SUSY breaking.  Our free parameters are specified at the
typical energy scale of an LC. The basic idea of this work is to take
today's low energy data, such as lower mass bounds and bounds on
leptonic dipole moments ($d_e$ and $a_\mu$), as constraints for a
parameter space scan. We then use the resulting, low energy compatible
points to check whether high energy experiments at an LC (in either
the $e^+e^-$ or the $e^-e^-$ mode) could provide additional
information on phases. We restrict ourselves to the following total,
unpolarized cross sections:
\ben \label{processes} \beqq
e^+e^- &\to \tilde{\chi}_i^0 \tilde{\chi}_j^0 \qquad i,j = 1,
 \dots, 4; \\
e^+e^- &\to \tilde{\chi}_i^- \tilde{\chi}_j^+ \qquad i,j = 1,2; \\
e^+e^- &\to \tilde{e}_i^- \tilde{e}_j^+ \qquad i,j = 1,2; \\
e^-e^- &\to \tilde{e}_i^- \tilde{e}_j^- \qquad i,j = 1,2 .
\eeqq \een

There is a complementarity between the leptonic dipole operators and
the high energy production amplitudes. Since several diagrams
involving neutralinos as well as charginos contribute coherently to
the low--energy observables, they can only give bounds on {\em
combinations} of phases. In contrast, high energy observables can be
used to investigate the different sectors of the theory separately. As
our aim is to study the impact of low energy compatible,
non--vanishing phases on the cross sections, we assign a significance
$\cS(f_1 f_2)$ to each final state, defined as difference in
production rates between a CP--conserving point (CPC--point: real
parameters, all phases identical to zero or $\pi$) in parameter space
and a CP--violating point (CPV--point: same absolute values of
parameters, but non--vanishing phases) normalized to the statistical
error of the cross section in the CPC--point. Since the phase
dependence of a given cross section might partly arise from
kinematical effects (kinematical masses depend on the phases), we also
introduce a second significance $\bar\cS(f_1 f_2)$, where the
CPV--point is chosen such that the masses of two neutralinos and one
chargino coincide with the CPC--point; this can be achieved by
adjusting the absolute values of the relevant dimensionful input
parameters.

We find that these significances can be very large for some reactions
of the types (\ref{processes}a) and (\ref{processes}d), but are
usually small for (\ref{processes}b) once the low--energy constraints
have been taken into account. Moreover, if the absolute values of the
input parameters, or three chargino/neutralino masses, are kept fixed
while the phases are varied randomly, there is no visible correlation
between these high--energy significances and $d_e$. On the other hand,
in most cases sizable high--energy significances are strongly
correlated with each other, and slightly less strongly correlated with
$a_\mu$.

Strictly speaking these significances only measure deviations from the
CP--conserving version of the MSSM. These deviations might also be
explained by some extension of the MSSM without invoking new sources
of CP violation. If some deviation from the CP--conserving MSSM is
observed, a more direct probe for CP violation in the production and
decay of superparticles thus becomes important. We therefore compute
the CP--odd polarization of charginos and neutralinos that is normal
to the production plane, and find that in all cases considered, it can
be sizable for neutralinos; a large CP--odd polarization of charginos
is possible only at large $|\mu|$ and small $\tan\beta$. In contrast
to earlier, related work \cite{barger1, barger} we emphasize a
detailed semi--analytical understanding of the observed effects;
isolate the measurements that hold the most promise; and analyze the
correlations between various phase sensitive observables.

The remainder of this paper is organized as follows. In
Section~\ref{sec:particlemixing} we briefly review the mass spectra
and mixing patterns of the sleptons, charginos and neutralinos. This
section also contains an overview of the relevant parameters. After
summarizing the relevant parts of the MSSM Lagrangian in
Section~\ref{sec:feynmanrules}, we present in
Section~\ref{sec:lowenergy} the analytical expressions for the SUSY
contributions to $d_e$ and $a_\mu$, and discuss briefly possible
scenarios for suppressing these leptonic dipole moments while keeping
some phases sizable. This Section also discusses numerical constraints
on these phases in three benchmark scenarios where selectrons as well
as the lighter neutralino and chargino eigenstates can be produced at
a 500 GeV $e^+e^-$ collider. Section~\ref{sec:highenergy} summarizes
the well--known results for total cross sections of the production
channels (\ref{processes}). We also give results for the components of
polarization vectors for reactions (\ref{processes}a,b).  The
significances are introduced in Section~\ref{sec:newobjects}. In
Section~\ref{sec:numerics} we show the most important results of our
detailed numerical analysis of the high energy observables.
Section~\ref{sec:conclusion} completes our work with a brief summary
of our findings and some conclusions.

%%%%%%%%%%%%%%%%%%%%%%%%%%%%%%%%%%%%%%%%%%%%%%%%%%%%%%%%%%%%%%%%%
%%%%%%%%%%%%%%%%%%%%%%%%%%%%%%%%%%%%%%%%%%%%%%%%%%%%%%%%%%%%%%%%%
\section{Particle mixing}
\label{sec:particlemixing}
%%%%%%%%%%%%%%%%%%%%%%%%%%%%%%
\setcounter{footnote}{0} \setcounter{equation}{0}
\renewcommand{\theequation} {\thesection.\arabic{equation}}
%%%%%%%%%%%%%%%%%%%%%%%%%%%%%%%%%%%%%%%%%%%%%%%%%%%%%%%%%%%%%%%%%
%%%%%%%%%%%%%%%%%%%%%%%%%%%%%%%%%%%%%%%%%%%%%%%%%%%%%%%%%%%%%%%%%

\subsection{Slepton mixing}

As mentioned in the Introduction, we will assume that flavor mixing is
negligible in the slepton sector. This can e.g. be motivated by the
very tight experimental constraints on branching ratios for lepton
flavor violating decays like $\mu \rightarrow e \gamma, \, \mu
\rightarrow 3e$ etc. The simplest way to satisfy these bounds on
flavor changing processes is to assume that soft SUSY breaking
parameters in the slepton sector are the same for the first and second
generation, as is the case in most models that attempt to describe
SUSY breaking by a small number of parameters (which are usually
defined at a high energy scale). The only relevant mixing in the
slepton sector then occurs between $SU(2)$ doublet sleptons $\slep_L$
and singlets $\slep_R$. The squared mass matrix $\cM^2_\slep$ in the
basis $(\slep_L, \slep_R)$ is given by \cite{haberkane}
\beq \label{m2sl}
\cM ^2_\slep = \left( \begin{array}{cc} X_\slep & Z_\slep \\
 Z_\slep^\star & Y_\slep\end{array} \right).
\eeq
The elements of this matrix are defined as
\ben \label{slel} \beqq
X_\slep &= m_l^2 + m_{\slep_L}^2 + \frac{1}{2} \left( \mz^2 - 2\mw^2
\right) \cbz, \\ 
Y_\slep &= m_l^2 + m_{\slep_R}^2 + \left( \mw^2 - \mz^2 \right) \cbz, \\
|Z_\slep| &= m_l |A_l^\star + \mu\tb|, \\
\arg(Z_\slep^\star) &= \phi_\slep = \arg\left( -A_l - \mu^\star \tb \right),
\eeqq \een
where $m_l$ is the mass of the charged lepton $l$, $m^2_{\slep_{L,R}}$
and $A_l$ are soft SUSY breaking parameters, which we assume to be the
same for the first and second generation, $\mu$ is the Higgsino mass
parameter, and $\tb$ is the ratio of vacuum expectation values (vevs)
of the two neutral Higgs fields. In general, $\mu \equiv |\mu| e^{i
\phi_\mu}$ and $A_l \equiv |A_l| e^{i \phi_A}$ can be complex, while
all other parameters appearing in eqs.(\ref{slel}) are real.

$\cM_\slep^2$ can be diagonalized by a unitary transformation
\beq \label{slepdiag}
U_\slep^\dagger \cM_\slep^2 U_\slep = {\rm diag} \left( m^2_{\slep_1},
m^2_{\slep_2} \right),
\eeq
with the mass ordering $m^2_{\slep_1}\leq m^2_{\slep_2}$ by
convention. The diagonalization matrix  $U_\slep$ can be parameterized
as
\beq \label{Uslep}
U_\slep = \left( \begin{array}{cc} \cos\theta_\slep &
-\sin\theta_\slep e^{-i\phi_\slep} \\ \sin\theta_\slep e^{i\phi_\slep}
& \cos\theta_\slep \end{array} \right),
\eeq
where $-\pi/2 \leq \theta_\slep \leq \pi/2$ and $0 \leq \phi_\slep \leq
2\pi$. Defining
\ben \label{mbar} \beqq
\bar{M}_\slep^2 & \equiv \frac{m^2_{\slep_2} + m^2_{\slep_1} } {2} =
\frac {X_\slep + Y_\slep} {2}, \\
\Delta_{\slep} & \equiv m^2_{\slep_2} - m^2_{\slep_1} = \sqrt{
(X_\slep - Y_\slep)^2 + 4|Z_\slep|^2 }, 
\eeqq \een
the slepton mass eigenvalues and mixing angles are given as 
\ben \label{slepmass} \beqq
m^2_{\slep_{1,2}} &= \bar{M}_\slep^2 \mp \frac{\displaystyle\Delta_{\tilde
    l}} {\displaystyle 2}; \\
\sin2\theta_\slep &= -2 \frac {\displaystyle |Z_\slep| } {\displaystyle
\Delta_{\tilde l} }; \qquad \cos2\theta_\slep = \frac{\displaystyle
X_\slep - Y_\slep } {\displaystyle \Delta_{\tilde l} }.
\eeqq \een
Eqs.(\ref{slepmass}b) and (\ref{slel}c) show that slepton left-right
mixing is suppressed by the corresponding lepton mass, but is enhanced
for large $\tan\beta$ and large $|\mu|$.

As sneutrinos are only present as components of left handed
superfields in the MSSM, there is no partner to mix with and the mass
simply reads
\beq \label{msnu}
m_{\tilde{ \nu}_l}^2 = m_{\tilde{l}_L}^2 + \frac{1}{2} \cos2\beta
\mz^2. 
\eeq

%%%%%%%%%%%%%%%%%%%%%%%%%%%%%%%%%%%%%%%%%%%%%%%%%%%%%%%%%%%%%%%%%
\subsection{Chargino mixing}

The Dirac mass matrix for charginos mixes the $SU(2)$ gaugino $\tilde
w^\pm$ and the charged Higgsinos $\tilde h^\pm$. In the basis $(\tilde
w^-, \tilde h^-)$ it is given by \cite{haberkane}
\beq \label{mc}
\cM_C = \left( \begin{array}{cc} M_2 & \sqrt{2} \mw \cb \\ \sqrt{2}
\mw \sb & \mu \end{array} \right),
\eeq
where the soft breaking mass parameter $M_2$ for $SU(2)$ gauginos is
taken to be real and positive; this can be achieved without loss of
generality by appropriate field redefinitions. This complex mass
matrix is asymmetric and hence has to be diagonalized by a biunitary
transformation
\beq \label{chardiag}
U_R \cM_C U_L^\dagger = {\rm diag} \left( m_{\tilde{\chi}_1^\pm},
m_{\tilde{\chi}_2^\pm} \right),
\eeq 
with the mass ordering $m_{\tilde{\chi}_1^\pm} \leq
m_{\tilde{\chi}_2^\pm}$ as convention. The mixing matrices may be
written as \cite{Choi:2000ta}
\ben \label{uchar} \beqq
U_L &= \left( \begin{array}{cc} \cos\phi_L & \sin\phi_L e^{-i\beta_L}
\\ -\sin\phi_L e^{i\beta_L} & \cos\phi_L \end{array}\right), \\
U_R &= \left( \begin{array}{cc} e^{i\gamma_1} & 0 \\ 0 & e^{i\gamma_2}
\end{array} \right)
\left( \begin{array}{cc} \cos\phi_R & \sin\phi_R e^{-i\beta_R} \\
-\sin\phi_R e^{i\beta_R} &\cos\phi_R \end{array} \right),
\eeqq \een
with $-\pi/2 \leq \phi_{L,R} \leq \pi/2$ and $0 \leq \gamma_{1,2},
\beta_{L,R} \leq 2\pi$. $\gamma_1$ and $\gamma_2$ denote two possible
Dirac phases which have to be introduced to ensure that the mass
eigenvalues of $\cM_C$ are positive and real. The parameters of $U_L$
and $U_R$ can be determined from $\cM_C^\dagger \cM_C$ and $\cM_C
\cM_C^\dagger$, respectively. Introducing the quantity
\beqq \label{deltac}
\Delta_C &= \left\{ \left( M_2^2 - |\mu|^2 \right)^2 + 4 \mw^4 \cos^2
(2\beta) + 4 \mw^2 \left( M_2^2 + |\mu|^2 \right) + 8 \mw^2 |\mu|\cos\phi_\mu
M_2 \sbz \right\}^{1/2} \nonumber \\
&= m^2_{\tilde{\chi}_2^\pm} - m^2_{\tilde{\chi}_1^\pm},
\eeqq
the squared mass eigenvalues are:
\beq \label{charmass} 
m^2_{\tilde{\chi}_{1,2}^\pm} = \frac{1}{2} \left( M_2^2 + |\mu|^2 +
2\mw^2 \mp \Delta_C \right),
\eeq
while the mixing angles can be computed from
\ben \label{charmix} \beqq
\cos2\phi_L &= \frac{\displaystyle -M_2^2 + |\mu|^2 + 2\mw^2 \cbz}
{\displaystyle \Delta_C }, \\
\sin2\phi_L &= \frac{-2 \sqrt{2} \mw} {\Delta_C} \left( M_2^2 \cos^2\beta +
|\mu|^2 \sin^2\beta + M_2 |\mu| \cos \phi_\mu \sbz \right)^{1/2}, \\
\cos2\phi_R &= \frac{\displaystyle -M_2^2 + |\mu|^2 - 2\mw^2 \cbz }{
\displaystyle \Delta_C }, \\
\sin2\phi_R &= \frac{-2 \sqrt{2} \mw} {\Delta_C} \left( M_2^2 \sin^2 \beta +
|\mu|^2 \cos^2 \beta + M_2 |\mu| \cos \phi_\mu \sbz \right)^{1/2},
\eeqq \een
and the phases are
\ben \label{charphase} \beqq
\tan\beta_L &= \frac{ -\displaystyle |\mu| \sin\phi_\mu }{
\displaystyle |\mu| \cos\phi_\mu + \ctb M_2}; \\
\tan\beta_R &= \frac{\displaystyle |\mu| \sin\phi_\mu} {\displaystyle
|\mu| \cos\phi_\mu + \tb M_2}; \\
\cot\gamma_1 &= \frac{\displaystyle \mw^2 |\mu| \cos\phi_\mu \sbz +
M_2 \left( m^2_{\tilde{\chi}_1^\pm} - |\mu|^2 \right) } {\displaystyle
\mw^2 |\mu| \sin\phi_\mu \sin 2\beta }; \\
\cot\gamma_2 &= -\frac{\displaystyle |\mu| \cos\phi_\mu \left(
m_{\tilde{\chi}_2^\pm}^2 - M_2^2 \right) + \mw^2M_2 \sbz} {\displaystyle
|\mu| \sin\phi_\mu \left( m_{\tilde{\chi}_2^\pm}^2 -M_2^2 \right)}.
\eeqq \een

\subsection{Neutralino mixing}
\label{subsec:pertuneut}
The neutralino mass matrix $\cM_N$ mixes the neutral components
of both Higgsinos $\tilde h^0_{u,d}$ with hypercharge $\pm 1/2$, the
$U(1)_Y$ gaugino $\tilde B$ and the neutral $SU(2)$ gaugino $\tilde
W_3$. The mass matrix in the basis $(\tilde B, \tilde W_3, \tilde
h^0_d, \tilde h_u^0)$ reads \cite{haberkane}
\beq \label{mn}
\cM_N = \left( \begin{array}{cccc}
    M_1 & 0 &-\mz\cb\sw & \mz\sb\sw \\
    0 & M_2 &\mz\cb\cw &-\mz\sb\cw \\
    -\mz\cb\sw & \mz\cb\cw & 0 &-\mu \\
    \mz\sb\sw & -\mz\sb\cw & -\mu & 0
\end{array}\right).
\eeq
The $U(1)_Y$ gaugino mass parameter $M_1 \equiv |M_1| e^{i \phi_1}$ is
in general complex. This symmetric mass matrix is diagonalized by a
unitary transformation
\beq \label{neutdiag}
N^T \cM_N N = {\rm diag} (m_{\tilde{\chi}^0_1}, m_{\tilde{\chi}^0_2},
m_{\tilde{\chi}^0_3}, m_{\tilde{\chi}^0_4}), 
\eeq 
i.e. the $n-$th mass eigenstate\footnote{When written as a row vector
$\tilde \chi_n^0$ in the $(\tilde B, \, \tilde W_3, \, \tilde h_d^0,
\, \tilde h_u^0)$ basis, the mass eigenstate satisfies ${\cal M}_N
(\tilde \chi_n^0)^\dagger = m_{\tilde \chi_n^0} (\tilde \chi_n^0)^T$
(no sum over $n$), i.e. it is not an eigenvector of ${\cal M}_N$ in
the usual sense.} is given by the complex conjugate of the $n-$th
column of $N$. Although the neutralino mass matrix can be
diagonalized analytically even for complex parameters
\cite{Choi:2001ww}, the general expressions are too lengthy to
reproduce here. Of course, a numerical computation of $N$ is
straightforward. However, in order to qualitatively understand mixing
in the neutralino sector, a perturbative diagonalization of the mass
matrix (\ref{mn}) is often sufficient. Here $\mz$ is considered to be
a small parameter compared to $|M_1|, \, M_2$ and $|\mu|$. Keeping all
terms up to first order in $\mz$, as well as a few ${\cal O}(M_Z^2)$
terms that will be important later, one finds for the masses and
eigenvectors:
\ben \label{neutdiagapp} \beqq
m_{\tilde \chi_1^0} &\simeq |M_1| + \delta m_1, \ \tilde \chi_1^0 
\simeq e^{i\phi_1/2} \left(1,\, \delta_{12}, \, \delta_{13},\, \delta_{14} 
\right) / N_1; \\
m_{\tilde \chi_2^0} &\simeq M_2 + \delta m_2, \ \ \tilde \chi_2^0
\simeq \left( \delta_{21}, \, 1,\, \delta_{23},\, \delta_{24} \right)
/ N_2; \\ 
m_{\tilde \chi_3^0} &\simeq |\mu|, \ \ \tilde \chi_3^0 \simeq
         \frac{e^{i(\phi_\mu+\pi)/2}}{\sqrt{2}} \left( \delta_{31},\, \delta_{32},\,
1,  \, 1 \right);  \\
m_{\tilde \chi_4^0} &\simeq |\mu|, \ \ \tilde \chi_4^0 \simeq
\frac{e^{i\phi_\mu/2 }}{\sqrt{2}} \left( \delta_{41},\, \delta_{42},\,1, \, -1 \right).
\eeqq \een
In eqs.(\ref{neutdiagapp}) we have assumed the ordering $|M_1| < M_2 <
|\mu|$. If these three mass parameters are ordered differently, the
eigenstates in eqs.(\ref{neutdiagapp}) are no longer labeled in order
of increasing mass. Note that the eigenvalues do not receive ${\cal
O}(\mz)$ corrections. However, mixing between gauginos and Higgsinos
is generated at this order, and $\tilde B - \tilde W_3$ mixing is
generated at order $M_Z^2$. These mixings are described by the complex
quantities $\delta_{ij}$ in eqs.(\ref{neutdiagapp}); they are given
by:
\ben \label{deltaij} \beqq
\delta_{12} &= - \frac {M_Z^2 \sin\theta_W \cos\theta_W \left[ |M_1|^2
+ M_1^\ast M_2 + \sin(2\beta) \left( \mu^\ast M_1^\ast + M_2 \mu
\right) \right]} { \left( M_2^2 - |M_1|^2 \right) \left( |\mu|^2 -
|M_1|^2 \right) }; \\
\delta_{13} &= \frac {\mz \sw \left( M_1^\ast \cb + \mu \sb \right)} 
{|\mu|^2 - |M_1|^2}; \\
\delta_{14} &= - \frac{ \mz \sw \left( M_1^\ast \sb + \mu \cb \right)} 
{|\mu|^2 - |M_1|^2}; \\
\delta_{21} &= \frac{ M_Z^2 \sin\theta_W \cos\theta_W \left[ M_2^2 +
M_1 M_2 + \sin(2\beta) \left( M_2 \mu^\ast + M_1 \mu \right) \right] }
{ \left( M_2^2 - |M_1|^2 \right) \left( |\mu|^2 - M_2^2 \right) }; \\
\delta_{23} &= - \frac{ \mz \cw \left( M_2 \cb + \mu \sb \right) } 
{ |\mu|^2 - M_2^2 }; \\
\delta_{24} &=  \frac{ \mz \cw \left( M_2 \sb + \mu \cb \right) } 
{ |\mu|^2 - M_2^2 }; \\
\delta_{31} &= \frac{ \mz \sw (\sb - \cb) } { \left( |\mu|^2
- |M_1|^2 \right) } \left( M_1 - \mu^\ast \right); \\
\delta_{32} &= \frac{ \mz \cw (\sb - \cb) } { \left( |\mu|^2
- M_2^2 \right) } \left( \mu^\ast - M_2 \right); \\
\delta_{41} &= -\frac{ \mz \sw (\sb + \cb) } { \left( |\mu|^2
- |M_1|^2 \right) } \left( M_1 + \mu^\ast \right); \\
\delta_{42} &= \frac{ \mz \cw (\sb + \cb) } {\left( |\mu|^2
- M_2^2 \right) } \left( \mu^\ast + M_2 \right). 
\eeqq \een
$N_1, \ N_2$ in eqs.(\ref{neutdiagapp}a,b) are normalization
constants, which differ from unity at ${\cal O}(M_Z^2)$. Note that the
phases of the $0-$th order eigenstates have been factored out in
eqs.(\ref{neutdiagapp}); this gives more symmetric looking expressions
for the $\delta_{ij}$. Finally, the ${\cal O}(M_Z^2)$ mass shifts of
the gaugino--like states are given by
\ben \label{deltami} \beqq
\delta m_1 &= - \frac {M_Z^2 \sin^2 \theta_W} {|\mu|^2 - |M_1|^2}
\left[ |M_1| + |\mu| \sin2\beta \cos(\phi_1+\phi_\mu) \right]; \\
\delta m_2 &= - \frac {M_Z^2 \cos^2 \theta_W} {|\mu|^2 - M_2^2}
\left[ M_2 + |\mu| \sin2\beta \cos\phi_\mu \right].
\eeqq \een

Eqs.(\ref{deltaij}) and (\ref{deltami}) show that the expansion will
break down if $|\mu| - |M_1|$ or $|\mu| - M_2$ becomes close to $M_Z$
in absolute value. In other words, even if the unknown dimensionful
parameters in the mass matrix (\ref{mn}) are all $\gg M_Z$, there can
still be strong Higgsino--gaugino mixing if some of these parameters
have similar absolute values.

%%%%%%%%%%%%%%%%%%%%%%%%%%%%%%%%%%%%%%%%%%%%%%%%%%%%%%%%%%%%%%%%%
\subsection{Relevant parameters}
%%%%%%%%%%%%%%%%%%%%%%%%%%%%%%%%%%%%%%%%%%%%%%%%%%%%%%%%%%%%%%%%%

The mixing patterns in the part of the SUSY spectrum which will be
relevant for the remainder of our work depend on 10 SUSY parameters
(plus some SM parameters whose values are already known
accurately). Some of these parameters ($m_{\tilde l_L}, \ m_{\tilde
l_R}, \ |A_l|,\ \phi_A,\ |M_1|, \ \phi_1$) only enter in a single
sector (sleptons and neutralinos, respectively), while $M_2$ appears
in both the chargino and neutralino mass matrices, and $|\mu|, \
\phi_\mu$ and $\tan\beta$ affect all three sectors. Therefore the
mixing patterns in the separate sectors are partly correlated to each
other. In particular, choosing the parameters of the neutralino mass
matrix completely determines the chargino mass matrix as well.
Moreover, increasing $|\mu|$ suppresses gaugino--Higgsino mixing, but
enhances $\tilde l_L - \tilde l_R$ mixing. Finally, taking $\tan \beta
\gg 1$ again enhances $\tilde l_L - \tilde l_R$ mixing, but reduces
the impact of all phases on the physical masses.

%%%%%%%%%%%%%%%%%%%%%%%%%%%%%%%%%%%%%%%%%%%%%%%%%%%%%%%%%%%%%%%%%
\section{Interaction Lagrangian}
\label{sec:feynmanrules}
%%%%%%%%%%%%%%%%%%%%%%%%%%%%%%
\setcounter{footnote}{0}
\setcounter{equation}{0}

\renewcommand{\theequation} {\thesection.\arabic{equation}}

In order to make our paper self--contained, and to fix the notation,
this section is devoted to a short collection of the relevant pieces
of the interaction Lagrangian expressed in terms of physical mass
eigenstates.

%%%%%%%%%%%%%%%%%%%%%%%%%%%%%%%%%%%%%%%%%%%%%%%%%%%%%%%%%%%%%%%%%
\subsection{Interactions involving SM gauge bosons}
%%%%%%%%%%%%%%%%%%%%%%%%%%%%%%%%%%%%%%%%%%%%%%%%%%%%%%%%%%%%%%%%%

First of all, the well--known SM coupling between charged leptons and
gauge bosons is given by
\beq \label{lagsm}
\cL_{l\bar{l}\gamma,Z} = e \bar{l} \gamma^\mu \left( A_\mu
Q^{\alpha,l}_\gamma P_\alpha + Z_\mu Q^{\alpha,l}_Z P_\alpha \right) l, 
\eeq 
where $e$ is the QED coupling constant, and $P_\alpha, \ \alpha \in
\{+,-\} \equiv \{R,L\}$, are standard chirality projection operators,
defined as
\beq \label{projectors}
P_\pm = \frac{1\pm\gamma_5}{2}.
\eeq
The linear charges $Q_{\gamma,Z}^{\alpha,l}$ in eq.(\ref{lagsm}) are
\ben \label{charsm} \beqq
Q_\gamma^{+,l} &= Q_\gamma^{-,l} = 1;\\
Q_Z^{-,l} &= \frac{-1}{\sw\cw} \left( \swq - \frac{1}{2} \right); \\
Q_Z^{+,l} &= -\tw. 
\eeqq\een
Slepton $L-R$ mixing does not affect the couplings between sleptons
and photons. Moreover, in case of selectrons $L-R$ mixing can safely
be neglected for high--energy applications. The vertices with two
charged sleptons and one gauge boson are defined via the
momentum--space Lagrangian
\beq \label{lagslep}
\cL_{\tilde{l}_i\tilde{l}_j\gamma,Z} = e \left( A_\mu
  Q^{\tilde{l},ij}_\gamma + Z_\mu Q^{\tilde{l},ij}_Z \right) 
  (k_i + k_j)^\mu \tilde{l}_i(k_i) \tilde{l}_j(k_j)^\star,
\eeq 
where $i,j \in \{R,L\}$. The corresponding linear charges
$Q^{\tilde{l},ij}_{\gamma,Z}$ are
\ben \label{charslep} \beqq
Q^{\tilde{l},ij}_\gamma &= \delta_{ij}; \\
Q^{\tilde{l},ij}_Z &= -\delta_{ij} \left[ \tw - \frac{1}{2\cw\sw}
\delta_{iL} \right].
\eeqq \een

The couplings between physical Majorana neutralinos and the $Z$ boson
are given by
\beq \label{lagneut}
\cL_{\tilde{\chi}_i^0 \tilde{\chi}_j^0 Z} = \frac {e} {2\cw\sw} Z_\mu
\overline{\tilde{\chi}^0_i} \gamma^\mu Q^{\alpha,ij}_{\tilde{\chi}^0}
P_\alpha \tilde{\chi}_j^0,
\eeq 
where the linear charges $Q^{\alpha,ij}_{\tilde{\chi}^0}$ are defined
as
\beq \label{charneut}
Q^{+,ij}_{\tilde{\chi}^0} = -(Q^{-,ij}_{\tilde{\chi}^0})^\star =
\frac{1}{2} \left( N_{3i} N_{3j}^\star - N_{4i} N_{4j}^\star \right)
\equiv Z_{ij}.
\eeq 
The first equality in eq.(\ref{charneut}) follows from the Majorana
nature of the neutralinos. Of course there is no neutralino--photon
coupling.

Finally, the interactions between neutral gauge bosons and charginos
are given by
\beq \label{lagchar}
\cL_{\tilde{\chi}_i^\pm \tilde{\chi}_j^\mp \gamma,Z} =
e\overline{ \tilde{\chi}_i^-} \gamma^\mu \left(
Q_{\tilde{\chi}^\pm,\gamma}^{\alpha,ij} P_\alpha A_\mu +
Q_{\tilde{\chi}^\pm,Z}^{\alpha,ij} P_\alpha Z_\mu \right)
\tilde{\chi}_j^-,
\eeq
with
\ben \label{charchar} \beqq
Q_{\tilde{\chi}^\pm,\gamma}^{-,ij} &=
Q_{\tilde{\chi}^\pm,\gamma}^{+,ij} = \delta_{ij}; \\
Q_{\tilde{\chi}^\pm,Z}^{\pm,ij} &= \frac{-1}{\cw\sw} \left( \swq
\delta_{ij} - (W_\pm)_{ij} \right).
\eeqq \een
The matrices $(W_\pm)_{ij}$ can be obtained from the chargino mixing
matrices via
\beq 
(W_\pm)_{ij} = (U_\pm)_{i1} (U_\pm)_{j1}^\star + \frac{1}{2}
(U_\pm)_{i2} (U_\pm)_{j2}^\star \;\;\;\; [+,-=R,L]\;\;\;,
\eeq
and read explicitly in terms of chargino mixing angles and phases
as
\ben \label{wdef} \beqq
(W_-) &=
\left(\begin{array}{cc}
\frac{3}{4} + \frac{1}{4} \cos2\phi_L & -\frac{1}{4} \sin2\phi_L
e^{-i\beta_L} \\
-\frac{1}{4} \sin2\phi_L e^{i\beta_L} & \frac{3}{4} - \frac{1}{4}
\cos2\phi_L \end{array} \right); \\
(W_+)&=
\left(\begin{array}{cc}
\frac{3}{4} + \frac{1}{4} \cos2\phi_R & -\frac{1}{4} \sin2\phi_R
e^{i(\gamma_1-\beta_R-\gamma_2)} \\
- \frac{1}{4} \sin2\phi_R e^{-i(\gamma_1-\beta_R-\gamma_2)} &
\frac{3}{4} - \frac{1}{4} \cos2\phi_R 
\end{array}\right).
\eeqq \een
%

%%%%%%%%%%%%%%%%%%%%%%%%%%%%%%%%%%%%%%%%%%%%%%%%%%%%%%%%%%%%%%%%%
\subsection{Slepton interactions with a chargino or neutralino}
%%%%%%%%%%%%%%%%%%%%%%%%%%%%%%%%%%%%%%%%%%%%%%%%%%%%%%%%%%%%%%%%%

The neutralino--slepton--lepton vertices receive contributions from
both gauge and Yukawa interactions:
\beq \label{lagslepneut}
\cL_{\tilde{l}l\tilde{\chi}_i^0} = \frac{e}{\sqrt{2}\sw} \bar{l}
\left( G^\alpha_{ij} + Y^\alpha_{ij} \right) P_\alpha \tilde{\chi}_i^0
\tilde{l}_j + h.c.,
\eeq
with
\ben \label{GY} \beqq
G^-_{ij} &= -2 \tw N_{1i} ( U_{\tilde{l}} )_{2j}; \\
G^+_{ij} &= \left( \tw N_{1i}^\star + N_{2i}^\star \right) (
U_{\tilde{l}} )_{1j}; \\
Y_{ij}^- &= -\sqrt{2} Y_l N_{3i} ( U_{\tilde{l}} )_{1j}; \\
Y_{ij}^+ &= -\sqrt{2} Y_l N_{3i}^\star ( U_{\tilde{l}} )_{2j}.
\eeqq \een
Here the dimensionless, rescaled Yukawa coupling $Y_l$ is given by
\beq \label{defY}
Y_l = \frac{m_l} {\sqrt{2}\mw\cb}.
\eeq
Note that we have to keep terms $\propto Y_l$, as well as a
non--trivial sleptonic mixing matrix $U_\slep$, when computing
leptonic dipole moments. On the other hand, for high--energy
applications $Y_e$ can be set to zero, which implies $Y_{ij}^\pm = 0$
in case of selectrons. In the same limit $L-R$ mixing can be
neglected, in which case the gauge contributions $G^\pm_{ij}$
simplify to
\ben \label{defG} \beqq
G^-_{ij} &= -2\tw N_{1i} \delta_{Rj};\\
G^+_{ij} &= \left( \tw N_{1i}^\star + N_{2i}^\star \right) \delta_{Lj}.
\eeqq \een

The couplings between chargino, sneutrino and lepton also receive
gauge and Yukawa contributions:
\beq \label{lagsneutchar}
\cL_{\tilde{\nu}_l l \tilde{\chi}_i^\pm} = \frac{e}{\sw}
\overline{\tilde{\chi}^-_i} N_{\alpha,i} P^\alpha l
\tilde{\nu}_l^\star  + h.c.;
\eeq 
where 
\beq \label{defN}
N_{\alpha,i} = -\delta_{\alpha L} (U_R)_{i1} + \delta_{\alpha R}
(U_L)_{i2} Y_l.
\eeq 
As before, the term $\propto Y_e$ in eq.(\ref{defN}) can be dropped in
collider physics applications, but it has to be kept when computing
the leptonic dipole moments.

%%%%%%%%%%%%%%%%%%%%%%%%%%%%%%%%%%%%%%%%%%%%%%%%%%%%%%%%%%%%%%%%%
\section{Low energy constraints}
\label{sec:lowenergy}
%%%%%%%%%%%%%%%%%%%%%%%%%%%%%%
\setcounter{footnote}{0}
\setcounter{equation}{0}
\renewcommand{\theequation} {\thesection.\arabic{equation}}
%%%%%%%%%%%%%%%%%%%%%%%%%%%%%%%%%%%%%%%%%%%%%%%%%%%%%%%%%%%%%%%%%

\subsection{Experimental constraints}

In this paper we are only interested in purely leptonic processes. We
therefore ignore the (quite stringent) experimental constraints on the
electric dipole moments of the neutron and mercury atom. The main
reason for this choice is that leptonic processes suffer much less
from uncertainties due to non--perturbative strong interactions. For
example, ref.\cite{Abel} finds that different models relating electric
dipole moments of quarks to that of the neutron or Hg atom differ by
typically a factor of two. Since large phases in the hadronic sector
can be tolerated if there are cancellations between different
contributions, which have different hadronic matrix elements, a
conservative interpretation of the experimental bounds on $d_n$ tends
to give \cite{kane} significantly weaker constraints on model
parameters than the bound on the electric dipole moment of the
electron does, even if one assumes some connection between the
CP--violating phases in the squark and slepton sectors. The only
CP--violating (more exactly, T--violating) low--energy quantity of
relevance to us is therefore the electric dipole moment of the
electron $d_e$. Given our assumption of flavor universality of the
soft breaking terms in the slepton sector, at least as far as the
first and second generation are concerned, the bound on the electric
dipole moment of the muon \cite{pdg} need not be considered
separately: since $(d_l)_{\rm SUSY} \propto m_l$, all combinations of
parameters that satisfy the constraint on the SUSY contribution to
$d_e$ will be at least five orders of magnitude below the maximal
allowed SUSY contribution to $d_\mu$.

On the other hand, our assumption of universal sleptonic soft breaking
terms for the first two generations also implies \cite{Graesser} that
the measurement \cite{brown} of the anomalous magnetic moment of the
muon, $a_\mu \equiv (g_\mu-2)/2$, gives a {\em tighter} constraint on
SUSY parameters than the anomalous magnetic moment of the electron
does. The reason is that for universal soft breaking masses the SUSY
contribution to these leptonic magnetic moments is essentially
proportional to the squared mass of the lepton, and the experimental
errors satisfy \cite{pdg,brown} $\delta a_\mu / m^2_\mu < \delta a_e /
m^2_e$. The second low--energy quantity of relevance to us is
therefore $a_\mu$. Note that the SUSY contributions to $a_\mu$ and
$d_e$ show very similar dependences on the absolute values of the
relevant parameters; however, $d_e$ receives nonvanishing
contributions only in the presence of nontrivial phases, while the
contribution to $|a_\mu|$ becomes maximal if all phases are zero or
$\pi$.

The SM prediction for $d_e$ is negligible. The current experimental
measurement \cite{pdg}
\beq \label{deexp}
(d_e)_{\rm exp} = (0.069\pm0.074)\times 10^{-26}\; e \cdot {\rm cm}
\eeq
can therefore directly be translated into a $2 \sigma$ range for the
supersymmetric contribution to $d_e$:
\beq \label{derange}
-0.079 \times 10^{-26} \; e \cdot {\rm cm} \leq (d_e)_{\rm SUSY} \leq
0.217 \times 10^{-26} \; e \cdot{\rm cm}.
\eeq

The interpretation of the most recent measurement \cite{brown} of
$a_\mu$,
\beq \label{amuexp}
(a_\mu)_{\rm exp} = (11659208 \pm6) \times 10^{-10},
\eeq
is less clear. The reason is that non--perturbative hadronic terms do
contribute to $a_\mu$, at about the $10^{-8}$ level. In principle this
contribution can be calculated from experimental data using dispersion
relations \cite{davier,kinoshita}. Unfortunately calculations based on
different data do not quite agree, although the discrepancy has become
smaller after the recent release of corrected data by the CMD--2
collaboration \cite{cmd2}. Using $e^+e^-$ annihilation data as input
tends to give an SM prediction which falls a little short of the
experimental value (\ref{amuexp}). A recent analysis which includes
all existing $e^+e^-$ data \cite{hoecker_new} finds
\beq \label{amuee}
(a_\mu)_{\rm SM} = (11659180.9 \pm 8.0) \times10^{-10}.
\eeq
Adding all errors in quadrature, this gives a $\sim 2.7 \sigma$
discrepancy. On the other hand, using data from $\tau$
decays gives \cite{hoecker_new}
\beq \label{amutau}
(a_\mu)_{\rm SM} = (11659195.6 \pm 6.8) \times 10^{-10},
\eeq
which is only $\sim 1.4 \sigma$ below the measurement
(\ref{amuexp}). Since even the $e^+e^-$ data lead to a less than
3$\sigma$ discrepancy between the prediction for and measurement of
$a_\mu$, we do not want to claim evidence for a non--vanishing SUSY
contribution. In order to be conservative, we construct the upper
limit of the ``2$\sigma$ allowed'' range for $(a_\mu)_{\rm SUSY} =
(a_\mu)_{\rm exp} - (a_\mu)_{\rm SM}$ by using the lower value
(\ref{amuee}), reduced by two combined standard deviations, as our
estimate of $(a_\mu)_{\rm SM}$. Similarly, the lower end of this
``2$\sigma$ range'' is obtained when $(a_\mu)_{\rm SM}$ is estimated
by adding two standard deviations to the higher value
(\ref{amutau}). This gives:
\beq \label{amurange}
-5.7 \times 10^{-10} \leq (a_\mu)_{\rm SUSY} \leq 47.1 \times
10^{-10} .
\eeq
The upper bound in (\ref{amurange}) constrains the SUSY parameter space
only for large values of $\tan\beta$, but the lower bound is significant
also for moderate $\tan\beta$.

%============================================================================
\subsection{Analytical results}

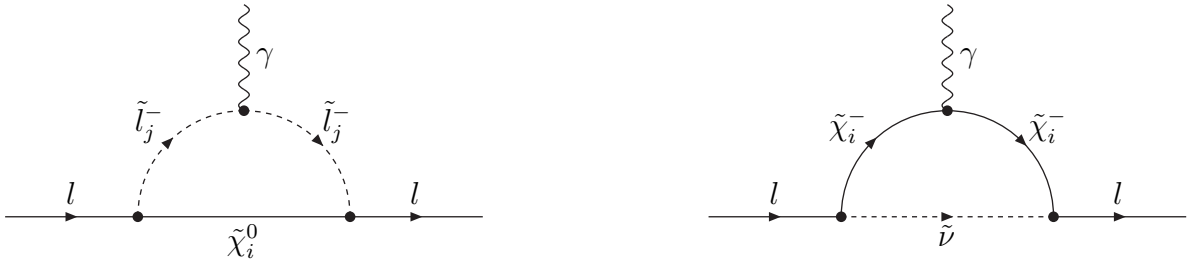
\begin{figure}[ht]
\begin{minipage}{0.45\textwidth}
\begin{picture}(190,100)
\ArrowLine(10,15)(60,15)
\ArrowLine(140,15)(190,15)
\Line(60,15)(140,15)
\Vertex(60,15){2}
\Vertex(140,15){2}
\DashArrowArcn(100,15)(40,90,0){2}
\DashArrowArcn(100,15)(40,180,90){2}
\Vertex(100,55){2}
\Photon(100,55)(100,95){2}{5}
\Text(35,20)[bc]{$l$}
\Text(165,20)[bc]{$l$}
\Text(100,12)[tc]{$\tilde{\chi}_i^0$}
\Text(70,43)[br]{$\tilde{l}_j^-$}
\Text(130,43)[bl]{$\tilde{l}_j^-$}
\Text(105,75)[lc]{$\gamma$}
\end{picture}\end{minipage}\hfill
\begin{minipage}{0.45\textwidth}
\begin{picture}(190,100)
\ArrowLine(10,15)(60,15)
\DashArrowLine(60,15)(140,15){2}
\ArrowLine(140,15)(190,15)
\Vertex(60,15){2}
\Vertex(140,15){2}
\ArrowArcn(100,15)(40,90,0)
\ArrowArcn(100,15)(40,180,90)
\Vertex(100,55){2}
\Photon(100,55)(100,95){2}{5}
\Text(35,20)[bc]{$l$}
\Text(165,20)[bc]{$l$}
\Text(100,12)[tc]{$\tilde{\nu}$}
\Text(70,43)[br]{$\tilde{\chi}_i^-$}
\Text(131,43)[bl]{$\tilde{\chi}_i^-$}
\Text(105,75)[lc]{$\gamma$}
\end{picture}\end{minipage}

\caption{SUSY contributions to leptonic dipole operators}
\label{dipolediagrams}
\end{figure}

The supersymmetric one--loop contributions to lepton dipole moments are
shown in Fig.~\ref{dipolediagrams}. The left diagram depicts the
neutralino contribution while the right one contains the chargino
contribution. Using the interaction Lagrangians given in
Section~\ref{sec:feynmanrules} we find for the chargino contribution
to the electric dipole moment of the electron
\beq \label{dechar}
\left(\frac{d_e}{e}\right)_{\rm SUSY}^{\tilde{\chi}^\pm} = \frac {1}{96\pi^2}
\sum_{i=1}^2 \frac {2} {m_{\tilde{\chi}_i^\pm}} f_1(x_i) \imag(
c_{Li}^\star c_{Ri} ).
\eeq
The chargino loop contribution to the magnetic dipole moment of the
muon is
\beq \label{amuchar}
(a_\mu)_{\rm SUSY}^{\tilde{\chi}^\pm} = \frac {1} {192\pi^2}
\sum_{i=1}^2 \left\{ \frac {8 m_\mu} {m_{\tilde{\chi}_i^\pm}} f_1(x_i)
\real ( c_{Li}c_{Ri}^\star ) + \frac {m_\mu^2}
{m_{\tilde{\chi}_i^\pm}^2} f_3(x_i) (|c_{Li}|^2+|c_{Ri}|^2) \right\}.
\eeq
The corresponding results for the neutralino contribution read
\beq \label{deneut}
\left(\frac{d_e}{e}\right)_{\rm SUSY}^{\tilde{\chi}^0} = \frac {-1} {96\pi^2}
\sum_{i=1}^4 \sum_{\alpha=1}^2 \frac {1} {m_{\tilde{\chi}_i^0}}
f_2(y_{i\alpha}) \imag( n_{Li\alpha}^\star n_{Ri\alpha});
\eeq
\beq \label{amuneu}
(a_\mu)_{\rm SUSY}^{\tilde{\chi}^0} = \frac {-1} {192\pi^2}
\sum_{i=1}^4 \sum_{\alpha=1}^2 \left\{ \frac {4m_\mu}
{m_{\tilde{\chi}_i^0}} f_2(y_{i\alpha}) \real( n_{Li\alpha}^\star
n_{Ri\alpha} ) + \frac {m_\mu^2} {m_{\tilde{e}_\alpha}^2}
f_3(y_{i\alpha}) ( |n_{Li\alpha}|^2 + |n_{Ri\alpha}|^2 )\right\}.
\eeq
The variables $x_i$ and $y_{i\alpha}$ are defined as
\beq \label{exy}
x_i = \frac {m^2_{\tilde{\chi}^\pm_i}} {m^2_{\tilde \nu}}, \;\;\;\;\;\;\;\;
y_{i\alpha} = \frac{m_{\tilde l_\alpha}^2}{m_{\tilde \chi^0_i}^2},
\eeq
and the loop functions $f_i$ are
\ben \label{efi} \beqq
f_1(z) &= \frac {3z} {2(z-1)^3} ( z^2 - 4z + 3 + 2\log z )\, ; \\
f_2(z) &= \frac {3} {(z-1)^3} ( z^2 - 1 - 2 z\log z )\, ;\\
f_3(z) &= \frac {2z} {(z-1)^4} ( z^3 - 6z^2 + 3z + 2 + 6z\log z )\, .
\eeqq \een
These functions are normalized such that $f_i(1) = 1, \ i=1, 2,
3$. Finally the coupling coefficients $c_{Ai}$ and $n_{Ai\alpha}$ can
be written as:
\ben \label{ecn} \beqq
c_{Li} &= -\frac {e}{\sw} \left( U_R \right)_{i1}; \\
c_{Ri} &= \frac {e}{\sw} Y_l \left( U_L \right)_{i2}; \\
n_{Li\alpha} &= \frac {e}{\sqrt{2}\sw} \left[ (N_{2i} + \tw N_{1i})
\left( U_{\tilde l} \right)_{L\alpha}^\star - \sqrt{2} Y_l N_{3i}
\left( U_{\tilde l} \right)_{R\alpha}^\star \right]; \\
n_{Ri\alpha} & = -\frac {e} {\sqrt{2}\sw} \left[ 2\tw N_{1i}^\star
\left( U_{\tilde l} \right)_{R\alpha}^\star + \sqrt{2} Y_l N_{3i}^\star
\left( U_{\tilde l}\right)_{L\alpha}^\star \right].
\eeqq\een
Our results agree with those of refs.  \cite{kane},\cite{dipolcheck};
the neutralino contribution in ref.\cite{bartl} seems to have some
misprints.

The analytic expressions of Sec.~2.2 can be used to rewrite both
chargino contributions in terms of the loop functions $f_i$ and the
basic SUSY parameters:
\beq \label{deana}
\left(\frac{d_e}{e}\right)_{\rm SUSY}^{\tilde{\chi}^\pm} = -\frac {m_e}{48\pi^2} \frac
{e^2} {\swq} \frac {\tb |\mu| M_2\sin\phi_\mu} {\Delta_C} \sum_{i=1}^2
(-1)^i \frac {f_1(x_i)} {m_{\tilde{\chi}_i^\pm}^2},
\eeq 
\beqa \label{amuana}
(a_\mu)_{\rm SUSY}^{\tilde{\chi}^\pm} &=& - \frac{m_\mu^2}{96\pi^2}
\frac {e^2} {\swq} \left\{ 2\sum_{i=1}^2 \frac {f_1(x_i)} {m_{\tilde
\chi_i^\pm}^2} - \frac {1} {4} ( 1 + Y_\mu^2 ) \sum_{i=1}^2 \frac
{f_3(x_i)} {m_{\tilde \chi_i^\pm}^2} \right. \nonumber\\
&+& \left. 2 \left[ M_2^2 + |\mu|^2 + 2\tb M_2 |\mu| \cos\phi_\mu +
2 \mw^2 \cbz \right] \sum_{i=1}^2 \frac {(-1)^i f_1(x_i)} {\Delta_C
m_{\tilde{\chi}_i^\pm}^2} \right.\\
&-& \left. \frac {1}{4} \left[ ( M_2^2 - |\mu|^2 )( 1 - Y_\mu^2 ) +
2\mw^2 \cbz (1 + Y_\mu^2) \right] \sum_{i=1}^2 \frac{(-1)^i f_3(x_i)} {
\Delta_C m_{\tilde{\chi}_i^\pm}^2 } \right\},\nonumber 
\eeqa
where $\Delta_C$ has been defined in eq.(\ref{deltac}). Together with
eqs.(\ref{efi}) these expressions explicitly show that the chargino
contributions to the leptonic dipole moments decouple like
$1/m^2_{\tilde \chi^\pm}$ for $m^2_{\tilde \chi^\pm} \gg m^2_{\tilde
\nu}$, and like $1/m^2_{\tilde \nu}$ in the opposite limit
$m^2_{\tilde \nu} \gg m^2_{\tilde \chi^\pm}$. For completeness we have
included terms $\propto Y_\mu^2$, even though eq.(\ref{defY}) shows
that $Y_\mu^2 \ll 1$; if these terms are neglected, $( a_\mu )_{\rm
SUSY}^{\tilde \chi^\pm} \propto m_\mu^2$, as advertised earlier.

Analogous statements also hold for the neutralino contributions, but
because of the more complicated nature of neutralino mixing we were not
able to find simple exact analytic expressions for these
contributions. However, with the help of eqs.(\ref{neutdiagapp}) and
(\ref{deltaij}) one can derive an approximate expression for the
neutralino loop contribution to $d_e$:
\beqq \label{deneutapp}
\left(\frac{d_e}{e}\right)^{\tilde \chi^0}_{\rm SUSY} &\simeq - \frac {e^2 m_e}
{96 \pi^2} \left\{ \frac { \left| A_e^\ast + \mu \tan\beta \right| } 
{\cos^2 \theta_W |M_1| } \cdot \frac { f_2(m^2_{\tilde e_R}/|M_1|^2) -
f_2(m^2_{\tilde e_L}/|M_1|^2) } { m^2_{\tilde e_L} - m^2_{\tilde e_R}
} \cdot \sin \left( \phi_1 - \phi_{\tilde e} \right)
 \right. \nonumber \\ & \left.
\hspace*{19mm} + \, \frac { \tan\beta \sin \left( \phi_\mu + \phi_1 \right) }
{ \cos^2 \theta_W |M_1 \mu| \left( |\mu|^2 - |M_1|^2 \right) }
\left[ |\mu|^2 \left( f_2(m^2_{\tilde e_R}/|M_1|^2) - \frac{
f_2(m^2_{\tilde e_L}/|M_1|^2) } {2} \right) 
\right. \right. \nonumber \\ & \left. \left.
\hspace*{73mm} - |M_1|^2 \left( f_2(m^2_{\tilde e_R}/|\mu|^2) - \frac
{f_2(m^2_{\tilde e_L}/|\mu|^2)} {2} \right) \right]
\right. \nonumber \\ & \left.
\hspace*{19mm} +\, \frac { \tan\beta \sin \phi_\mu \left[ |\mu|^2
f_2(m^2_{\tilde e_L}/M_2^2) - M_2^2 f_2(m^2_{\tilde e_L}/|\mu|^2) 
\right] } {2 \sin^2 \theta_W M_2 |\mu| \left( |\mu|^2 - M_2^2 \right)
}  \right\}.  
\eeqq
In the first line of eq.(\ref{deneutapp}) we have used an approximate
treatment of selectron mixing, which is quite sufficient for the given
purpose.\footnote{The electric dipole moment is chirality violating
and hence proportional to the Yukawa coupling. Therefore slepton
mixing, which is proportional to the Yukawa coupling, cannot be
neglected.} The last line in eq.(\ref{deneutapp}), which involves the
$SU(2)$ gauge interactions, has a very similar structure as the
chargino loop contribution (\ref{deana}); however, the overall factor
in front of the neutralino contribution is four times smaller than
that of the chargino contribution. Note also that eq.(\ref{deneutapp})
does not contain contributions $\propto \sin \phi_1$, which in our
convention measures the relative phase between $M_1$ and $M_2$; only
the phase of $M_1$ relative to either the phase $\phi_{\tilde e}$ in
selectron mixing or to the phase of $\mu$ is relevant.

%=======================================================================
\subsection{Numerical analysis}
\label{sec:lowenernum}

As well known \cite{oldde, petcov, Kizukuri:nj,kane,nath}, the
experimental bound (\ref{deexp}) on $d_e$ provides stringent
constraints on MSSM parameter space. For example, the chargino
contribution (\ref{deana}) to $d_e$ can be estimated to be
\beq \label{deest}
\left( d_e \right)_{\rm SUSY}^{\tilde \chi^\pm} \sim 3 \cdot
10^{-24} \cdot \tan\beta \sin \phi_\mu \cdot
\left( \frac {100 \, {\rm GeV} } {m_{\rm SUSY}} \right)^2  \, e \cdot
{\rm cm} , 
\eeq
where $m_{\rm SUSY}$ stands for the relevant sparticle (sneutrino or
chargino, whichever is heavier) mass scale. The chargino contribution
by itself can therefore satisfy the experimental constraint
(\ref{derange}) only for very small phase $\phi_\mu$ and/or very large
sparticle masses. For sparticle masses not much above 100 GeV, one
would need phases of order $10^{-3}$ ($10^{-2}$) or less in the
chargino (neutralino or slepton) mass matrices; if $\tan\beta \gg 1$,
this constraint would become even stronger. Such small phases are
unlikely to lead to measurable effects in high--energy collider
experiments \cite{barger1, barger}. Alternatively one can postulate
that sparticle masses are very large \cite{Kizukuri:nj}. Since gaugino
masses are coupled to parameters in the Higgs sector via one--loop
renormalization group equations, whereas a similar coupling between
first generation sfermion masses and the Higgs sector only exists at
two--loop level \cite{RGE}, naturalness arguments favor models with
large slepton masses and relatively modest gaugino masses. The
estimate (\ref{deest}) indicates that first generation slepton masses
well above 1 TeV would be required if the relevant phases are ${\cal
O}(1)$. In that case these sleptons would be beyond the reach of the
next linear $e^+e^-$ collider, which will have center--of--mass energy
$\sqrt{s} \lsim 1$ TeV. Moreover, since FCNC constraints would then
also indicate very large masses for second generation sleptons (recall
that we assume them to be exactly degenerate with the first
generation), a possible excess in $a_\mu$ could not be accommodated
within the MSSM.

We therefore focus on the third possibility for satisfying the
constraint (\ref{derange}), where the different contributions to $d_e$
largely cancel \cite{nath,kane}; that is, the neutralino contribution
must cancel the chargino contribution. Here we quantitatively analyze
this possibility for three scenarios; later we will analyze
high--energy observables that are sensitive to phases within the same
scenarios.

In all cases we assume that the ratio of $M_2$ and $|M_1|$ is similar
to that in models with gaugino mass unification at the GUT scale,
which predicts \cite{RGE} $|M_1| \simeq 0.5 M_2$ at the weak scale.
Similarly, we take values for the soft breaking masses of $SU(2)$
singlet and doublet sleptons that are consistent with the assumption
of universal scalar masses at the GUT scale. Recall that we assume
degenerate first and second generation soft breaking parameters in the
slepton sector:
\ben \label{degen} \beqq 
m_{\tilde{e}_L} &= m_{\tilde{\mu}_L} \equiv m_{\tilde{l}_L}; \\
m_{\tilde{e}_R} &= m_{\tilde{\mu}_R} \equiv m_{\tilde{l}_R}; \\
A_e &= A_\mu \equiv A.
\eeqq \een
The assumption of universal scalar masses at the GUT scale implies
\cite{RGE} that 
\beq \label{slepmasses}
m^2_{\tilde l_L} \simeq m^2_{\tilde l_R} + 0.46 M_2^2
\eeq
at the weak scale. Finally, we are interested in scenarios where at
least $\tilde l_R, \, \tilde l_L, \, \tilde \chi_1^\pm$ and $\tilde
\chi_2^0$ can be pair--produced at a ``first stage'' linear collider
operating at $\sqrt{s} = 500$ GeV.

This leads us to consider three different scenarios, which we call B1,
B2 and B3. Scenario B1 has $|\mu| = M_2$, i.e. is characterized by
strong mixing between $SU(2)$ gauginos and Higgsinos; this will occur
in both the chargino and neutralino sector. Contrariwise, B2 has
$|\mu|^2 \gg M_2^2$, i.e. all Higgsino--gaugino mixing is
suppressed. In these two cases we take a relatively large value of
$|A|$, which enhances slepton $L-R$ mixing for small $\tan\beta$; we
will see shortly that this increases the possibility of cancellations
between the chargino and neutralino contributions to $d_e$. On the
other hand, $\tilde e_L - \tilde e_R$ mixing, while important for the
calculation of $d_e$, in all cases remains negligible as far as
selectron production at high energies is concerned. Case B3, which is
almost\footnote{The agreement becomes exact for the ``benchmark
value'' $\tan\beta = 10$ and vanishing phases.} identical to the
much--studied Snowmass ``benchmark point SPS1A'' \cite{sps1a}, has
intermediate gaugino--Higgsino mixing, as well as slightly reduced
slepton masses. In all three cases we take four different values of
$\tan\beta$. Moreover, we allow the three relevant phases $\phi_1, \,
\phi_\mu$ and $\phi_A$ to float freely, i.e. we pick random values for
these phases. (Recall that we work in a phase convention where $M_2$
is real and positive.) These three scenarios are summarized in
Table~1. Of course, we respect all relevant limits from direct
searches for superparticles at colliders, in particular at LEP2
\cite{pdg}.

\begin{center}
\begin{table}[ht!]
\begin{center}
\begin{tabular}{|c|c|c|c|c|c|c|c|c|c|}\hline
\rule[0pt]{0pt}{13pt}& $|M_1|$ & $M_2$ & $m_{\tilde l_L}$ & $m_{\tilde
l_R}$ & $|A|$ & $|\mu|$ & $\tan\beta$ & $\phi_1,\, \phi_\mu,\,
\phi_A$\\ \hline
B1 & 100 & 200 & 235 & 180 & 500 & 200 & 3, 6,9, 12 & $ \in
[-\pi,\pi]$ \\ \hline
B2 & 100 & 200 & 235 & 180 & 500 & 500 & 3, 6, 9, 12 & $\in
[-\pi,\pi]$ \\ \hline
B3 & 102.2 & 191.8 & 198.7 & 138.2 & 255.5 & 343.2 & 5, 10, 15, 20 &
$\in [-\pi,\pi]$ \\ \hline
\end{tabular}
\end{center}
\caption{The three scenarios studied in this paper. All dimensionful
parameters are in GeV.}
\label{scenarios}
\end{table}
\end{center}

\setcounter{footnote}{0}

For simplicity and limited space of representation we only show
results for two choices of $\tan\beta=3$ or 12 in scenarios B1 and B2,
and for $\tan\beta=10$ or 20 in case B3. Results for the other cases
are qualitatively similar and can be obtained by extrapolation from
these extreme cases. Note that the numerical results shown below are
projections of a three--dimensional parameter space onto
two--dimensional planes. Hence it should be kept in mind that each
correlation in the $\phi_x-\phi_y$ plane has been obtained by scanning
over the entire allowed range for $\phi_z$. By far the strongest
restriction on parameter space comes from $d_e$: at least 99.4\% of
all randomly chosen points in a given run violate the constraint
(\ref{derange}); the success rate at large $\tan\beta$ is even
smaller.\footnote{This indicates that rather severe fine-tuning is
required \cite{barger} to obtain the necessary cancellations if all
phases are indeed independent quantities. To put it differently, one
faces the challenge to construct models that ``naturally'' explain the
required correlations between these phases. We will not attempt to do
this here.} In the following we will quote upper bounds on
$|\phi_\mu|$ that result from the constraint (\ref{derange}). A
similar band around $\phi_\mu = \pi$ exists for small $\tan\beta$ and
large $|\mu|$.

\begin{figure}[ht!]
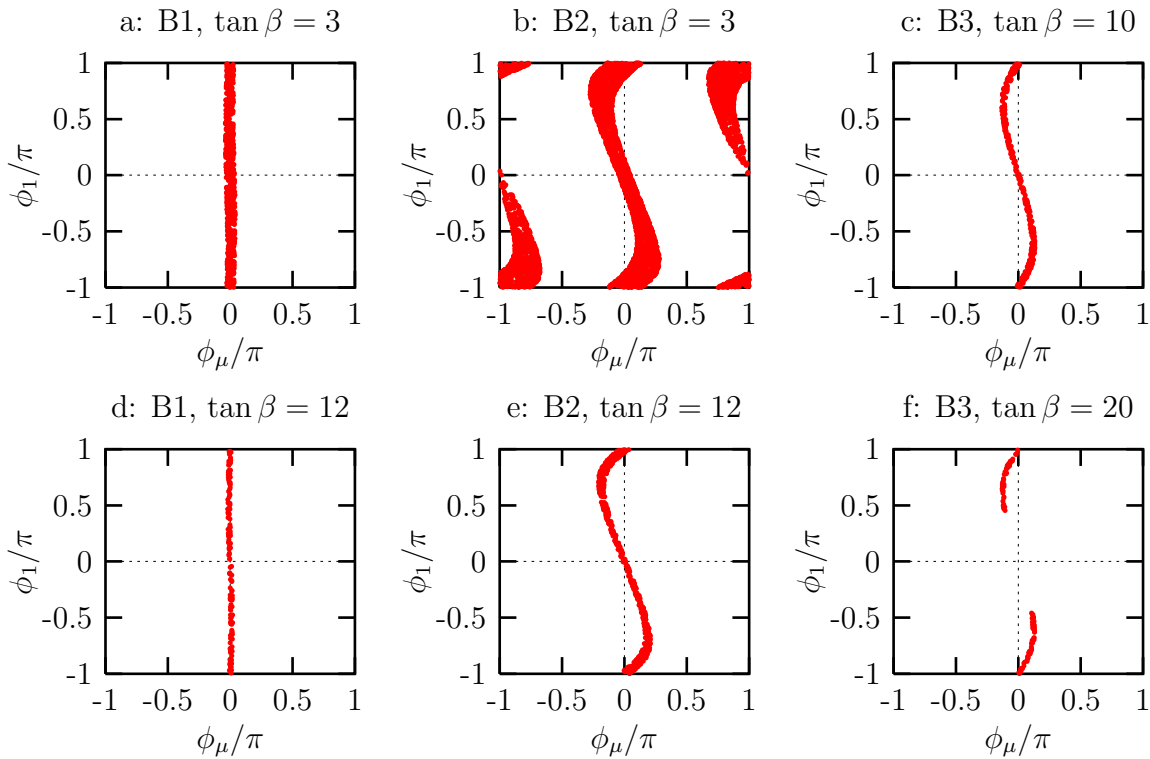

\begin{minipage}[ht!]{0.3\textwidth}
\input{phmuph1B1tan3_tex.tex}
\end{minipage}
\begin{minipage}[ht!]{0.3\textwidth}
\input{phmuph1B2tan3_tex.tex}
\end{minipage}
\begin{minipage}[ht!]{0.3\textwidth}
\input{phmuph1B3tan10_tex.tex}
\end{minipage}\hfill\\
\begin{minipage}[ht!]{0.3\textwidth}
\input{phmuph1B1tan12_tex.tex}
\end{minipage}
\begin{minipage}[ht!]{0.3\textwidth}
\input{phmuph1B2tan12_tex.tex}
\end{minipage}
\begin{minipage}[ht!]{0.3\textwidth}
\input{phmuph1B3tan20_tex.tex}
\end{minipage}\hfill
\caption{Combinations of $\phi_\mu$ and $\phi_1$ that are allowed for
at least one $\phi_A \in [-\pi,\pi]$.} 
\label{phimuphi1}
\end{figure}

Fig.~\ref{phimuphi1} shows allowed combinations of the phases
$\phi_\mu$ and $\phi_1$. We observe very strong constraints on
$\phi_\mu$ in scenario B1, which become stronger as $\tan\beta$
increases. Scenario B2 allows much larger values of $|\phi_\mu|$,
which moreover do not decrease much with increasing $\tan\beta$, while
scenario B3 is intermediate between these two. This behavior can be
understood from eqs.(\ref{deana}) and (\ref{deneutapp}). We saw that
the contributions involving $SU(2)$ gauge interactions have very
similar structure in both cases, but the chargino loop contribution is
bigger by a factor $\sim 4$ than this part of the neutralino
contribution. The potentially most important cancellation therefore
occurs between the chargino contribution [more exactly: the total
contribution involving $SU(2)$ interactions, which is however always
dominated by the chargino contribution] and the neutralino
contributions involving $U(1)_Y$ interactions.

Scenario B1 has $|\mu| = M_2$, i.e. very strong mixing between
Higgsinos and $SU(2)$ gauginos. Eq.(\ref{deneutapp}) no longer gives
an accurate estimate of the neutralino contribution in this limit, but
we expect it to remain qualitatively correct; note that it gives a
finite answer (involving the derivative of the function $f_2$) in this
case. In particular, the contributions involving the $SU(2)$ gauge
coupling would be much bigger than those involving $U(1)_Y$
interactions if the relevant phases had similar magnitude; in other
words, a significant cancellation can only occur if $|\phi_\mu|$ is
well below $|\phi_1|$. Furthermore, for this choice of parameters a
strong internal cancellation occurs between the two contributions from
$U(1)_Y$ interactions that grow $\propto \tan\beta$, i.e. between the
first line and the following two lines on the r.h.s. of
eq.(\ref{deneutapp}). As a result we find $|\phi_\mu| \leq \pi/30$
even for $\tan\beta = 3$. Moreover, the dominant contribution from
$U(1)_Y$ interactions in this scenario involves $A$, i.e. is
independent of $\tan\beta$, whereas the contribution from $SU(2)$
interactions increases $\propto \tan\beta$. The upper bound on
$|\phi_\mu|$ therefore scales essentially like $\cot\beta$. The
importance of $\phi_A$ in this scenario also explains why there is
almost no correlation between the allowed values of $\phi_\mu$ and
$\phi_1$. Moreover, in this scenario values of $\phi_\mu$ near $\pi$
are excluded by the lower bound (\ref{amurange}) on $(a_\mu)_{\rm
SUSY}$.

Eq.(\ref{deana}) shows that increasing $|\mu|$ while keeping all other
parameters the same decreases the chargino contribution to $d_e$;
according to eq.(\ref{deneutapp}) it also decreases the neutralino
contributions that involve $SU(2)$ interactions, but it actually {\em
increases} the neutralino contribution that is sensitive to $\tilde
e_L - \tilde e_R$ mixing, i.e. the first line in
eq.(\ref{deneutapp}).\footnote{In principle one can therefore have
large cancellations between chargino and neutralino contributions even
for $M_2 \simeq |\mu|$, if $M_2 \simeq m_{\tilde \nu} \gg |M_1|,
m_{\tilde e_R}$. However, if $M_2$ and $m_{\tilde \nu}$ are as in
scenario B1, this would require values of $m_{\tilde e_R}$ well below
the direct search limit of $\sim 100$ GeV.} Much larger values of
$|\phi_\mu|$ therefore now become possible. For the given choice of
parameters the coefficient of the neutralino contribution $\propto
\sin(\phi_1 - \phi_{\tilde e})$ is still about 5 times smaller than
the coefficient of $\sin\phi_\mu$ in the chargino contribution,
leading to an upper limit of $\sim \pi/4$ on $|\phi_\mu|$. Since these
two coefficients have the same sign, cancellations obtain only if
$\phi_1+\phi_\mu$ and $\phi_\mu$ have opposite signs. Note that both of
these contributions are (essentially) $\propto \tan\beta$. The upper
bound on $|\phi_\mu|$ is therefore now almost independent of
$\tan\beta$. However, one needs increasingly more perfect
cancellations as $\tan\beta$ increases; moreover, the relative
importance of the phase $\phi_A$ diminishes with increasing
$\tan\beta$, since its contribution to $\tilde e_L - \tilde e_R$
mixing is not enhanced in this limit. These two considerations explain
why the width of the allowed band decreases essentially like
$\cot\beta$ for large $\tan\beta$.

The increase of $|\mu|$ when going from scenario B1 to B2 also reduces
the supersymmetric contribution to $a_\mu$. For $\tan\beta = 3$ we
therefore now also find an allowed band with $\phi_\mu \simeq \pi$;
however, this band disappears at $\tan\beta \sim 10$. Note that the
phase $\phi_1$ enters $a_\mu$ mostly in the combination\footnote{For
$|\mu|\tan\beta \gg |A|$, $\cos(\phi_1 - \phi_{\tilde \ell})
\simeq \cos(\phi_1 + \phi_\mu)$ as well.} $\cos(\phi_1+\phi_\mu)$.
This means that $\phi_1 \simeq 0$ will give positive (negative)
contributions to $a_\mu$ if $\phi_\mu \simeq 0 \ (\pi)$. In other
words, for values of $\phi_1$ near zero the $U(1)_Y$ interactions
contributes with equal sign to $a_\mu$ as the (usually leading)
$SU(2)$ interactions do, whereas $\phi_1 \simeq \pi$ leads to a
partial cancellation between $U(1)_Y$ and $SU(2)$ contributions.
$\phi_\mu \simeq \pi$ therefore remains allowed to slightly higher
values of $\tan\beta$ if $\phi_1 \simeq \pi$ as well.

If $|\mu|$ is increased by another factor of $\sim \sqrt{5}$, chargino
and neutralino loop contributions to $d_e$ can be of the same size, in
which case no upper limit can be given on either $|\phi_1+\phi_\mu|$ or
$|\phi_\mu|$ separately \cite{kane}, although a strong
(anti--)correlation between these two phases still has to hold. If
$|\mu|$ is increased even further, the neutralino contribution becomes
dominant. In that case $\phi_\mu$ could take any value (after scanning
over the other phases), but significant absolute constraints on the
combination $\phi_1+\phi_\mu$ would emerge that hold even after
scanning over all $\phi_A$ and $\phi_\mu$. However, most models of
supersymmetry breaking prefer \cite{RGE} values of $|\mu|$ that are
not much larger than $M_2$. We therefore do not discuss scenarios with
$|\mu| \gg M_2$ any further.

Scenario B3 has a significantly smaller value of $|\mu|$ than scenario
B2, although it is larger than in B1. The absolute upper bound on
$|\phi_\mu|$ is therefore reduced to $\sim \pi/8$. The allowed bands in
figs.~\ref{phimuphi1}c,f are narrower than in \ref{phimuphi1}b,e due
to the larger values of $\tan\beta$ and slightly smaller slepton
masses; both effects tend to increase the SUSY contributions to $d_e$,
requiring correspondingly more perfect cancellations. Note also that
for $\tan\beta = 20$ values of $\phi_1$ near zero give $a_\mu$ above
the range (\ref{amurange}), i.e. in this case neutralino and chargino
contributions to $a_\mu$ must not add constructively. Parameter sets
with $\phi_\mu$ near $\pi$ are only allowed for $\tan\beta \lsim 5$.

\begin{figure}[ht!]
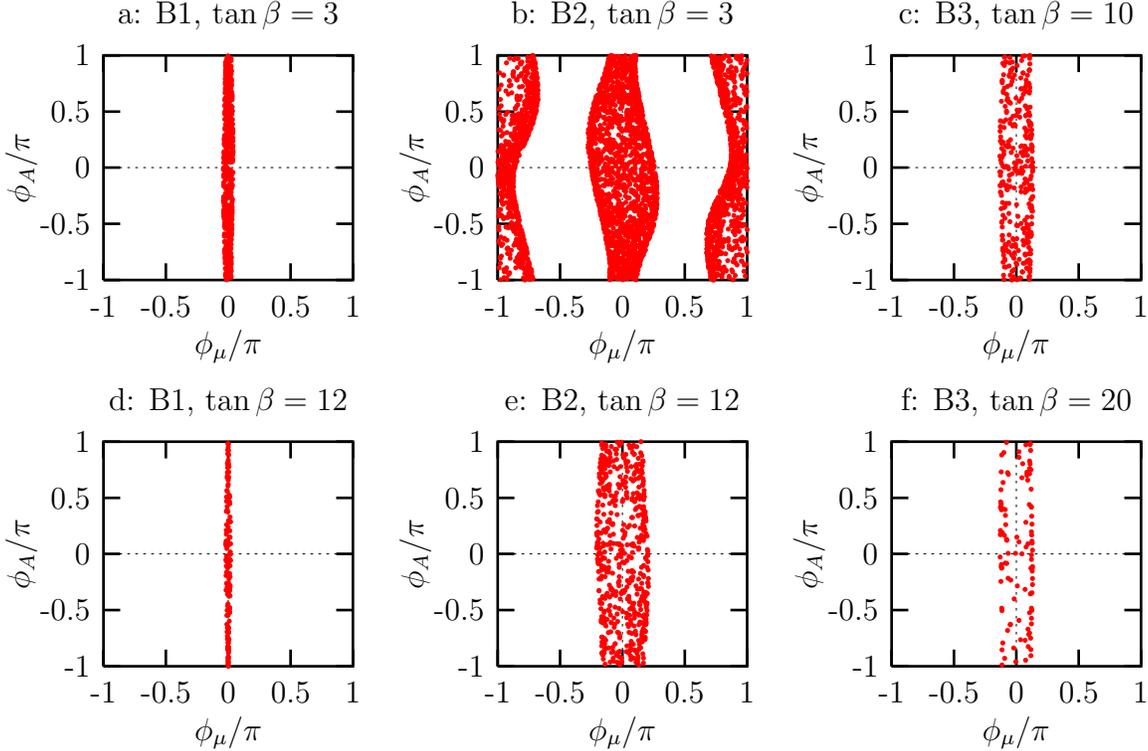

\begin{minipage}[ht!]{0.3\textwidth}
\input{phmuphaB1tan3_tex.tex}
\end{minipage}
\begin{minipage}[ht!]{0.3\textwidth}
\input{phmuphaB2tan3_tex.tex}
\end{minipage}
\begin{minipage}[ht!]{0.3\textwidth}
\input{phmuphaB3tan10_tex.tex}
\end{minipage}\hfill\\
\begin{minipage}[ht!]{0.3\textwidth}
\input{phmuphaB1tan12_tex.tex}
\end{minipage}
\begin{minipage}[ht!]{0.3\textwidth}
\input{phmuphaB2tan12_tex.tex}
\end{minipage}
\begin{minipage}[ht!]{0.3\textwidth}
\input{phmuphaB3tan20_tex.tex}
\end{minipage}\hfill
\caption{Combinations of $\phi_\mu$ and $\phi_A$ that are allowed for 
at least one value of $\phi_1\in [-\pi,\pi]$. }
\label{phimuphia}
\end{figure}

The allowed regions in the $(\phi_\mu, \phi_A)-$plane are shown in
Fig.~\ref{phimuphia}. Most of our scenarios have $|\mu \tan \beta|$
significantly above $|A|$, in which case the value of $\phi_A$ is not
very important. Even if $\phi_A$ is important, as in scenario B1,
there is little correlation between $\phi_A$ and $\phi_\mu$, since
$\phi_A$ only enters in the combination $\phi_{\tilde e} - \phi_1$,
and $\phi_1$ is scanned in Fig.~\ref{phimuphia}. In all cases the
bound on $|\phi_\mu|$ is slightly weaker for $\phi_A \simeq 0$ than
for $\phi_A \simeq \pm \pi$, since in the former case $A$ and $\mu$
add (mostly) constructively to the mixing of selectrons, thereby
increasing the first line on the r.h.s. of eq.(\ref{deneutapp}).

\begin{figure}[ht!]
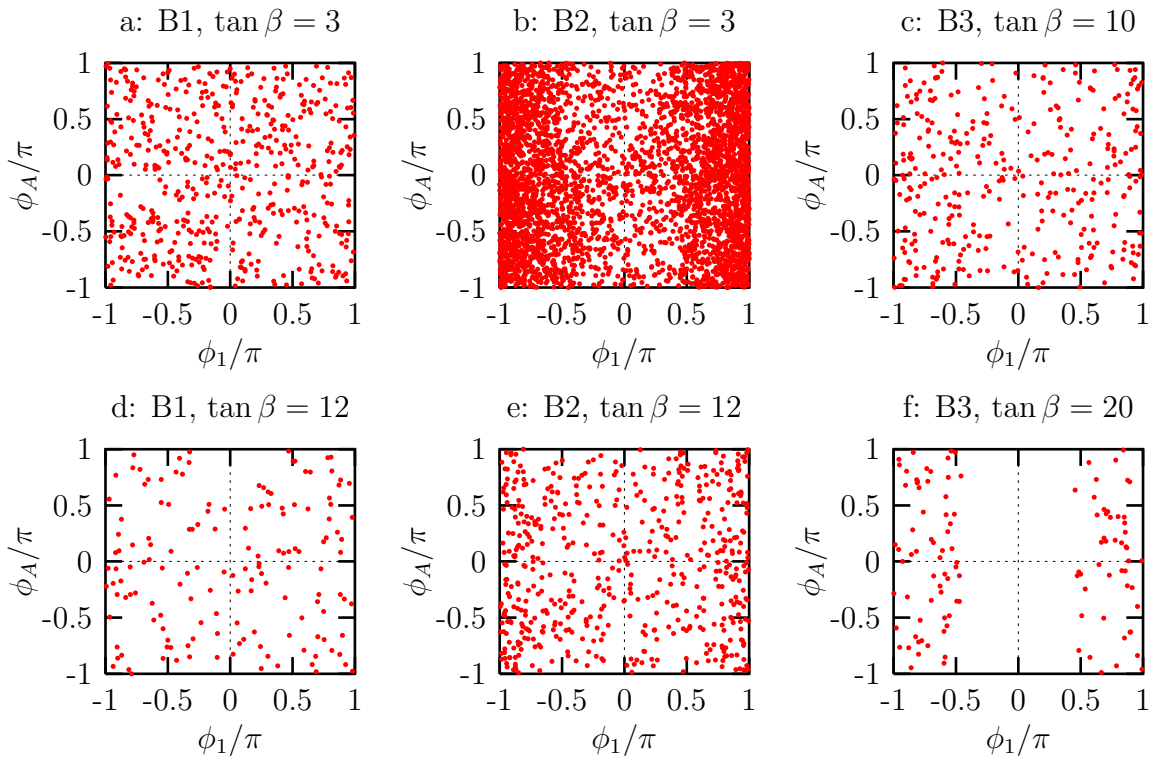

\begin{minipage}[ht!]{0.3\textwidth}
\input{ph1phaB1tan3_tex.tex}
\end{minipage}
\begin{minipage}[ht!]{0.3\textwidth}
\input{ph1phaB2tan3_tex.tex}
\end{minipage}
\begin{minipage}[ht!]{0.3\textwidth}
\input{ph1phaB3tan10_tex.tex}
\end{minipage}\hfill\\
\begin{minipage}[ht!]{0.3\textwidth}
\input{ph1phaB1tan12_tex.tex}
\end{minipage}
\begin{minipage}[ht!]{0.3\textwidth}
\input{ph1phaB2tan12_tex.tex}
\end{minipage}
\begin{minipage}[ht!]{0.3\textwidth}
\input{ph1phaB3tan20_tex.tex}
\end{minipage}\hfill
\caption{Combinations of $\phi_1$ and $\phi_A$ that are allowed for
at least one value of $\phi_\mu \in [-\pi,\pi]$. }
\label{phi1phia}
\end{figure}

We see from figs.~\ref{phimuphi1} and \ref{phimuphia} that in all
cases the entire range of values of $\phi_A$ and $\phi_1$ is allowed
by the $d_e$ constraint for some combinations of the other
phases. Fig.~\ref{phi1phia} shows that there is little correlation
between the allowed ranges of these two phases. Indeed, the $d_e$
constraint allows all combinations of these two phases, for some value
of $\phi_\mu$. On the other hand, in case B3 with $\tan\beta = 20$ the
$a_\mu$ constraint (\ref{amurange}) excludes $|\phi_1| \lsim \pi/2$,
see Fig.~\ref{phimuphi1}f.

In Sec.~7.3 we will study correlations between low-- and high--energy
observables. To that end it is instructive to see how the low--energy
observables correlate with the SUSY phases in the experimentally
allowed region of parameter space. We saw above that $\phi_\mu$ is
tightly constrained, whereas $\phi_A$ and $\phi_1$ are not. Since
$\phi_A$ does not affect high--energy observables, the most
interesting correlations are those between low--energy observables and
$\phi_1$, after scanning over $\phi_A$ and $\phi_\mu$.

\begin{figure}[ht!]
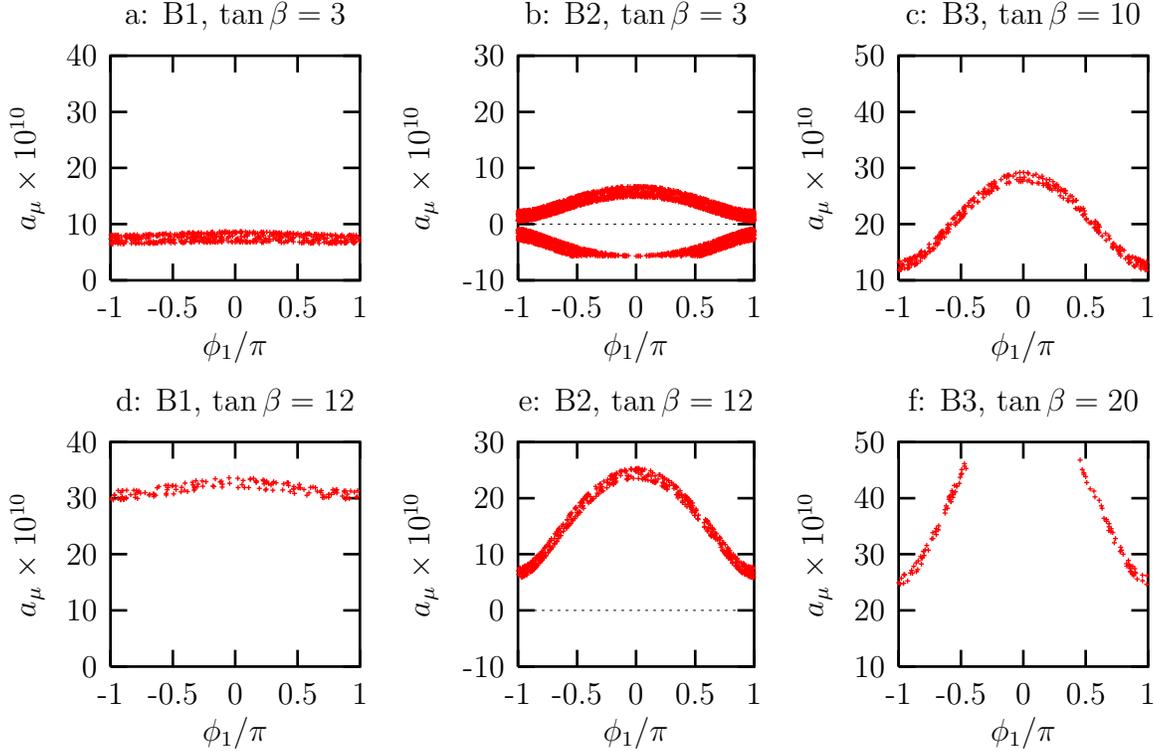

\begin{minipage}[ht!]{0.3\textwidth}
\input{amuph1b1t3_tex.tex}
\end{minipage}
\begin{minipage}[ht!]{0.3\textwidth}
\input{amuph1b2t3_tex.tex}
\end{minipage}
\begin{minipage}[ht!]{0.3\textwidth}
\input{amuph1b3t10_tex.tex}
\end{minipage}\hfill\\
\begin{minipage}[ht!]{0.3\textwidth}
\input{amuph1b1t12_tex.tex}
\end{minipage}
\begin{minipage}[ht!]{0.3\textwidth}
\input{amuph1b2t12_tex.tex}
\end{minipage}
\begin{minipage}[ht!]{0.3\textwidth}
\input{amuph1b3t20_tex.tex}
\end{minipage}\hfill
\caption{$(a_\mu)_{\rm SUSY}$ vs. $\phi_1$ after scanning over
$\phi_\mu$ and $\phi_A$.} 
\label{amuphi1}
\end{figure}

We find that there is no correlation between $d_e$ and $\phi_1$ (not
shown), whereas in scenarios B2 and B3 $a_\mu$ shows a behavior
$\propto a \cos\phi_1 + b$ with a finite scatter, see
Fig.~\ref{amuphi1}. This difference originates from the requirement of
very strong cancellations in $d_e$ discussed above. In particular, the
phases $\phi_1$ and $\phi_\mu$ have to be correlated such that the
leading terms $\propto \sin(\phi_1+\phi_\mu)$ and $\propto
\sin\phi_\mu$ cancel each other, to an accuracy determined by the size
of (subleading) terms $\propto \sin\phi_A$ as well as by the
experimental error on $d_e$. This completely removes the correlation
between $d_e$ and $\sin\phi_1$ that one might naively expect from
eq.(\ref{deneutapp}). The phase--dependent neutralino loop
contributions to $(a_\mu)_{\rm SUSY}$ can be read off from
eq.(\ref{deneutapp}) by replacing $m_e$ by $2m_\mu^2$ in the overall
factor, and all $\sin$ by $\cos$; in addition, there are
phase--independent contributions of comparable size. Since for our
examples $\phi_\mu$ is constrained to be small (or near $\pi$),
$|\cos\phi_\mu| \simeq 1$ and one finds a $\cos-$like dependence of
$a_\mu$ on $\phi_1$, as already stated. The crucial observation is
that the $\phi_1$--dependent and $\phi_\mu$--dependent terms do {\em
not} cancel in this case, so the `naive' dependence of $a_\mu$ on
$\phi_1$ survives. We note in passing that $(a_\mu)_{\rm SUSY} = 0$
can usually not be achieved for a given choice of the absolute values
of the SUSY parameters once we have required large cancellations in
$d_e$, i.e. we cannot choose the phases such that there are large
cancellations in both $d_e$ and $(a_\mu)_{\rm SUSY}$. A better
measurement of, and more accurate SM prediction for, $a_\mu$ therefore
has higher potential to further constrain the SUSY phases than
improved measurements of $d_e$.\footnote{Of course, experimentally
establishing a nonvanishing value of $d_e$ would be of the greatest
importance, since it would require physics beyond the SM. However,
while it would require some SUSY phase to be non--zero, it would not
further reduce the allowed ranges of any one of these phases after
scanning over the other two phases.}

%%%%%%%%%%%%%%%%%%%%%%%%%%%%%
%
% amu versus phi_1 plots
%
%%%%%%%%%%%%%%%%%%%%%%%%%%%%

%%%%%%%%%%%%%%%%%%%%%%%%%%%%%%%%%%%%%%%%%%%%%%%%%%%%%%%%%%%%%%%%%
%%
%%
\section{High Energy Observables}
\label{sec:highenergy}
We are now ready to analyze the impact of SUSY phases on high energy
observables. After defining the relevant kinematical quantities for
the $2 \rightarrow 2$ processes under consideration, we briefly
present 
%the most important steps in 
the calculation of the corresponding unpolarized total cross
sections. All these processes have already been discussed in the
literature: results for $\tilde{e}^-\tilde{e}^+$ and
$\tilde{e}^-\tilde{e}^-$ production results can be found in
\cite{oldslep,baertata,slepprod, thomas} and \cite{polslepprod},
whereas results for $\tilde{\chi}^-_i \tilde{\chi}_j^+$ and
$\tilde{\chi}^0_i \tilde{\chi}_j^0$ production are given in
\cite{oldinos, baertata, Choi:2000ta} and \cite{charneuprod}. We
nevertheless list our results here in order to provide a
self--contained presentation and to illustrate consistency with
previous works.
\clearpage
%%%%%%%%%%%%%%%%%%%%%%%%%%%%%%
\setcounter{footnote}{0} \setcounter{equation}{0}
\renewcommand{\theequation} {\thesection.\arabic{equation}}
%%%%%%%%%%%%%%%%%%%%%%%%%%%%%%%%%%%%%%%%%%%%%%%%%%%%%%%%%%%%%%%%%
\subsection{Kinematics}
%%%%%%%%%%%%%%%%%%%%%%%%%%%%%%%%%%%%%%%%%%%%%%%%%%%%%%%%%%%%%%%%%
\begin{figure}[h!]
\begin{minipage}{0.2\textwidth}\end{minipage}\hfill
\begin{minipage}{0.5\textwidth}
\begin{picture}(200,120)
\SetOffset(40,0)
\LongArrow(14,60)(62,60)
\LongArrow(126,60)(78,60)
\Vertex(70,60){1}
\CArc(70,60)(5,0,360)
\LongArrow(74,64)(108,98)
\LongArrow(66,56)(32,22)
\DashLine(70,66)(70,100){3}
\DashLine(130,60)(164,60){3}
\ArrowArc(70,60)(30,0,45)
\Text(1,58)[tc]{$e^-(p_1,\sigma_1)$}
\Text(120,50)[cc]{$e^+,e^-(p_2,\sigma_2)$}
\Text(111,98)[bl]{$b (k_1,\lambda_1)$}
\Text(30,22)[tr]{$c( k_2,\lambda_2)$}
\Text(103,73)[lc]{$\theta$}
\Text(147,63)[bc]{$e_z$}
\Text(65,83)[rc]{$e_x$}
\LongArrow(138,73)(152,73)
\LongArrow(50,79)(50,93)
\end{picture}
\end{minipage}
\hfill
\begin{minipage}{0.2\textwidth}\end{minipage}
\caption{Kinematical situation}
\label{kinematic}
\end{figure}
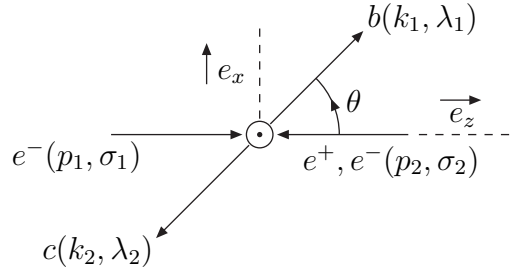

The kinematical situation is illustrated by Fig.~\ref{kinematic}. The
momenta and helicities of the incoming (first) electron and positron
(second electron) are denoted by $p_1^\mu$ and $\sigma_1$, and
$p_2^\mu$ and $\sigma_2$, respectively. The momenta of the produced
superparticles, generically labeled by $b$ and $c$, are denoted by
$k_1^\mu$ and $k_2^\mu$. In case of fermions being produced their
helicities are denoted by $\lambda_1$ and $\lambda_2$.

Working in the center of mass (CMS) frame, we define the $z-$axis of
the coordinate system such that $\vec{p}_1$ points in $+z$
direction. The event plane is then completed by the momentum
$\vec{k}_1$ of particle $b$ and defines the $(x,z)$ plane of the
coordinate system. The scattering angle $\theta$ is defined as the
angle between $\vec{p}_1$ and $\vec{k}_1$. The nominal range for
$\theta$, which we use when going from the differential to the total
cross section, extends from $0$ to $\pi$. However, if the final state
consists of two identical particles, physically $\theta$ has to be
$\leq \pi/2$; we therefore have to multiply the cross section for the
production of identical particles with a factor of 1/2. Notice that
our convention implies vanishing azimuthal angle $\phi$. This
definition of the $(x,z)$ plane is convenient since we are only
interested in total cross sections for unpolarized $e^\pm$
beams.\footnote{A nontrivial dependence on the azimuthal angle would
arise only if we considered transversely polarized $e^\pm$ beams,
and/or were interested in the kinematical distribution of the decay
products of the produced superparticles $b$ and $c$.} Of course, the
phase space integration, which should be performed in a lab--fixed
coordinate system, still gives a factor of $2\pi$ from the integration
over the azimuthal angle. Explicit expressions for the momenta
$p_i^\mu$ and $k_i^\mu$ can be found in \ref{app:kine}.
%%%%%%%%%%%%%%%%%%%%%%%%%%%%%%%%%%%%%%%%%%%%%%%%%%%%%%%%%%%%%%%%%
\subsection{Cross section for $e^-e^+\to \tilde{e}_i^-\tilde{e}_j^+$}
%%%%%%%%%%%%%%%%%%%%%%%%%%%%%%%%%%%%%%%%%%%%%%%%%%%%%%%%%%%%%%%%%
% selectron pair
\begin{figure}[ht!]
\begin{minipage}{0.45\textwidth} 
\begin{picture}(190,90)
\SetOffset(20,-5)
\ArrowLine(10,15)(60,50)
\ArrowLine(60,50)(10,85)
\DashArrowLine(160,15)(110,50){3}
\DashArrowLine(110,50)(160,85){3}
\Photon(60,50)(110,50){3}{5}
\Text(85,55)[bc]{$\gamma$, Z}
\Text(5,17)[cc]{$e^-$}
\Text(5,87)[cc]{$e^+$}
\Text(165,17)[lc]{$\tilde{e}_j^+$}
\Text(165,87)[lc]{$\tilde{e}_i^-$}
\end{picture}
\end{minipage}\hfill+\hfill
\begin{minipage}{0.45\textwidth} 
\begin{picture}(190,90)
\SetOffset(0,-5)
\ArrowLine(10,15)(85,15)
\DashArrowLine(85,15)(160,15){3}
\ArrowLine(85,85)(10,85)
\DashArrowLine(160,85)(85,85){3}
\Line(85,15)(85,85)
\Text(89,50)[lc]{$\tilde{\chi}_k^0$}
\Text(5,17)[cc]{$e^-$}
\Text(5,87)[cc]{$e^+$}
\Text(165,17)[lc]{$\tilde{e}_i^-$}
\Text(165,87)[lc]{$\tilde{e}_j^+$}
\end{picture}
\end{minipage}
\caption{Diagrams for $e^+e^-\to\tilde{e}_i^-\tilde{e}_j^+$}
\label{sleslepro}
\end{figure}
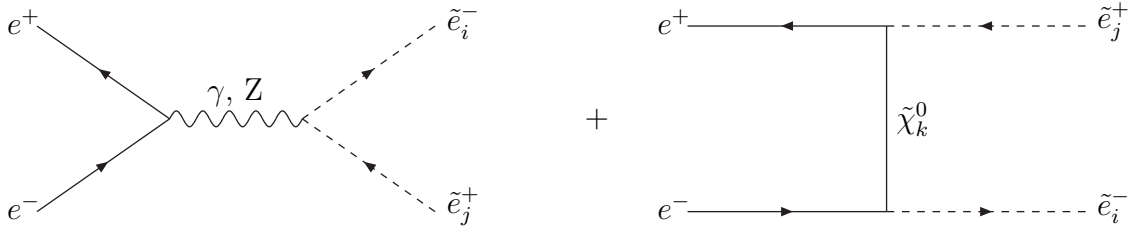

Fig.~\ref{sleslepro} shows the $s$- and $t$-channel diagram contributing
to selectron pair production. By introducing a dimensionless $Z$ boson
propagator
\beq
D_Z=\frac{s}{s-\mz^2+i\mz\Gamma_Z},
\eeq
and bilinear charges $Z_{ij}^\pm$
\ben \label{Z} \beqq
Z_{LL}^-&=1+\frac{(\swq-\frac{1}{2})^2}{\swq\cwq}D_Z,\;\;\;
\;
Z_{RR}^-=1+\frac{\swq-\frac{1}{2}}{\cwq}D_Z,\\
Z_{LL}^+&=1+\frac{\swq-\frac{1}{2}}{\cwq}D_Z,\;\;\;\;\;\;\;\;
\;
Z_{RR}^+=1+\frac{\swq}{\cwq}D_Z,\\
Z_{LR}^\pm&=Z_{RL}^\pm=0,
\eeqq\een
the gauge contribution to the helicity amplitudes can be written as
\beq
\cM_{ij}^{\sigma_1\sigma_2,G}=\frac{e^2}{s}\bar{v}(p_2,\sigma_2)Z^\alpha_{ij}
P_\alpha u(p_1,\sigma_1)(k_i-k_j)^\mu. 
\eeq
The neutralino contribution is
\beq
\cM_{ij}^{\sigma_1 \sigma_2, \tilde{\chi}_k^0} = -\bar{v}
(p_2,\sigma_2) K^{j\star}_{-\alpha k} P^\alpha (\pslash_1 - \kslash_1
+ m_{\tilde{\chi}_k^0} ) D_t^k K^i_{\beta k} P^\beta u(p_1,\sigma_1).
\eeq
The coefficients $K^i_{\alpha k}$ are given by
\ben \label{defK} \beqq
K^L_{Lk} &= \frac {e} {\sqrt{2}\cw\sw} (\cw N_{2k} + \sw N_{1k} ), \\
K^R_{Rk} &= \frac{-2e}{\sqrt{2}\cw} N_{1k}^\star, \\
K^R_{Lk} &= K^L_{Rk} = 0,
\eeqq\een
and the neutralino propagators are
\beq \label{tuprop}
D^k_{t,u} = \frac{1} {(t,u) - m_{\tilde{\chi}_k^0}^2},
\eeq
where $t = (p_1 - k_1)^2$ and $u = (p_1 - k_2)^2$. By introducing a
shorthand notation for the helicity amplitudes
\beq
\langle \sigma_1 \sigma_2 \rangle_{ij} = \cM_{ij}^{\sigma_1
\sigma_2,G} + \cM_{ij}^{\sigma_1 \sigma_2,\tilde{\chi}_k^0}, 
\eeq
and using the explicit expressions for helicity amplitudes given in
\ref{app:heli} and the definition of the neutralino functions in
Eqs.(\ref{neufun}) we find six non-vanishing helicity amplitudes
($\theta$ is the angle between the momenta of the incident $e^-$ and
the produced $\tilde e^-$):
\ben \beqq 
\langle ++ \rangle_{RL} &= -2e^2 M_{LR}^\star(s,t); \\
\langle -- \rangle_{LR} &= 2e^2 M_{RL}(s,t) ;\\
\langle +- \rangle_{RR} &= -e^2 \lambda_{RR}^{\frac{1}{2}} \sin\theta
\left( N_{RR}(s,t) + Z^+_{RR} \right) ;\\
\langle +- \rangle_{LL} &= -e^2 \lambda_{LL}^{\frac{1}{2}} \sin\theta
Z^+_{LL} ;\\
\langle -+ \rangle_{RR} &= -e^2 \lambda_{RR}^{\frac{1}{2}} \sin\theta
Z^-_{RR} ;\\
\langle -+ \rangle_{LL} &= -e^2 \lambda_{LL}^{\frac{1}{2}} \sin\theta
\left( N_{LL}(s,t) + Z^-_{LL} \right).
\eeqq\een
Here, the kinematical factors $\lambda^{\frac{1}{2}}_{ij} \equiv
\lambda^{\frac{1}{2}}_{\tilde e_i \tilde e_j}$ are as in eqs.(A.3).
As usual, the unpolarized cross section can be obtained by averaging
over initial helicities. After integrating over the azimuthal angle,
we obtain
\ben \beqq
\frac{ d\sigma_{LL} } {d\cos\theta} &= \frac
{\lambda_{LL}^{\frac{1}{2}}} {128\pi s} \left( | \langle +-
\rangle_{LL}|^2 + | \langle -+ \rangle_{LL}|^2\right); \\
\frac{ d\sigma_{RR} } {d\cos\theta} &= \frac
{\lambda_{RR}^{\frac{1}{2}}} {128\pi s} \left( | \langle +-
\rangle_{RR} |^2 + | \langle -+ \rangle_{RR} |^2 \right); \\
\frac{ d\sigma_{LR} } {d\cos\theta} &= \frac
{\lambda_{LR}^{\frac{1}{2}}} {128\pi s} | \langle -- \rangle_{LR}|
^2;\\
\frac{ d\sigma_{RL} } {d\cos\theta} &= \frac
{\lambda_{RL}^{\frac{1}{2}}} {128\pi s} | \langle ++ \rangle_{RL} |^2.
\eeqq\een 
Finally, for these and all following reactions the total, unpolarized
cross sections may be obtained by performing the remaining integration
over the scattering angle:
\beq
\sigma_{ij} = \int_{-1}^1 d\cos\theta \left( \frac {d\sigma_{ij}}
{d\cos\theta} \right).
\eeq
%%%%%%%%%%%%%%%%%%%%%%%%%%%%%%%%%%%%%%%%%%%%%%%%%%%%%%%%%%%%%%%%%
% selectron selectron
\subsection{Cross section for $e^-e^-\to \tilde{e}_i^-\tilde{e}_j^-$}
\setcounter{footnote}{0}

\begin{figure}[ht!]
\begin{minipage}{0.45\textwidth} 
\begin{picture}(190,90)
\SetOffset(0,-5)
\ArrowLine(10,15)(85,15)
\DashArrowLine(85,15)(160,15){3}
\ArrowLine(10,85)(85,85)
\DashArrowLine(85,85)(160,85){3}
\Line(85,15)(85,85)
\Text(89,50)[lc]{$\tilde{\chi}_k^0$}
\Text(5,17)[cc]{$e^-$}
\Text(5,87)[cc]{$e^-$}
\Text(165,17)[lc]{$\tilde{e}_j^-$}
\Text(165,87)[lc]{$\tilde{e}_i^-$}
\end{picture}
\end{minipage}
\hfill+\hfill
\begin{minipage}{0.45\textwidth} 
\begin{picture}(190,90)
\SetOffset(0,-5)
\ArrowLine(10,15)(85,15)
\DashLine(85,15)(115,43){3}
\DashArrowLine(115,43)(160,85){3}
\ArrowLine(10,85)(85,85)
\DashLine(85,85)(115,57){3}
\DashArrowLine(115,57)(160,15){3}
\Line(85,15)(85,85)
\Text(89,50)[lc]{$\tilde{\chi}_k^0$}
\Text(5,17)[cc]{$e^-$}
\Text(5,87)[cc]{$e^-$}
\Text(165,17)[lc]{$\tilde{e}_j^-$}
\Text(165,87)[lc]{$\tilde{e}_i^-$}
\end{picture}
\end{minipage}
\caption{Diagrams for $e^-e^-\to\tilde{e}_i^-\tilde{e}_j^-$}
\label{sleaslepro}
\end{figure}
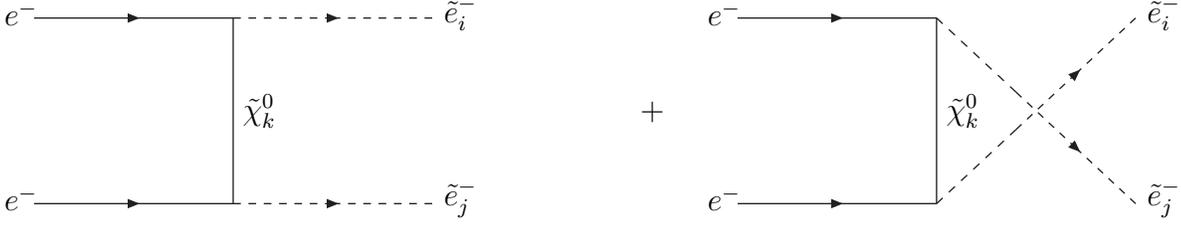

The $t$- and $u$-channel diagrams contributing to $\tilde{e}^-_i
\tilde{e}_j^-$ pair production are shown in Fig.~\ref{sleaslepro}. The
corresponding invariant amplitude can be written as
\beq\begin{array}{ll}
\cM_{ij}^{\sigma_1\sigma_2} &= -K_{\alpha k}^j K_{\beta k}^i
\bar{v}(p_2,\sigma_2) \times \\[0.1ex] \\
 &\left\{ \delta_{\beta,-\alpha} \left[ \frac {\pslash_1-\kslash_1}
{t-m^2_{\tilde{\chi}_k^0}} + \frac {\pslash_1-\kslash_2}
{u-m^2_{\tilde{\chi}_k^0}} \right] P_\alpha + \delta_{\alpha\beta}
m_{\tilde{\chi}_k^0} \left[ \frac {1} {t-m^2_{\tilde{\chi}_k^0}} +
\frac {1} {u-m^2_{\tilde{\chi}_k^0}} \right] P_\alpha \right\}
u(p_1,\sigma_1). 
\end{array}
\eeq
Using the results of
\ref{app:heli} we evaluate the helicity amplitudes and find
\beqa
\cM_{ij}^{\sigma_1\sigma_2} = \langle \sigma_1\sigma_2 \rangle_{ij} &=&
-\frac {s} {2} \sin\theta \lambda_{ij}^{\frac{1}{2}}
\delta_{\sigma_2,-\sigma_1} \left( K^i_{\sigma_1 k} K^j_{-\sigma_1 k}
D_t^k - K^i_{-\sigma_1 k} K^j_{\sigma_1 k} D_u^k \right)
\nonumber \\
&+& m_{\tilde{\chi}_k^0} \sqrt{s} K_{\sigma_1 k}^i K_{\sigma_1 k}^j
\sigma_1 \delta_{\sigma_1\sigma_2} \left( D_t^k + D_u^k \right).
\eeqa
Rewriting these results in terms of neutralino functions as defined in
Eqs.(\ref{neufun}) we find four non--vanishing helicity amplitudes
($\theta$ is the angle between the momenta of an incident $e^-$ and a
produced $\tilde e^-$; it does not matter which initial and final
state particles are chosen, since the cross section is invariant under
$\theta \rightarrow \pi - \theta$):
\ben \beqq
\langle ++ \rangle_{RR} &= 2 e^2 \left[ M_{RR}^\star(s,t) +
M_{RR}^\star(s,u) \right]; \\
\langle -- \rangle_{LL} &= -2 e^2 \left[ M_{LL}(s,t)+ M_{LL}(s,u)
\right]; \\
\langle -+ \rangle_{LR} &= e^2 \lambda_{LR}^{\frac{1}{2}} \sin\theta
N_{LR}(s,t); \\
\langle +- \rangle_{RL} &= -e^2 \lambda_{LR}^{\frac{1}{2}} \sin\theta
N_{LR}(s,u).
\eeqq\een
After calculating the polarization averaged squared matrix elements
and including the phase space factor, the differential cross sections
are:
\ben \beqq
\frac {d\sigma_{LL}} {d\cos\theta} &= \frac
{\lambda_{LL}^{\frac{1}{2}}} {256\pi s} | \langle -- \rangle_{LL}
|^2;\\
\frac {d\sigma_{RR}} {d\cos\theta} &= \frac
{\lambda_{RR}^{\frac{1}{2}}} {256\pi s} \langle ++ \rangle_{RR}
|^2; \\
\frac {d\sigma_{LR}} {d\cos\theta} &= \frac
{\lambda_{LR}^{\frac{1}{2}}} {128\pi s} \left(| \langle -+ \rangle_{LR}
|^2 + |\langle +- \rangle_{RL}|^2 \right).
\eeqq\een 
Note that $\sigma_{LR}$ and $\sigma_{RL}$ are not physically 
distinguishable in this case, unlike for $e^+ e^-$ annihilation.
%%%%%%%%%%%%%%%%%%%%%%%%%%%%%%%%%%%%%%%%%%%%%%%%%%%%%%%%%%%%%%%%%
\subsection{Cross section for $e^-e^+\to \tilde{\chi}_i^-\tilde{\chi}_j^+$}
\begin{figure}[ht!]
% chargino
\begin{minipage}{0.45\textwidth} 
\begin{picture}(190,90)
\SetOffset(20,-5)
\ArrowLine(10,15)(60,50)
\ArrowLine(60,50)(10,85)
\ArrowLine(160,15)(110,50)
\ArrowLine(110,50)(160,85)
\Photon(60,50)(110,50){3}{5}
\Text(85,55)[bc]{$\gamma$, Z}
\Text(5,17)[cc]{$e^-$}
\Text(5,87)[cc]{$e^+$}
\Text(165,17)[lc]{$\tilde{\chi}_j^+$}
\Text(165,87)[lc]{$\tilde{\chi}_i^-$}
\end{picture}
\end{minipage}\hfill+\hfill
\begin{minipage}{0.45\textwidth}
\SetOffset(0,-5) 
\begin{picture}(190,90)
\ArrowLine(10,15)(85,15)
\ArrowLine(85,15)(160,15)
\ArrowLine(85,85)(10,85)
\ArrowLine(160,85)(85,85)
\DashArrowLine(85,15)(85,85){3}
\Text(89,50)[lc]{$\tilde{\nu}_e$}
\Text(5,17)[cc]{$e^-$}
\Text(5,87)[cc]{$e^+$}
\Text(165,17)[lc]{$\tilde{\chi}_i^-$}
\Text(165,87)[lc]{$\tilde{\chi}_j^+$}
\end{picture}
\end{minipage}
\caption{Diagrams for $e^-e^+\to \tilde{\chi}_i^-\tilde{\chi}_j^+$}
\label{charprod}
\end{figure}
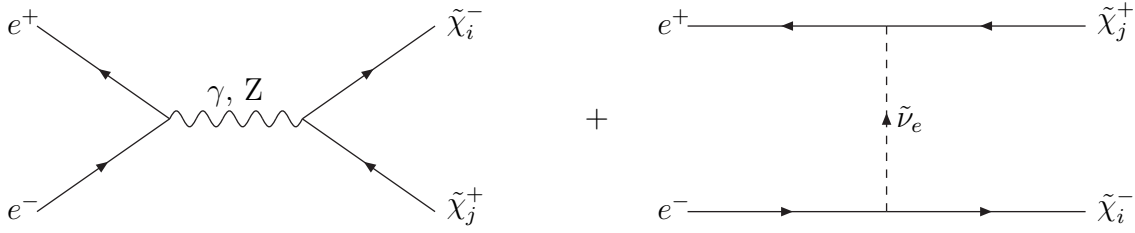

Fig.~\ref{charprod} shows the $s$- and $t$-channel contributions to
$\tilde{\chi}_i^-\tilde{\chi}_j^+$ pair production. After a
Fierz--rearrangement of the $\tilde{\nu}$ contribution, the invariant
amplitude can be written as
\beq \label{charprodampl}
\cM_{\sigma_1\sigma_2;\lambda_1\lambda_2}^{ij} = \frac {-e^2} {s}
\bar{v}(p_2,\sigma_2) \gamma_\mu P^\alpha u(p_1,\sigma_2)
Q_{\alpha\beta}^{ij}\bar{u}_i(k_1,\lambda_1) \gamma^\mu P^\beta
v_j(k_2,\lambda_2), 
\eeq 
where we introduced the bilinear charges (with $x=\swq$)
\ben \label{bichar11} \beqq
Q_{LL}^{11} &= 1 + \frac {D_Z(2x-1)} {2x(1-x)} \left( x
-\frac{3}{4} - \frac{1}{4} \cos2\phi_L \right), \\
Q_{RR}^{11} &= 1 + \frac {D_Z} {1-x} \left( x - \frac{3}{4} -
\frac{1}{4} \cos2\phi_R \right), \\
Q_{LR}^{11} &= 1 + \frac {D_Z(2x-1)} {2x(1-x)} \left ( x -
\frac{3}{4} - \frac{1}{4} \cos2\phi_R \right) + \frac
{s D_t^{\tilde{\nu}}} {4x} ( 1 + \cos2\phi_R ), \\
Q_{RL}^{11} &= 1 + \frac {D_Z} {1-x} \left ( x - \frac{3}{4} -
\frac{1}{4} \cos2\phi_L\right );
\eeqq \een
\ben \label{bichar22} \beqq
Q_{LL}^{22} &= 1 + \frac {D_Z(2x-1)} {2x(1-x)} \left ( x -
\frac{3}{4} + \frac{1}{4} \cos2\phi_L\right ), \\
Q_{RR}^{22} &= 1 + \frac {D_Z} {1-x} \left( x - \frac{3}{4} +
\frac{1}{4} \cos2\phi_R \right) ,\\
Q_{LR}^{22} &= 1 + \frac {D_Z(2x-1)} {2x(1-x)} \left( x -
\frac{3}{4} + \frac{1}{4} \cos2\phi_R \right) + \frac
{s D_t^{\tilde{\nu}}} {4x} ( 1 - \cos2\phi_R), \\
Q_{RL}^{22} &= 1 + \frac {D_Z} {1-x} \left( x - \frac{3}{4} +
\frac{1}{4} \cos2\phi_L\right );
\eeqq\een
\ben \label{bichar12} \beqq
Q_{LL}^{12} = \left(Q_{LL}^{21}\right)^\star & = \frac {D_Z(2x-1)}
{8x(1-x)} \sin2\phi_L e^{-i\beta_L} ,\\
Q_{RR}^{12} = \left( Q_{RR}^{21} \right)^\star &= \frac {D_Z}
{4(1-x)} \sin2\phi_R e^{i(\gamma_1-\beta_R-\gamma_2)}, \\
Q_{LR}^{12} = \left(Q_{LR}^{21} \right)^\star &= \left( \frac
{D_Z(2x-1)} {8x(1-x)} - \frac {s D_t^{\tilde{\nu}}} {4x} \right)
\sin2\phi_R e^{i(\gamma_1-\beta_R-\gamma_2)} ,\\
Q_{RL}^{12} = \left(Q_{RL}^{21} \right)^\star &= \frac {D_Z}
{4x(1-x)} \sin2\phi_L e^{-i\beta_L}.
\eeqq\een 
Here the sneutrino propagator $D_t^{\tilde \nu}$ is defined
analogously to the neutralino propagators (\ref{tuprop}). Using the
results of \ref{app:heli} we find for a generic helicity amplitude
($\theta$ is the angle between the momenta of the incident $e^-$ and
the produced $\tilde \chi^-$):
\beqa \label{gen2fampl}
\langle \sigma_1, -\sigma_1; \lambda_1 \lambda_2 \rangle_{ij} &=&
\frac {-e^2} {2} \sum_\beta Q_{\sigma_1 \beta}^{ij} \left\{ \lambda_1
\delta_{\lambda_1 \lambda_2} \sqrt{ 1 - \eta_{\beta\lambda_1}^2 }
\sin\theta \right. \\
&+& \left. \delta_{\lambda_1,-\lambda_2} \sqrt{ ( 1 + \beta \lambda_1
\eta_{\beta \lambda_1}) ( 1 + \beta \lambda_1 \eta_{-\beta\lambda_1} )
} \left( \cos\theta + \lambda_1 \sigma_1 \right) \right\}, \nonumber
\eeqa
where the kinematical quantities $\eta_\pm$ are defined in
eq.(\ref{etadef}). 
%Explicitly (chargino indices suppressed):
%%
%\ben \label{exp2fampl} \beqq
%\langle +- ; ++ \rangle &= -\frac {e^2} {2} \left( \sqrt{ 1 - \eta_+^2
%} Q_{++} + \sqrt{ 1 - \eta_-^2 } Q_{+-} \right) \sin\theta \\
%%
%\langle +- ; +- \rangle &= -\frac {e^2} {2} \left( \sqrt{ ( 1 + \eta_+
%) ( 1 + \eta_- ) } Q_{++} + \sqrt{ ( 1 - \eta_+ ) ( 1 - \eta_- )}
%Q_{+-} \right) ( 1 + \cos\theta ) \\
%%
%\langle +- ; -+ \rangle &= \frac {e^2} {2} \left( \sqrt{ ( 1 - \eta_+
%) ( 1 - \eta_- ) } Q_{+-} + \sqrt{ ( 1 + \eta_+ ) ( 1 + \eta_- ) }
%Q_{+-} \right) ( 1 - \cos\theta )\\
%%
%\langle +- ;-- \rangle &= \frac {e^2} {2} \left( \sqrt{ 1 - \eta_-^2 }
%Q_{++} + \sqrt{ 1 - \eta_+^2 } Q_{+-} \right) \sin\theta \\
%%
%\langle -+ ; ++ \rangle &= -\frac {e^2} {2} \left( \sqrt{ 1 - \eta_+^2
%} Q_{-+} + \sqrt{ 1 - \eta_-^2 } Q_{--} \right) \sin\theta \\
%%
%\langle -+ ; +- \rangle &= \frac {e^2} {2} \left( \sqrt { ( 1 + \eta_+
%) ( 1 + \eta_- ) } Q_{-+} + \sqrt{ ( 1 - \eta_+ ) ( 1 - \eta_- ) }
%Q_{--} \right) ( 1 - \cos\theta ) \\
%%
%\langle -+ ;-+ \rangle &= -\frac {e^2} {2} \left( \sqrt{ ( 1 - \eta_+
%) ( 1 - \eta_- ) } Q_{-+} + \sqrt{ ( 1 + \eta_+ ) ( 1 + \eta_- ) }
%Q_{--} \right) ( 1 + \cos\theta )\\
%%
%\langle -+ ; -- \rangle &= \frac {e^2} {2} \left( \sqrt{ 1 - \eta_-^2
%} Q_{-+} + \sqrt{ 1 - \eta_+^2 } Q_{--} \right) \sin\theta.
%\eeqq\een 
%%
The unpolarized cross sections can be computed from
eq.(\ref{gen2fampl}) by averaging over initial helicities and summing
over the final ones:
\beq \label{dsigchar}
\frac {d\sigma_{ij}} {d\cos\theta} = \frac {\pi\alpha^2} {8s}
\lambda_{ij}^{\frac{1}{2}} \left\{ \left[ ( 1 - \Delta_{ij}^2 ) +
\lambda_{ij} \cos^2\theta \right] Q_1^{ij} + 8 \mu_i \mu_j Q_2^{ij} +
2 \lambda_{ij}^{\frac{1}{2}} \cos\theta Q_3^{ij} \right\},
\eeq
where $4\pi\alpha=e^2, \ \mu_i = m_{\tilde \chi_i}/\sqrt{s}$ and
$\Delta_{ij} = \mu_i^2 - \mu_j^2$. The new quartic charges $Q_n^{ij}$
are given by:
\ben \label{quartic} \beqq
Q_1^{ij} &= | Q^{ij}_{++} |^2 + | Q^{ij}_{+-} |^2 + | Q^{ij}_{-+} |^2
+ | Q^{ij}_{--} |^2 ;\\
Q_2^{ij} &= \real \left( Q_{++}^{ij} Q_{+-}^{ij\star} + Q_{--}^{ij}
Q_{-+}^{ij\star} \right); \\
Q_3^{ij} &= | Q^{ij}_{++} |^2 - | Q^{ij}_{+-} |^2 - | Q^{ij}_{-+} |^2
+ | Q^{ij}_{--}|^2.
\eeqq\een 
%
%%%%%%%%%%%%%%%%%%%%%%%%%%%%%%%%%%%%%%%%%%%%%%%%%%%%%%%%%%%%%%%%%
\subsection{Cross section for $e^-e^+\to \tilde{\chi}_i^0\tilde{\chi}_j^0$}
% neutralino
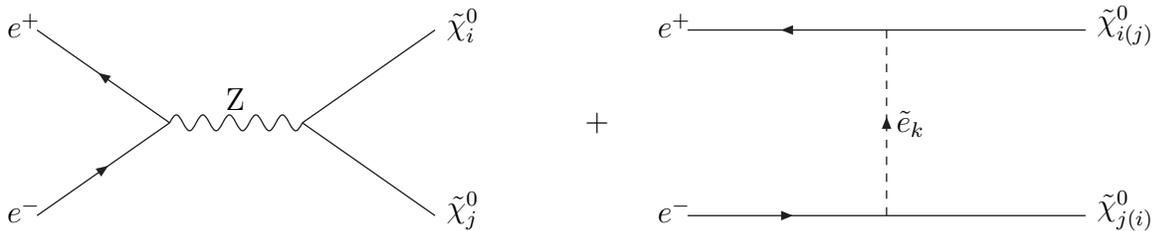
\begin{figure}[ht!]
\begin{minipage}{0.45\textwidth} 
\begin{picture}(190,90)
\SetOffset(20,-5)
\ArrowLine(10,15)(60,50)
\ArrowLine(60,50)(10,85)
\Line(160,15)(110,50)
\Line(110,50)(160,85)
\Photon(60,50)(110,50){3}{5}
\Text(85,55)[bc]{Z}
\Text(5,17)[cc]{$e^-$}
\Text(5,87)[cc]{$e^+$}
\Text(165,17)[lc]{$\tilde{\chi}_j^0$}
\Text(165,87)[lc]{$\tilde{\chi}_i^0$}
\end{picture}
\end{minipage}\hfill+\hfill
\begin{minipage}{0.45\textwidth} 
\begin{picture}(190,90)
\SetOffset(0,-5)
\ArrowLine(10,15)(85,15)
\Line(85,15)(160,15)
\ArrowLine(85,85)(10,85)
\Line(160,85)(85,85)
\DashArrowLine(85,15)(85,85){3}
\Text(89,50)[lc]{$\tilde{e}_k$}
\Text(5,17)[cc]{$e^-$}
\Text(5,87)[cc]{$e^+$}
\Text(165,17)[lc]{$\tilde{\chi}_{j(i)}^0$}
\Text(165,87)[lc]{$\tilde{\chi}_{i(j)}^0$}
\end{picture}
\end{minipage}
\caption{Diagrams for $e^-e^+\to \tilde{\chi}_i^0\tilde{\chi}_j^0$}
\label{neuprod}
\end{figure}

In Fig.~\ref{neuprod} the $s-$ and $t-$channel contributions to $
\tilde{\chi}_i^0 \tilde{\chi}_j^0$ production are shown; the
additional, destructively interfering $u-$channel diagram is indicated
by the exchanged indices in parentheses. Applying a Fierz
rearrangement on both the $t-$ and $u-$channel diagram, and re--ordering
the $u-$channel amplitude, we obtain the invariant amplitude
\beq \cM_{\sigma_1 \sigma_; \lambda_1 \lambda_2}^{ij} = \frac {-e^2}
{s} \bar{v}(p_2,\sigma_2) \gamma_\mu P^\alpha u(p_1,\sigma_2)
Q_{\alpha\beta}^{ij} \bar{u}(k_1,\lambda_1) \gamma^\mu P^\beta
v(k_2,\lambda_2).
\eeq
Here, the bilinear charges $Q_{\alpha\beta}^{ij}$ are given by
($x=\swq$)
\ben \label{bineut} \beqq
Q_{LL}^{ij} &= \frac {D_Z} {2x(1-x)} (2x-1) Z_{ij}^\star - s D_u^L
g_{Lij}, \\
Q_{RR}^{ij} &= -\frac {D_Z} {1-x} Z_{ij} - s D_u^R g_{Rij}^\star,\\
Q_{LR}^{ij} &= -\frac {D_Z} {2x(1-x)} (2x-1) Z_{ij} + s D_t^L
g_{Lij}^\star ,\\
Q_{RL}^{ij} &= \frac {D_Z} {1-x} Z_{ij}^\star + s D_t^R g_{Rij},
\eeqq\een 
with 
\ben \label{gLR} \beqq
g_{Lij} &= \frac {1} {4x} \left( N_{2i}^\star + \tw N_{1i}^\star
\right) \left( N_{2j} + \tw N_{1j} \right), \\
g_{Rij} &= \frac {1} {1-x} N_{1i}^\star N_{1j}.
\eeqq\een 
The selectron propagators are defined as
\beq \label{selprop}
D_{t,u}^{L,R} = \frac{1} {(t,u) - m^2_{\tilde e_{L,R}} }.
\eeq
Since this amplitude has the same structure as the amplitude for
$\tilde{\chi}_i^-\tilde{\chi}_j^+$ production,
eq.~(\ref{charprodampl}), we can directly translate the result
(\ref{gen2fampl}) from this calculation; we just have to replace the
bilinear charges. We can also use the result (\ref{dsigchar}) for the
unpolarized, differential cross section, but we have to include a
statistical factor:
\beq 
\frac {d\sigma_{ij}} {d\cos\theta} = 2^{-\delta_{ij}} \frac{ \pi
\alpha^2 } {8s} \lambda_{ij}^{\frac{1}{2}} \left\{ \left[ ( 1 -
\Delta_{ij}^2 ) + \lambda_{ij} \cos^2\theta \right] Q_1^{ij} + 8 \mu_i
\mu_j Q_2^{ij} + 2 \lambda_{ij}^{\frac{1}{2}} \cos\theta Q_3^{ij}
\right\}.
\label{dsigmaneutr}
\eeq
Of course, now the bilinear charges of eqs.(\ref{bineut}) have to be
used when evaluating the quartic charges defined in
eqs.(\ref{quartic}).

\subsection{$P_N$ for 2 fermion production}
\label{subsec:polarvector}
\setcounter{footnote}{0}

We will see in Sec.~7 that some of the cross sections calculated in
the previous subsections depend quite sensitively on the CP--violating
phases $\phi_1$ and/or $\phi_\mu$. Nevertheless, if measurements of
these cross sections establish a deviation from the CP--conserving
MSSM, one will need to measure some CP--violating asymmetries in order
to convince oneself that the observed deviation is indeed due to
non--vanishing phases, rather than due to some extension of the
MSSM. We will see that this is possible only for the production of
fermionic final states. Consider
\beq \label{orgproc}
e^-(\vec{p}_1, \vec{s}_1) e^+(\vec{p}_2, \vec{s}_2)
\rightarrow \tilde \chi_i(\vec{k}_1, \vec{\tilde s}_1) \bar{\tilde{\chi}}_j
(\vec{k}_2, \vec{\tilde s}_2) \, .
\eeq
The momenta $\vec{p}_{1,2}$ and $\vec{k}_{1,2}$ have been defined in
Fig.~\ref{kinematic}, and $\vec s_{1,2}$ and $\vec{\tilde s}_{1,2}$ are the
spin vectors in the initial and final state, respectively. A CP
transformation on reaction (\ref{orgproc}) gives the CP--conjugate
process 
\beq \label{cpproc}
{\rm CP}: \ e^+(-\vec{p}_1, \vec{s}_1) e^-(-\vec{p}_2, \vec{s}_2)
\rightarrow \bar{\tilde{\chi}}_i(-\vec{k}_1, \vec{\tilde s}_1) \tilde \chi_j
(-\vec{k}_2, \vec{\tilde s}_2) \, .
\eeq
In the center--of--mass system, $\vec{p}_1 = -\vec{p}_2$ and
$\vec{k}_1 = - \vec{k}_2$. The initial state will therefore be
self--conjugate if $\vec{s}_1 = \vec{s}_2$, in particular for
unpolarized beams. Comparing reactions (\ref{orgproc}) and
(\ref{cpproc}) one can introduce two CP--odd asymmetries even after
summing over the spins in the final state. One can define a rate
asymmetry for chargino production, essentially $\sigma(\tilde \chi_1^-
\tilde \chi_2^+) - \sigma( \tilde \chi_2^- \tilde \chi_1^+)$, as well
as an angular asymmetry for the production of two different
neutralinos, proportional to $d \sigma(\tilde \chi_i^0 \tilde
\chi_j^0, \theta) - d \sigma(\tilde \chi_i^0 \tilde \chi_j^0, \pi -
\theta)$. However, far from the $Z$ pole, both these asymmetries
vanish identically at the tree level. The reason is that they are odd
under a combined CP$\widetilde{\rm T}$ transformation, where the
``naive time reversal'' $\widetilde{\rm T}$ reverses the direction of
all 3--momenta, but does {\em not} exchange initial and final
state. Quantities that are odd under CP$\widetilde{\rm T}$ can be
non--zero only in the presence of absorptive phases, which can come
from nearly resonant $s-$channel propagators, or from loop corrections
if the kinematics allows the particles in the loop to be on--shell.

A CP--odd quantity can therefore only be non--zero in the absence of
absorptive phases, if it is also $\widetilde{\rm T}-$odd. This is true
for triple products of momentum and spin vectors. In general, the spin
of the final state fermions in (\ref{orgproc}) can be decomposed in
three components: $P_L^{i,ij}$ is the component of $\vec{\tilde s}_i$ in
direction of $\vec{k}_i$, averaged over many events (with fixed
$\theta$); $P_T^{i,ij}$ is orthogonal to $\vec{k}_i$, but lies {\em in} the
event plane; and $P_N^{i,ij}$ is orthogonal to $\vec{k}_i$ {\em and}
orthogonal to the event plane. The first two of these quantities are
$\widetilde{\rm T}-$even; however, since
\beq \label{pndef} P_N^{i,ij} = \langle \vec{\tilde s}_i \cdot ( \vec{p_1} \times
\vec{k}_i ) \rangle, \ i=1,2, \eeq
$P_N^{i,ij}$ is indeed $\widetilde{\rm T}-$odd; here $\langle \dots
\rangle$ denotes averaging over many events with fixed scattering
angle $\theta$.\footnote{Strictly speaking, $P_N^i$ is CP--odd only
for self--conjugate final states (any two neutralinos, or $\tilde
\chi_i^- \tilde \chi_i^+$). However, since at tree--level and away
from $s-$channel resonances T and $\widetilde{\rm T}$ transformations
are essentially the same, a non--vanishing $P_N^i$ in $\tilde \chi_1^-
\tilde \chi_2^+$ production can also be considered evidence for
CP--violation.} We will comment on the measurability of this quantity
when we present numerical results.

The normal components of the polarizations of $\tilde \chi_i$ and
$\tilde \chi_j$ can be computed using results of ref.\cite{2fpol}:
\ben  \label{pn1} \beqq
P_N^{i,ij} = & \frac { -2 \imag \left\{ \sum_{\sigma_1 \sigma_2} \left[
\langle \sigma_1 \sigma_2; ++ \rangle \langle \sigma_1 \sigma_2;-+
\rangle^\star + \langle \sigma_1 \sigma_2; +- \rangle \langle \sigma_1
\sigma_2; -- \rangle^\star \right] \right\} } {\sum_{\sigma_1\sigma_2}
| \langle \sigma_1 \sigma_2; ++ \rangle |^2 + | \langle \sigma_1
\sigma_2;+- \rangle |^2 + | \langle \sigma_1 \sigma_2; -+ \rangle |^2
+ | \langle \sigma_1 \sigma_2; -- \rangle |^2 }; \\
P_N^{j,ij}=& \frac{ 2 \imag \left\{ \sum_{\sigma_1\sigma_2} \left[
\langle \sigma_1 \sigma_2; ++ \rangle \langle \sigma_1 \sigma_2; +-
\rangle^\star + \langle \sigma_1 \sigma_2; -+ \rangle \langle \sigma_1
\sigma_2; -- \rangle^\star \right] \right\} } {
\sum_{\sigma_1\sigma_2} \left[| \langle \sigma_1 \sigma_2; ++ \rangle |^2 +
| \langle \sigma_1 \sigma_2; +- \rangle |^2 + | \langle \sigma_1
\sigma_2; -+ \rangle |^2 + | \langle \sigma_1 \sigma_2 ; -- \rangle
|^2 \right]}.
\eeqq\een 
After introducing a fourth quartic charge,
\beq \label{q4}
Q_4^{ij} = \imag \left( Q_{++}^{ij} Q_{+-}^{ij^\star} + Q^{ij}_{--}
Q_{-+}^{ij^\star} \right),
\eeq
and using eq.(\ref{gen2fampl}), eqs.(\ref{pn1}) simplify to
\ben \label{pnresult} \beqq
P_N^{i,ij} =& \frac { 4 \sin\theta \mu_j \lambda_{ij}^{\frac{1}{2}}
Q_4^{ij} } { \left[ 1 - \Delta_{ij}^2 + \lambda_{ij} \cos^2\theta
\right] Q_1^{ij} + 8\mu_i \mu_j Q_2^{ij} + 2\lambda_{ij}^{\frac{1}{2}
} \cos\theta Q_3^{ij}};\\
P_N^{j,ij} =& \frac{ -4 \sin\theta \mu_i \lambda_{ij}^{\frac{1}{2}}
Q_4^{ij} } { \left[ 1 - \Delta_{ij}^2 + \lambda_{ij} \cos^2\theta
\right] Q_1^{ij} + 8 \mu_i \mu_j Q_2^{ij} + 2
\lambda_{ij}^{\frac{1}{2}} \cos\theta Q_3^{ij}} 
= - P_N^{i,ij} \frac{\mu_i}{\mu_j}.
\eeqq  \een 
We see that $P_N^{i(j),ij}$ vanishes both at threshold (where
$\lambda_{ij} \rightarrow 0$) and far above threshold (where
$\mu_{j,i} \rightarrow 0$). Eqs.(\ref{bichar11}) and (\ref{bichar22})
show that all bilinear charges describing $\tilde \chi_i^- \tilde
\chi_i^+$ production are real. Likewise, eqs.(\ref{Z}) and (\ref{gLR})
show that the couplings appearing in the expressions (\ref{bineut}) of
the bilinear charges for neutralino pair production are real for final
states consisting of two identical neutralinos. We thus see that
$Q_4^{ij}$ and hence $P_N^{i(j),ij}$ can only be non--vanishing for
off--diagonal production modes ($i\neq j$). Moreover, the second
identity in eq.(\ref{pnresult}b) shows that there is only one
independent $P_N$ for each distinct $\tilde \chi_i \tilde \chi_j$
production channel, for a total of 7 independent CP--odd observables.

%%%%%%%%%%%%%%%%%%%%%%%%%%%%%%%%%%%%%%%%%%%%%%%%%%%%%%%%%%%%%%%%%
\subsection{Approximate results}
\label{subsec:approx}
\setcounter{footnote}{0} 

The results presented in the previous sections allow the exact
(tree--level) calculation of the phase dependences of the selectron
and neutralino production cross sections, and of $P_N$. However, it is
useful to get a qualitative understanding of where one can expect
strong sensitivity to the fundamental phases in the supersymmetric
Lagrangian. To this end we here discuss the behavior of the relevant
cross sections and polarization components using the approximate
diagonalization of the neutralino mass matrix described by
eqs.(\ref{neutdiagapp})--(\ref{deltami}).

We begin with the cross sections for selectron pair production. All
modes receive ${\cal O}(M_Z^0)$ contributions from the exchange of the
bino--like neutralino; in case of $\tilde e_L$ pair production the
exchange of the wino--like neutralino also contributes at order
$M_Z^0$. The $\tilde e^-_L \tilde e^-_L$ mode is the only one which
has a phase sensitivity to order $M_Z^0$, where the cross section is
sensitive to the relative phase between $M_1$ and $M_2$. The other
cross sections show phase sensitivity only at order $M_Z^2$. This is
due to both the exchange of heavier, Higgsino--like neutralinos which
develop gaugino components at ${\cal O}(M_Z)$, and due to the ${\cal
O}(M_Z^2)$ corrections to the gaugino components of the gaugino--like
neutralinos; the importance of these latter contributions explains why
we included the ${\cal O}(M_Z^2)$ quantities $\delta_{12}$ and
$\delta_{21}$, and included the normalization factors $N_{1,2}$, in
eqs.(\ref{neutdiagapp}) and (\ref{deltaij}). The ${\cal O}(M_Z^2)$
shifts $\delta m_{1,2}$ of the masses of the gaugino--like neutralinos
also affect the selectron production cross sections, either directly
(if the physical masses are allowed to vary with the phases), or
indirectly (if physical masses are kept fixed, in which case the
absolute values of the input parameters have to be varied along with
the phases); since ${\cal O}(M_Z^2)$ shifts of $|M_1|, \, M_2$ and
$|\mu|$ change the eigenstates $\tilde \chi_i^0$ only at ${\cal
O}(M_Z^3)$, we will ignore such indirect effects in the
following. Since $\tilde e_R$ does not have $SU(2)$ interactions,
$\sigma(\tilde e^-_R \tilde e^\pm_R)$ are at ${\cal O}(M_Z^2)$ only
sensitive to the phase combination $\phi_1 + \phi_\mu$, whereas the
other modes are also sensitive to\footnote{Recall that $\phi_2 \equiv
0$ in our convention.} $\phi_1$ and $\phi_\mu$. One should also bear
in mind that the phase sensitivity of the diagonal $\tilde e^+ \tilde
e^-$ production channels is further diluted by the presence of large
$s-$channel ($\gamma$ and $Z$ exchange) contributions, which do not
depend on any phase.

\begin{table}[ht!]
\begin{center}
\begin{tabular}{|c||c|c|c|}\hline
&\rule[-6.5pt]{0pt}{16.5pt}$\cos(\phi_\mu+\phi_1) $ & $\cos \phi_\mu$
& $\cos \phi_1 $\\ \hline
\rule[-8.5pt]{0pt}{22.5pt} $\tilde{e}^-_L \tilde{e}_L^+$ & $\sbz \frac
{M_Z^2 |M_1|} {M_2^2 |\mu|}$ & $\sbz \frac {M_Z^2} {M_2 |\mu|}$ & $
\frac {M_Z^2 |M_1|} {M_2 |\mu|^2} $ \\ \hline 
\rule[-8.5pt]{0pt}{22.5pt} $\tilde{e}^-_L \tilde{e}_R^+$ & $ \sbz
\frac {M_Z^2} {|M_1 \mu|}$ & $\sbz \frac {M_Z^2} {M_2 |\mu|}$ &
$\frac {M_Z^2 M_2}{ |M_1| |\mu|^2}$  \\ \hline
\rule[-8.5pt]{0pt}{22.5pt}$\tilde{e}^-_R \tilde{e}_R^+$ & $\sbz \frac
{M_Z^2 |M_1|} {|\mu|^3}$ & none & none \\ \hline \hline
\rule[-8.5pt]{0pt}{22.5pt}$\tilde{e}^-_L \tilde{e}_L^-$ & $\sbz
\frac{M_Z^2} {M_2 |\mu|}$ & $\sbz \frac {M_Z^2} {M_2 |\mu|}$ &
$ \frac {|M_1|} {M_2}$ \\ \hline
\rule[-8.5pt]{0pt}{22.5pt} $\tilde{e}^-_L \tilde{e}_R^-$ & $\sbz\frac
{M_Z^2 |M_1|} {M_2^2 |\mu|}$ & $\sbz \frac{M_Z^2} {M_2 |\mu|}$ &
$-\frac{M_Z^2 |M_1|} {M_2 |\mu|^2}$ \\ \hline
\rule[-8.5pt]{0pt}{22.5pt} $\tilde{e}^-_R \tilde{e}_R^-$ & $ \sbz
\frac {M_Z^2} {|M_1 \mu|}$ & none & none \\ \hline
\end{tabular} 
\caption{Phase dependence of the cross sections for selectron pair
production in $e^+e^-$ as well as $e^-e^-$ annihilation for fixed
physical neutralino masses. Each entry gives the dependence of the
coefficient of the indicated (combination of) phase(s) on the
supersymmetric parameters relative to the leading (phase--independent)
contribution to this cross section, under the assumption $|M_1|^2 < M_2^2
\ll |\mu|^2$. ``None'' means that the corresponding term does not exist
to ${\cal O}(M_Z^2)$. The cross section for $\tilde e^-_L \tilde
e^-_L$ production also has terms $\propto \cos(2 \phi_1 + \phi_\mu)$
and $\propto \cos(\phi_\mu - \phi_1)$, but with small coefficients
$\propto \sbz M_Z^2/ (|\mu|^3)$.}
\label{seltab}
\end{center} 
\end{table} 

The phase dependence of the selectron production cross sections for
fixed physical neutralino masses is summarized in
Table~\ref{seltab}. Here we show the coefficients of the various
phase--dependent terms that can appear, relative to the leading
(phase--independent) contribution to this cross section. We have
omitted numerical factors, including factors involving the weak mixing
angle. Nevertheless we can draw some conclusions from this
table. First, we notice that the dependence on the phase $\phi_\mu$
shown in the second and third columns vanishes like $1/\tan\beta$ for
$\tan\beta \gg 1$. The reason is that the dependence on this phase in
the neutralino mass matrix could be rotated\footnote{This rotation
does not introduce any phase in those parts of $f \tilde f \tilde
\chi$ vertices that come from gauge interactions, but {\em does}
introduce a phase in the Yukawa contribution to these vertices. Recall
that these Yukawa contributions can be ignored when calculating cross
sections, but have to be kept when computing leptonic dipole
moments. This explains why the $\phi_\mu$ dependence of $d_e$ and
$a_\mu$ is not suppressed at large $\tan\beta$.} into the
off--diagonal gaugino--Higgsino mixing entries $\propto
\cos\beta$. However, the dependence on the relative phase between the
two soft gaugino masses does not vary with $\tan\beta$.

Second, with the exception of the $\tilde e^-_L \tilde e^-_L$ mode,
all phase dependence vanishes as $|\mu| \rightarrow \infty$, but the
$|\mu|$ dependence varies for different modes. In particular, the
phase dependence of the diagonal mode $\tilde e^-_R \tilde e^+_R$
vanishes $\propto 1/|\mu|^3$ for large $|\mu|$, whereas all other
cross sections receive phase--dependent contributions that only fall
like $1/|\mu|$; however, for $\tan\beta \rightarrow \infty$ the
$|\mu|-$dependence of the total phase sensitivity becomes stronger, as
can be seen in the last column. In most cases the leading phase
dependence comes from the exchange of the lighter, gaugino--like,
neutralinos. The exception is the $\tilde e^-_R \tilde e^-_R$ mode,
where the the exchange of the heavier, Higgsino--like states
contributes at the same order.

Clearly the $LL$ mode will have the strongest phase dependence of all
$\tilde e^- \tilde e^-$ channels \cite{thomas}, and indeed of all
selectron production channels, since it already occurs at ${\cal
O}(M_Z^0)$, as noted earlier. The phase--dependent terms in $\tilde
e^-_L \tilde e^+_L$ and $\tilde e^-_L \tilde e^+_R$ production are of
similar size. For our choice of parameters the second mode is
preferable, since it is accessible at lower energies, and since the
cross section near threshold scales like $\sqrt{\lambda}$, rather than
like $\lambda^{3/2}$. Finally, the relative importance of
phase--sensitive and phase--insensitive terms in most selectron
production cross sections does not depend strongly on the beam
energy. We therefore expect the best statistical accuracy for the
determination of the relevant phases when the beam energy is chosen
such that the cross section being investigated is maximal.

So far we have kept the physical neutralino masses fixed, which means
that $|M_1|, \, M_2$ and $|\mu|$ have to be varied along with the
phases; we saw above that this affects the cross sections only at
${\cal O}(M_Z^2)$ relative to the leading term. If instead these input
parameters are held fixed, the physical neutralino masses will vary at
${\cal O}(M_Z^2)$. Of particular interest are the masses of the
gaugino--like states, whose exchange gives much bigger contributions
to the matrix elements than that of the Higgsino--like
neutralinos. The relevant mass shifts are given in
eqs.(\ref{deltami}). We see that these effects also vanish $\propto
1/\tan\beta$ for large $\tan\beta$. However, they only scale like
$1/|\mu|$ for large $|\mu|$. They will therefore dominate the total
phase dependence of the $\tilde e^+_R \tilde e^-_R$ production cross
section. For the other modes, the dependence on
$\cos(\phi_1+\phi_\mu)$ and on $\cos\phi_\mu$ that comes from the
variation of the masses of the gaugino--like neutralinos is
qualitatively the same as shown in Table 2, if we ignore factors
$|M_1|/M_2$. A more detailed analysis is therefore required to decide
which source of phase dependence dominates. However, in case of
$\tilde e^-_L \tilde e^-_L$ production the total phase dependence is
still dominated by the ${\cal O}(M_Z^0)$ term from bino--wino
interference.

We now turn to the cross sections for neutralino pair production in
$e^+ e^-$ annihilation, $\sigma_{ij} \equiv \sigma(e^+e^- \rightarrow
\tilde\chi_i^0 \tilde\chi_j^0)$. We first note that of the 10 distinct
cross sections, only four receive ${\cal O}(M_Z^0)$ contributions: the
cross sections $\sigma_{11}, \, \sigma_{12}$ and $\sigma_{22}$
describing the production of two gaugino--like neutralinos receive
large contributions from selectron exchange in the $t-$ or
$u-$channel, while $\sigma_{34}$ receives large contributions from
$Z$ exchange in the $s-$channel. The cross sections $\sigma_{33}$ and
$\sigma_{44}$ describing the production of two equal Higgsino--like
states receive non--vanishing contributions only at ${\cal O}(M_Z^4)$,
whereas the cross sections for the production of one Higgsino--like
and one gaugino--like state start at ${\cal O}(M_Z^2)$.

Only $\sigma_{12}$ has sensitivity to some phase (in this case,
$\phi_1$) at order $M_Z^0$. All other cross sections are sensitive to
phases only at order $M_Z^2$ or even $M_Z^4$. The strong phase
sensitivity of $\sigma_{12}$ can be traced to the $Q_2-$term in
eq.(\ref{dsigmaneutr}). It comes from the fact \cite{petcov} that the
production of two Majorana fermions is $P-$wave suppressed near
threshold if they have the same relative CP--phase, whereas any
difference in this phase leads to an $S-$wave contribution to the
cross section. This effect can be probed with optimal statistical
significance rather close to threshold, in this case for $\sqrt{s}$
not too much above $|M_1| + M_2$.

The ${\cal O}(M_Z^2)$ phase--dependent terms in the neutralino
production cross sections should be most easily observable in the
mixed ``gaugino--Higgsino'' final states, since here the cross
sections also only start at ${\cal O}(M_Z^2)$, as remarked above. Note
that the two Higgsino--like neutralinos are closely mass--degenerate
in the limit $|\mu^2| \gg M_Z^2$. This makes it very difficult to
experimentally distinguish between the production of $\tilde \chi_3^0$
and $\tilde \chi_4^0$. In the following discussion we therefore always
sum over these two Higgsino--like states. Once this has been done, we
again find that all terms involving $\phi_\mu$ come with a factor
$\sbz$, and are thus suppressed at large $\tan\beta$. These cross
sections also contain terms $\propto \cos(\phi_1 - \phi_\mu)$ and
$\cos(2 \phi_1 + \phi_\mu)$, which result from the
rephasing--invariant combinations of phases $\pm (\phi_1-\phi_2) -
(\phi_{2,1} + \phi_\mu)$ in our convention $\phi_2 = 0$. Altogether we
find the following phase--dependent terms in these two cross sections:
\ben \label{neutsigapp} \beqq
\sigma_{1 \tilde h} \equiv \sigma_{13} + \sigma_{14} \, : \ 
&\cos(\phi_\mu + \phi_1) \left( \frac{ \sbz |M_1| } {|\mu|}, \,
\frac{ \sbz |\mu M_1|} {s}  \right); \ 
\cos \phi_\mu \left( \frac{ \sbz M_2 } { |\mu| }, \,
\frac{ \sbz |M_1|^2 M_2 } {|\mu| s} \right); 
\nonumber \\
&\cos(\phi_1-\phi_\mu) \frac {\sbz |M_1| M_2^2} {|\mu| s}; \ 
\cos\phi_1 \left( \frac{ |M_1| M_2 } {|\mu|^2}, \,
\frac{ |M_1| M_2 } {s} \right) ; \\
\sigma_{2 \tilde h} \equiv \sigma_{23} + \sigma_{24} \, : \
&\cos(\phi_\mu + \phi_1) \left( \frac{ \sbz |M_1| } {|\mu|}, \,
\frac{ \sbz |M_1| M_2^2 } {|\mu| s} \right); \
\cos \phi_\mu \left( \frac{ \sbz  M_2 } { |\mu| }, \,
\frac{ \sbz |\mu| M_2} {s} \right); \nonumber \\
&\cos(2\phi_1+\phi_\mu) \frac {\sbz |M_1|^2 M_2 } {|\mu| s}; \
\cos\phi_1 \left( \frac{ |M_1| M_2 } {|\mu|^2}, \,
\frac{ |M_1| M_2} {s} \right) .
\eeqq \een
A common factor $\propto \alpha^2 M_Z^2 / (|\mu|^2 s) $,
characterizing the size of the leading phase--independent
contributions, has been factored out. Here we have listed the
$s-$dependent contributions coming from the terms $\propto Q_2$
separately, where present. Note that they usually have a different
dependence\footnote{In some cases these threshold terms seem to grow
with increasing $|\mu|$. However, $\sigma_{1 \tilde h} \ (\sigma_{2
\tilde h})$ is accessible only for $\sqrt{s} > |M_1| + |\mu| \
(\sqrt{s} > M_2 + |\mu|)$, i.e $|\mu|^2 / s < 1$ in the physical
region.} on $|\mu|$ than the terms that survive for $s \rightarrow
\infty$. We conclude from eqs.(\ref{neutsigapp}) that $\sigma_{1
\tilde h}$ might show a somewhat stronger overall phase dependence in
the region of parameter space allowed by low--energy data, since it
depends on the potentially large phase $\phi_1 + \phi_\mu$ through
terms with fewer powers of $|\mu|$ in the denominator than $\sigma_{2
\tilde h}$ does, whereas the dependence on $\cos\phi_1$ is
parametrically the same in both cases. This is fortunate, since the $1
\tilde h$ mode is accessible at lower energies. Finally, note that the
phase dependence of the neutralino masses affects these cross sections
only at ${\cal O}(M_Z^4)$.

We do not list the ${\cal O}(M_Z^2)$ phase dependent terms of the
cross sections that receive ${\cal O}(M_Z^0)$ contributions, since
these will clearly be much more difficult to measure.

The situation concerning chargino pair production is rather similar.
Here both diagonal modes start at ${\cal O}(M_Z^0)$, but receive
phase--dependent contributions only at ${\cal O}(M_Z^2)$. In case of
the off--diagonal mode\footnote{In principle $\tilde \chi_1^- \tilde
\chi_2^+$ production is now distinguishable from $\tilde \chi_1^+
\tilde \chi_2^-$ production. However, the two cross sections differ
only in the presence of an absorptive phase, i.e. after including loop
corrections.} both the cross section and the phase dependence starts
at ${\cal O}(M_Z^2)$; indeed, the phase dependence is very similar
to that in eq.(\ref{neutsigapp}b) with $\phi_1 \rightarrow 0$, since
the $U(1)_Y$ gaugino mass does not appear in the chargino mass matrix.

Finally the results for perturbative neutralino mixing may be applied
to the polarization vector components of the neutralinos produced in
$e^+e^-\to\tilde{\chi}_i^0\tilde{\chi}_j^0$ as calculated in
Sec.~\ref{subsec:polarvector}. Here we only discuss the normal
component as it is the only CP--odd quantity available if neutralino
decays are not included explicitly. Recall that a non--vanishing $P_N$
can only occur for final states consisting of two {\em different}
neutralinos. We find that the numerators in eqs.(\ref{pnresult})
receive ${\cal O}(M_Z^0)$ contributions only for the (12) mode; in
case of the $(1 \tilde h)$ and $(2 \tilde h)$ modes the numerator
starts at ${\cal O}(M_Z^2)$, just like the corresponding total cross
sections, and hence the denominators in (\ref{pnresult}). In all these
cases $P_N$ will therefore receive ${\cal O}(M_Z^0)$ contributions. On
the other hand, $P_N$ for the $\tilde h \tilde h$ [or $(34)$] mode
vanishes to ${\cal O} (M_Z^2)$; this final state is therefore of
little interest in the present context.  Explicitly, for (12)
production we find to ${\cal O}(M_Z^0)$:\footnote{Eqs.(\ref{pnresult})
show that $|P_N^{1,12}|$ is larger than $|P_N^{2,12}|$ by a factor
$M_2/|M_1| \simeq 2$. However, we assume that $\tilde \chi_1^0$ is the
LSP, and hence stable (if R--parity is conserved), so that its spin
cannot be measured.}
\beq \label{pn12app}
P_N^{2,12} = \frac{-4 \sqrt{\lambda_{12}} \frac{|M_1|}{\sqrt{s}}
\sqrt{1-z^2} \sin\phi_1 } {(1 - \Delta_{12}^2 + \lambda_{12}
z^2) \frac{\left(D_t^L\right)^2 + \left( D_u^L\right)^2} { D_t^L
D_u^L} - 8 \frac{|M_1| M_2}{s} \cos\phi_1 + 2
\sqrt{\lambda_{12}} z \frac{\left( D_t^L \right)^2 -
\left( D_u^L\right)^2} {D_t^L D_u^L}} ,
\eeq
where $z=\cos\theta$. As expected for a CP--odd quantity the dominant
dependence on $\phi_1$ is through a sine function, while the
denominator (basically the differential cross section discussed above)
contains a CP--even dependence on $\phi_1$ through a cosine.

Due to their mass degeneracy we have to average $P_N$ for the mixed
gaugino--Higgsino modes over the production of both Higgsino--like
neutralinos. Using the event numbers $N$ as weights, we obtain:
\beq \label{pnsum}
P_N^{i,i\tilde h} = \frac{N_{i3} P_N^{i,i3} +
  N_{i4} P_N^{i,i4}} {N_{i3}+ N_{i4}},\;\;\; i=1,2.
\eeq
This amounts to replacing the quartic charges in eq.~(\ref{pnresult}) by:
\beq
Q_k^{i\tilde h} = Q_k^{i3} + Q_k^{i4}.
\eeq
The calculation of the relevant quartic charges $Q_4^{i \tilde h}$ to
${\cal O}(M_Z^2)$ is now straightforward, if somewhat tedious. We find
the following terms, factoring out $\alpha^2 M^2_Z/|\mu|^2$:
\ben \label{pnhapp} \beqq
Q_4^{1 \tilde h} \, : \ &\sin(\phi_1 + \phi_\mu) \sbz; \ \sin \phi_\mu \frac
{\sbz |M_1| M_2} {|\mu|^2} ; \ \sin(\phi_1 - \phi_\mu) \frac {\sbz
M_2^2} {|\mu|^2}; \ \sin \phi_1 \frac {M_2}{|\mu|}; \\
Q_4^{2 \tilde h} \, : \ &\sin(\phi_1 + \phi_\mu) \frac {\sbz |M_1| M_2}
{|\mu|^2}; \, \sin \phi_\mu \sbz; \, \sin(\phi_\mu + 2 \phi_1) \frac
{\sbz |M_1|^2} {|\mu|^2}; 
\nonumber \\
&\sin \phi_1 \frac {|M_1|} {|\mu|}.
\eeqq \een
The terms in eqs.(\ref{pnhapp}) directly correspond to terms in $P_N$,
up to an additional factor of $|M_1|/\sqrt{s} \ (|\mu|/\sqrt{s})$ for
$P_N^{\tilde h, 1 \tilde h} \ (P_N^{2, 2 \tilde h})$, since the
dependence of the leading term in the denominator in
eq.(\ref{pnresult}) on SUSY parameters has already been factored
out. As expected, the phase dependence is through sine functions here,
and all terms that are sensitive to $\phi_\mu$ are suppressed at large
$\tan\beta$. The first term in $Q_4^{1 \tilde h}$ gives rise to a
contribution to $P_N^{\tilde h, 1 \tilde h}$ that remains finite as
$|\mu| \rightarrow \infty$, but vanishes $\propto 1/\tan\beta$ for
large $\tan\beta$. On the other hand, in the $(2 \tilde h)$ mode we
can measure the polarization of the lighter gaugino--like neutralino,
giving rise to an extra factor $|\mu|/\sqrt{s}$. We thus see that (as
long as $\sqrt{s} > |\mu| + M_2$) the second term in
eq.(\ref{pnhapp}b) gives a contribution to $P_N^{2, 2 \tilde h}$ that
{\em rises} with increasing $|\mu|$. However, for the range of $|\mu|$
of interest to us, this term is suppressed by the stringent upper
limit on $|\sin \phi_\mu|$, see Sec.~4.3; only in scenario B2 with
small $\tan\beta$ can it reach comparable magnitude as the last
term in eq.(\ref{pnhapp}b). This last term leads to a contribution to
$P_N^{2, 2 \tilde h}$ that approaches a constant for large $|\mu|$,
{\em and} remains finite for large $\tan\beta$.  We therefore conclude
that $(2 \tilde h)$ production should allow a somewhat more sensitive
direct probe of CP violation than $(1 \tilde h)$ production. Finally,
the normal components of the polarization vectors in $\tilde
\chi_1^\pm \tilde \chi_2^\mp$ production have similar structure as
eq.(\ref{pnhapp}b) with $\phi_1 \rightarrow 0$, but receive
additional contributions from the $Z$ coupling to the gaugino
component of the heavy chargino state $\tilde \chi_2^\pm$.

%%%%%%%%%%%%%%%%%%%%%%%%%%%%%%%%%%%%%%%%%%%%%%%%%%%%%%%%%%%%%%%%%
\section{Significances}
\label{sec:newobjects}
%%%%%%%%%%%%%%%%%%%%%%%%%%%%%%
\setcounter{footnote}{1} \setcounter{equation}{0}
\renewcommand{\theequation} {\thesection.\arabic{equation}} 

Our aim in this Section is to introduce objects quantifying the impact
of CP--odd phases on total cross sections, which are CP--even
quantities.  To this end we compare the difference in counting rates
between a CP--conserving point in parameter space (CPC: all phases
$\phi_i = 0$ or $\pi$) and a CP--violating one (CPV: identical
absolute values, but $\phi_i \neq 0$ and low--energy compatible) to
the statistical error at the CPC point. This determines the
significance $\cS$ with which a deviation from the cross section
predicted for the CPC point can be measured. It can be written as
\beq 
\cS = \frac{ \Delta N_{\rm CPC-CPV}} {\delta N_{\rm CPC}} = 
\frac{ N_{\rm CPC} - N_{\rm CPV} } {\sqrt{ N_{\rm CPC}} }.
\eeq 
Since there are two CP--conserving values ($0,\pi$) for each phase, we
have to deal with eight CPC points for each set of absolute values,
and hence the same number of significances is available for each
kinematical accessible cross section. The smallest of these evidently
determines the statistical significance with which the presence of
CP--violating phases can be inferred from this cross section for given
values of the absolute values of all SUSY parameters.  We therefore
define as our final measure of the sensitivity of a given cross
section to phases the significance
\beq \label{sdef}
\cS \left( f_i f_j \right) ={\rm
min}_n \left( \frac { |\sigma_{f_if_j}^{\rm CPV} - \sigma_{f_if_j}^{{\rm
CPC}_n}|} { \sqrt{\sigma_{f_i f_j}^{{\rm CPC}_n} } }\right)\times \sqrt{\cL},
\eeq 
where $\sigma_{f_if_j}$ is the total cross section for $e^- e^\pm \to
f_i f_j$, and $n = 1, \dots, 8$; we only include CPC points which are
low--energy compatible.\footnote{Since $\sigma_{f_i f_j}$ does not
depend on $\phi_A$, there are only four different values of
$\sigma_{f_i f_j}^{{\rm CPC}_n}$ for a given CPV point. However,
occasionally both $\phi_A = 0$ and $\phi_A = \pi$ have to be checked
to find a CPC point that is compatible with the bound on $a_\mu$. Of
course, the bound on $d_e$ is trivially satisfied by all CPC points.}
Finally, $\cL$ is the integrated luminosity, which is expected to be
different for the $e^+e^-$ and $e^-e^-$ options.

In the procedure outlined so far, the CPC and CPV points have the same
absolute values of $M_1, \, M_2$ and $\mu$. This means that these
points will in general have {\em different} physical neutralino and
chargino masses \cite{kneurmoul}. Recall that the phase dependence of
the $\tilde \chi$ masses is suppressed by $M_Z^2/(|\mu| m_{\tilde
\chi})$, see eq.(\ref{deltami}). Nevertheless, changes of several
percent are possible, in particular in the neutralino sector. This
could lead to similar changes in the cross sections through
kinematical factors (in $\tilde \chi$ production) or through
neutralino propagator factors (in $\tilde e$ production). Moreover,
these masses are often more easily measurable than the cross sections
which are the focus of this analysis.

We therefore introduce a second set of significances $\bar\cS$ where
CPC and CPV points have the same physical masses for $\tilde \chi_1^0,
\tilde \chi_3^0$ and $\tilde \chi_1^\pm$; in the limit of large
$\tilde \chi$ masses and for our choice $|\mu| \geq M_2 > |M_1|$,
these three masses essentially fix $|M_1|, \, |\mu|$ and $M_2$,
respectively. Note that we only have three (dimensionful) absolute
values that can be adjusted in the neutralino and chargino mass
matrices. We can therefore not guarantee that all chargino and
neutralino masses are the same in the CPC and CPV points. However,
after ensuring that these three $\tilde \chi$ masses are the same in
both points, the remaining variation of the other three $\tilde \chi$
masses between the CPC and CPV points is quite small. For technical
reasons we keep $|M_1|, \, M_2$ and $|\mu|$ fixed (at the values
listed in Table~1) for the CPC points, and adjust them at the CPV
points. Since the eight CPV points have four different $\tilde \chi$
mass spectra, a given set of phases now also produces several
different CPC points. The new significance can thus be written as
\beq \label{sbardef}
\bar{\cS} \left( f_i f_j \right) = {\rm min}_n \left(
\frac {|\sigma_{f_if_j}^{\overline{\rm CPV}_n} - \sigma_{f_i f_j}^{{\rm
CPC}_n}|} { \sqrt{\sigma_{f_i f_j}^{{\rm CPC}_n}} }\right) \times
\sqrt{\cL}.
\eeq

Our algorithm for calculating the significances can be summarized
as follows:
\begin{itemize}

\item[-] Select a CPV point. For a set of the absolute values of the
relevant SUSY parameters, as listed in Table~1 for our three scenarios
B1, B2 and B3, this amounts to randomly choosing values for the phases
$\phi_A, \, \phi_\mu$ and $\phi_1$. Repeat this step until a point
that is compatible with the low--energy constraints has been found.

\item[-] For each process, find the low--energy allowed CPC point that
minimizes $\cS(f_i f_j)$ as defined in eq.(\ref{sdef}). Note that
there are only eight CPC points for each scenario B1, B2 and B3 if
$\tan\beta$ is kept fixed; however, this procedure in general selects
different CPC points for different processes. This completes the
calculation of the $\cS$.

\item[-] Define four new CPV points $\overline{\rm CPV}_n$ by
adjusting $|M_1|, \, M_2$ and $|\mu|$ such that $m_{\tilde \chi_1^0},
\, m_{\tilde \chi_1^\pm}$ and $m_{\tilde \chi_3^0}$ are the same
in points $\overline{\rm CPV}_n$ and CPC$_n$.

\item[-] Calculate the $\bar{\cS}\left( f_i f_j \right)$ as in 
eq.(\ref{sbardef}).
\end{itemize}

Note that $\cS$ and $\bar\cS$ only measure {\em statistical}
significances. In addition there will be systematic uncertainties,
both from experiment and theory. We have little to say about
experimental systematic errors, except that we hope that they will be
small. A theoretical error is introduced since our cross sections can
only be predicted with finite precision. At tree--level these cross
sections are determined uniquely by the parameters listed in Table~1,
plus a few SM parameters that are already now known with high
precision.  However, explicit calculations for $\tilde \chi_1^\pm$
pair production show that quantum corrections can easily amount to
${\cal O}(10\%)$ \cite{charloop}. Some of these corrections can be
calculated unambiguously once the parameters listed in Table~1 are
specified, but the remaining corrections can still amount to several
percent. In particular, the lepton--slepton--gaugino ``gauge
couplings'' depend (logarithmically) on the squark mass scale
\cite{nondec}. The production of Higgsino--like charginos
\cite{charloop} and, presumably, neutralinos also depends on the
parameters appearing in third generation sfermion masses. These
corrections will only be calculable once the parameters of the
(presumably quite heavy) squark sector have been determined. Until
this has happened, out of two processes with roughly equal
significances as defined above, the process with a {\em smaller} cross
section should be preferred, since here a given significance
corresponds to a {\em larger} relative variation of the cross section
with the phases. 

\section{Numerical Analysis}
\label{sec:numerics}

We are now ready to present numerical results for our high--energy
observables. We will first discuss the impact of the CP--phases
on the (CP--even) cross sections, before turning to the (T--odd)
normal components of $\tilde \chi$ polarization vectors. Finally, in
Sec.~7.3 we will study correlations between phase--sensitive
quantities. 

\subsection{Cross sections}

As discussed in Sec.~4.3, we chose our SUSY parameters such that
selectron pair production as well as the production of two lighter
neutralinos or charginos is possible already at the first stage of a
future linear $e^+e^-$ collider (LC) operating at $\sqrt{s} = 500$
GeV, which is our default choice. However, in scenario B2 the
Higgsino--like states are not accessible at this energy. In this
scenario we therefore take $\sqrt{s} = 800$ GeV when discussing
reactions where at least one $\tilde\chi_3^0, \, \tilde\chi_4^0$ or
$\tilde\chi_2^\pm$ state is produced; note that all current LC designs
foresee an upgrade to at least that energy. A similar treatment is
used in scenario B3, except for the $\tilde \chi_1^0 \tilde
\chi_{3,4}^0$ final state, which is already accessible at $\sqrt{s} =
500$ GeV in this case.

In Table~\ref{xsectable} we show the maximal allowed cross sections
for the 19 different production channels discussed in Sec.~5, for our
three scenarios B1, B2 and B3, and the same choices of $\tan\beta$
employed in Sec.~4.3. Only combinations of phases that are allowed by
the low--energy constraints on $d_e$ and $a_\mu$ have been included in
the maximization. These cross sections have been calculated at
tree--level, as described in Secs.~5.2--5.5. We have also ignored
corrections due to initial--state radiation and beamstrahlung. These
effects are often larger than the dependence on CP--violating phases;
they should therefore certainly be included in any future experimental
analysis (along with radiative corrections, which will likely be known
well before the first LC commences operations). However, they are
largely independent of CP--phases, and should therefore not affect
our conclusions.

%%%%%%%%%%%%%%%%%%%%%%%%%%%%%
%
% cross sections
%
%%%%%%%%%%%%%%%%%%%%%%%%%%%%
\begin{center}
\begin{table}[ht!]
\begin{center}
\begin{tabular}{|c||c|c||c|c||c|c|}\hline
&\multicolumn{2}{|c|}{B1} & \multicolumn{2}{|c|}{B2} & \multicolumn{2}
{|c|}{B3}\\ \hline 
$\tan\beta$ & 3 & 12 & 3 & 12 & 10 &20 \\ \hline \hline
\rule[-6pt]{0pt}{18pt}$\tilde{e}^-_R \tilde{e}^-_R$ & 378 & 371 & 398 & 390
& 513 & 512 \\ \hline
\rule[-6pt]{0pt}{18pt}$\tilde{e}^-_L \tilde{e}^-_R$ & 79.8 & 79.0 & 80.3 & 75.1
& 181 & 182 \\ \hline
\rule[-6pt]{0pt}{18pt}$\tilde{e}^-_L \tilde{e}^-_L$ & 272 & 261 & 281 & 270 
& 523 & 378 \\ \hline \hline
\rule[-6pt]{0pt}{18pt}$\tilde{e}^-_R \tilde{e}^+_R$ & 180 & 172 & 182 & 
176 & 296 & 293 \\ \hline
\rule[-6pt]{0pt}{18pt}$\tilde{e}^-_L \tilde{e}^+_R$ & 106 & 104 & 96.5 & 
94.5 & 168 & 160 \\ \hline
\rule[-6pt]{0pt}{18pt}$\tilde{e}^-_L \tilde{e}^+_L$ & 8.3 & 7.2 & 8.0 & 
6.9 & 60.9 & 60.3 \\ \hline \hline
\rule[-6pt]{0pt}{18pt}$\tilde{\chi}^-_1 \tilde{\chi}^+_1$ & 250 & 212 & 144 
& 126 & 175 & 170 \\ \hline
\rule[-6pt]{0pt}{18pt}$\tilde{\chi}^-_1 \tilde{\chi}^+_2$ & 179 &
173 & 16.0$^\star$ & 7.5$^\star$ 
& 43.6$^\star$ & 38.7$^\star$ \\ \hline
\rule[-6pt]{0pt}{18pt}$\tilde{\chi}^-_2 \tilde{\chi}^+_2$ & -- & -- & -- & -- 
& 85.9$^\star$ & 89.4$^\star$ \\ \hline \hline
\rule[-6pt]{0pt}{18pt}$\tilde{\chi}^0_1 \tilde{\chi}^0_1$ & 201 & 197 & 236 &
231 & 271 & 271 \\ \hline
\rule[-6pt]{0pt}{18pt}$\tilde{\chi}^0_1 \tilde{\chi}^0_2$ & 130 & 120 & 140
& 132 & 159 & 161 \\ \hline
\rule[-6pt]{0pt}{18pt}$\tilde{\chi}^0_1 \tilde{\chi}^0_3$ & 46.8 &
41.2 & & & & 
\\ \cline{1-3}
\rule[-6pt]{0pt}{18pt}$\tilde{\chi}^0_1 \tilde{\chi}^0_4$ & 52.8 & 53.7 & 
\raisebox{1.5ex}[-1.5ex]{6.4$^\star$} &
\raisebox{1.5ex}[-1.5ex]{5.7$^\star$} & 
\raisebox{1.5ex}[-1.5ex]{20.1} & \raisebox{1.5ex}[-1.5ex]{19.7} \\
\hline 
\rule[-6pt]{0pt}{18pt}$\tilde{\chi}^0_2 \tilde{\chi}^0_2$ & 74.6 &
49.6 & 58.5 & 49 & 76.2 & 68.9 \\ \hline
\rule[-6pt]{0pt}{18pt}$\tilde{\chi}^0_2 \tilde{\chi}^0_3$ & 73.6 &
77.7 & & & & \\ \cline{1-3}
\rule[-6pt]{0pt}{18pt}$\tilde{\chi}^0_2 \tilde{\chi}^0_4$ & 27.1 & 22.8 &
\raisebox{1.5ex}[-1.5ex]{5.1$^\star$} &
\raisebox{1.5ex}[-1.5ex]{5.2$^\star$} & 
\raisebox{1.5ex}[-1.5ex]{22.3$^\star$} &
\raisebox{1.5ex}[-1.5ex]{21.4$^\star$} \\ \hline 
\rule[-6pt]{0pt}{18pt}$\tilde{\chi}^0_3 \tilde{\chi}^0_3$ & 0.26 &
0.43 & & & & \\ \cline{1-3}
\rule[-6pt]{0pt}{18pt}$\tilde{\chi}^0_3 \tilde{\chi}^0_4$ & 36.6 & 36.0 &
-- & -- & 38.3$^\star$ & 38.6$^\star$ \\ \cline{1-3}
\rule[-6pt]{0pt}{18pt}$\tilde{\chi}^0_4 \tilde{\chi}^0_4$ & -- & -- & & & &
\\ \hline
\end{tabular}
\end{center}
\caption{Maximal values of the total cross sections [in fb] for
unpolarized $e^\pm$ beams, for the scenarios defined in Table
1. ``--'' means that the corresponding mode is not accessible. In
scenarios B2 and B3 we have summed over the production of the heavy
Higgsino--like neutralinos, as described in the text. The beam energy
is 500 GeV in most cases, but has been raised to 800 GeV for the
production of $\tilde \chi_2^\pm$ and $\tilde \chi^0_{3,4}$ states in
scenarios B2 and B3, as indicated by the asterisk. Note that the
charge--conjugate mode is included, if it is distinct from the listed
one.}
\label{xsectable}
\end{table}
\end{center}

We saw in Secs.~2.3 and 5.7 that the two heaviest, Higgsino--like
neutralinos are close in mass if $|\mu| > M_2$ and $|\mu|^2 \gg
M_Z^2$; the degeneracy between these states is only lifted at
\mbox{${\cal O}(M_Z^2/[|\mu|^2-M_2^2])$} (as well as by radiative
corrections, which however are sizable only in the presence of large
$A-$terms in the stop sector \cite{Higgsinoloop}). Numerically, we
find that the relative difference between $m_{\tilde \chi_4^0}$ and
$m_{\tilde \chi_3^0}$ ranges from 24 to 35\% in scenario B1, but only
from 0.2 to 3.5\% (0.1 to 7.5\%) in B2 (B3). Since the production of
nearly degenerate particles is difficult to distinguish
experimentally, we simply sum over the production of $\tilde \chi_3^0$
and $\tilde \chi_4^0$ in scenarios B2 and B3; in particular, we only
give results for a single process of heavy Higgsino--like neutralino
pair production in these cases. Recall that we used the same treatment
in Sec.~5.7, eqs.(\ref{neutsigapp}) and (\ref{pnsum})--(\ref{pnhapp}).

As well known \cite{polslepprod,charneuprod} many of our cross
sections can be enhanced by factors of a few if both beams are
polarized. Moreover, the discussion of Sec.~5.7 indicates that the
greatest sensitivity to phases comes (through $\phi_1$) from the
interference of $SU(2)$ and $U(1)_Y$ interactions; these contributions
will be suppressed if one chooses $e^-_R$ beams, since $e^-_R$ is a
singlet under $SU(2)$. However, the sensitivity to other combinations
of phases is enhanced for different choices of beam polarizations. We
therefore only show results for unpolarized beams, with the understanding
that in many cases the cross section (phase sensitivity) could be
enhanced by up to a factor of 4 (2) if fully polarized beams were
available.

We see from Table 3 that the cross sections for selectron pair
production are generically bigger at $e^-e^-$ colliders than at
$e^+e^-$ colliders \cite{slepprod}. This difference is only partially
compensated by the higher $e^+e^-$ luminosity; we assume $\int {\cal
L}dt = 500$ (100) fb$^{-1}$ for $e^+e^-$ ($e^-e^-$) collisions. We use
these relatively conservative values since we do not include
efficiency factors. These are expected to reduce the actually
available event samples by factors of a few, the precise values
depending on both the process under consideration and the sparticle
spectrum. Moreover, at $e^-e^-$ colliders the diagonal,
chirality--conserving modes have higher cross section than the
off--diagonal, chirality--violating mode; recall that the latter is
$P-$wave suppressed near threshold, and vanishes for vanishing gaugino
masses. At $e^+e^-$ colliders the diagonal selectron production modes
are $P-$wave suppressed; this explains the rather small cross sections
for $\tilde e_L^+ \tilde e_L^-$ production. Finally, the selectron
production cross sections are highest in scenario B3, since the
selectron masses are somewhat smaller than in the other two cases;
this effect is particularly significant for $\tilde e^-_L \tilde
e^+_L$ production, which is a $P-$wave process quite close to
threshold. The strong $\tan\beta$ dependence of the maximal $\tilde
e^-_L \tilde e^-_L$ production cross section in this scenario follows
from the fact that the region near $\phi_1 = \phi_\mu = 0$ is excluded
by the $a_\mu$ constraint for $\tan\beta = 20$, see Fig.~2f.

The biggest cross sections at $e^+e^-$ collisions are those for
$\tilde e_R^+ \tilde e_R^-, \, \tilde \chi_1^+ \tilde \chi_1^-$ and
$\tilde \chi_1^0 \tilde \chi_1^0$ production. However, the latter
leads to an invisible, and hence undetectable, final state if
$\tilde\chi_1^0$ is a stable LSP; we will therefore not analyze it any
further. The cross sections for producing two heavy charginos or
neutralinos are suppressed both by phase space and by their
Higgsino--like nature. However, the production of one light and one
heavy $\tilde \chi$ state is possible in all three cases. Since, as
discussed in Sec.~5.7, these cross sections are non--vanishing only in
the presence of gaugino--Higgsino mixing, they fall with increasing
$|\mu|$. However, even in scenario B2 one will have several thousand
events containing these Higgsino--like states. For the other channels,
typically several tens of thousands of events will be available,
meaning that the cross sections could be measured with statistical
uncertainty of 1\% or less.

The maximal possible values of the significances $\cS$ and $\bar\cS$
of eqs.(\ref{sdef}) and (\ref{sbardef}) that can be found in our three
scenarios are summarized in Table~\ref{signitable}. The $\tilde e^-_L
\tilde e^-_L$ mode shows the strongest phase dependence of all
selectron production channels, i.e. the highest significance, largely
independent of $|\mu|$ and $\tan\beta$; the $\tan\beta$ dependence of
$\cS$ in scenario B3 is due to the fact that the point $\phi_1 =
\phi_\mu = 0$ is excluded by the $a_\mu$ constraint at $\tan\beta =
20$, but still allowed at $\tan\beta=10$, as shown in
Fig.~\ref{phimuphi1}c. The mixed $\tilde e^-_L \tilde e^+_R$ mode is
the for our purposes most promising selectron production mode at
$e^+e^-$ colliders. It would allow to unambiguously detect (at more
than five statistical standard deviations) the presence of
CP--violating phases over much of the allowed parameter space,
although the effect diminishes with increasing $|\mu|$ and increasing
$\tan\beta$ (except in case B3, for the reason given above). For both
these modes $\cS$ and $\bar\cS$ give very similar results. Except for
scenario B1 with strong Higgsino--gaugino mixing, $\tilde e_R$ pair
production at both $e^+e^-$ and $e^-e^-$ colliders is much less
promising, especially if the physical masses of $\tilde \chi_1^0, \,
\tilde \chi_1^\pm$ and $\tilde \chi_3^0$ are held fixed, i.e. for
$\bar\cS$. All these features can be understood from the discussion of
Table~\ref{seltab} in Sec.~5.7.

The small phase sensitivity of the $\tilde e^-_L \tilde e^+_L$ mode
relative to the $\tilde e^-_L \tilde e^+_R$ mode can partly be
explained by the smaller cross section of the former mode; recall that
the significances scale with the square root of the number of
events. In addition, closer inspection of the matrix elements shows
that in case of $\tilde e^-_L \tilde e^+_L$ production, the terms
$\propto \cos \phi_1$ and $\propto \cos(\phi_1+\phi_\mu)$ are
suppressed by extra factors $\sin^2 \theta_W$ and $\sin^4\theta_W$
relative to the leading phase--independent terms; for the $\tilde
e^-_L \tilde e^+_R$ mode the corresponding relative factors are $1$
and $\sin^2\theta_W$, respectively.

Turning to chargino modes, we observe that they are sensitive to
phases only in scenario B2, with large $|\mu|$, and for small
$\tan\beta$. The only relevant phase here is $\phi_\mu$.  Recall from
the discussion of Sec.~4.3 that the maximal allowed value of this
phase scales like $|\mu|^2$. This means that the maximal deviation of
$|\cos \phi_\mu|$ from unity scales like $|\mu|^4$. In case of $\tilde
\chi_1^+ \tilde \chi_1^-$ production the main phase sensitivity comes
from $m_{\tilde \chi_1^\pm}$, which gives an extra factor $\sin2\beta
/ |\mu|$. Altogether the maximal $\cS(\tilde \chi_1^+ \tilde
\chi_1^-)$ therefore scales like $|\mu|^3 \sin 2\beta$; this
reproduces the numerical behavior in scenarios B2 and B3, with small
gaugino--Higgsino mixing. A similar argument also holds for the mixed
$\tilde \chi_1^- \tilde \chi_2^+$ mode. However, in this case the cross section
itself vanishes in the absence of gaugino--Higgsino mixing. This means
that now the phase--dependent terms are of the same order in $M_W$ as
the phase--independent ones. Moreover, significant phase dependence
now also comes from the $Z \tilde \chi_1^- \tilde \chi_2^+$ coupling,
not only from the chargino masses. Hence both definitions of the
significance now give very similar results. Finally, the very strong
$\tan\beta$ dependence of these significances in case B2 is due to the
fact that values of $\phi_\mu$ near $\pi$ are only allowed for small
$\tan\beta$ in this case, see Fig.~\ref{phimuphi1}b.

In contrast to the chargino modes, some neutralino modes are promising
for all scenarios we considered. This is true in particular for the
$(12)$ mode. We saw in Sec.~5.7 that in this case both the total cross
section and the phase dependence (on $\phi_1$) already start at ${\cal
O}(M_Z^0)$, i.e. they are {\em not} suppressed for large $|\mu|$ or
large $\tan\beta$. Indeed, we find that this mode often allows
somewhat better sensitivity than the celebrated $\tilde e^-_L \tilde
e^-_L$ mode. The mixed gaugino--Higgsino modes also do well,
especially for not too large values of $|\mu|$. As expected from the
discussion of eqs.(\ref{neutsigapp}), the $(1 \tilde h)$ mode is
somewhat more promising than the $(2 \tilde h)$ mode. The rather good
phase sensitivity of the $(22)$ mode at first seems surprising, given
that the phase dependence only enters at ${\cal O}(M_Z^2)$, whereas
the cross section is ${\cal O}(M_Z^0)$. However, closer inspection of
the sensitivities for the $(22)$ and $(1 \tilde h)$ modes shows that
the relative factor between them is in fact ${\cal O}(|M_1|/M_Z)$,
which is close to unity in our case. Note that the relatively large
size of the $(22)$ cross section facilitates its precise measurements
and therefore increases the significances. However, as remarked at the
end of Sec.~6 we still consider the mixed $(1 \tilde h)$ final state
to be more promising, since it will be less sensitive to systematic
uncertainties.

%%%%%%%%%%%%%%%%%%%%%%%%%%%%%%%
%
% significances
%
%%%%%%%%%%%%%%%%%%%%%%%%%%%%%%%
\begin{center}
\begin{table}[ht!]
\begin{center}
\begin{tabular}{|c||c|c|c|c||c|c|c|c||c|c|c|c|}\hline
&\multicolumn{4}{|c||}{B1} & \multicolumn{4}{|c||}{B2} & 
\multicolumn{4}{|c|}{B3} \\ \hline
$\tan\beta$ & \multicolumn{2}{|c|}{3} & \multicolumn{2}{|c||}{12} &
\multicolumn{2}{|c|}{3} & \multicolumn{2}{|c||}{12} & \multicolumn{2}{|c|}{10}
& \multicolumn{2}{|c|}{20} \\ \hline
\rule[-4pt]{0pt}{16pt}&${\cal S}$ & $\bar{\cal S}$ & ${\cal S}$ & 
$\bar{\cal S}$ & ${\cal S}$ & $\bar{\cal S}$ & ${\cal S}$ & $\bar{\cal S}$
& ${\cal S}$ & $\bar{\cal S}$ & ${\cal S}$ & $\bar{\cal S}$ \\ \hline
\rule[-6pt]{0pt}{18pt}$\tilde{e}^-_R \tilde{e}^-_R$ & 3.7 & 17.0 & 0.8 & 5.0 &
2.9 & 1.0 & 0.8 & 0.4 & 0.5 & 1.1 & 0.3 & 0.8 \\ \hline
\rule[-6pt]{0pt}{18pt}$\tilde{e}^-_L \tilde{e}^-_R$ & 3.0 & 10 & 2.8 &
4.7 & 0.9 & 2.5 & 0.8 & 1.3 & 2.7 & 4.2 & 2.9 & 4.1 \\ \hline
\rule[-6pt]{0pt}{18pt}$\tilde{e}^-_L \tilde{e}^-_L$ & 61 & 60 & 61 & 60 & 59 &
57 & 59 & 59 & 90 & 90 & 136 & 136 \\ \hline \hline
\rule[-6pt]{0pt}{18pt}$\tilde{e}^-_R \tilde{e}^+_R$ & 10 & 27 & 2.2 & 7.8 &
6.7 & 1.1 & 1.8 & 0.5 & 4.3 & 2.6 & 3.0 & 2.1 \\ \hline
\rule[-6pt]{0pt}{18pt}$\tilde{e}^-_L \tilde{e}^+_R$ & 43 & 68 & 32 & 39 & 16
& 16 & 11 & 12 & 20 & 23 & 22 & 24 \\ \hline
\rule[-6pt]{0pt}{18pt}$\tilde{e}^-_L \tilde{e}^+_L$ & 1.9 & 3.3 & 1.5 & 0.9 &
1.2 & 1.3 & 0.5 & 0.7 & 3.3 & 4.0 & 3.5 & 3.8 \\ \hline \hline
\rule[-6pt]{0pt}{18pt}$\tilde{\chi}^-_1 \tilde{\chi}^+_1$ & 0.4 & 0.9
& $<0.1$ &  2.5 & 25 & 1.6 & 2.8 & 0.2 & 1.3 & 0.3 & 0.6 & 0.6 \\ \hline
\rule[-6pt]{0pt}{18pt}$\tilde{\chi}^-_1 \tilde{\chi}^+_2$ & $<0.1$ &
1.8 & $<0.1$ & 6.4 & 70$^\star$ & 70$^\star$ & 3.5$^\star$ &
3.5$^\star$ & 2.4$^\star$ & 1.7$^\star$ & 1.4$^\star$ & 2.9$^\star$ \\
\hline 
\rule[-6pt]{0pt}{18pt}$\tilde{\chi}^-_2 \tilde{\chi}^+_2$ & -- & -- & -- & --
& -- & -- & -- & -- & 1.4$^\star$ & 1.6$^\star$ & 0.7$^\star$ &
1.5$^\star$ \\ \hline \hline 
%\rule[-6pt]{0pt}{18pt}$\tilde{\chi}^0_1 \tilde{\chi}^0_1$ & 66 & 112 & 38 & 44
%& 2.3 & 6.0 & 1.5 & 2.5 & 7.2 & 9.5 & 7.4 & 9.0 \\ \hline
\rule[-6pt]{0pt}{18pt}$\tilde{\chi}^0_1 \tilde{\chi}^0_2$ & 41 & 46 & 34 & 32 
& 81 & 81 & 92 & 92 & 100 & 100 & 94 & 94 \\ \hline
\rule[-6pt]{0pt}{18pt}$\tilde{\chi}^0_1 \tilde{\chi}^0_3$ & 56 & 73 & 30 & 29
& & & & & & & & \\ \cline{1-5}
\rule[-6pt]{0pt}{18pt}$\tilde{\chi}^0_1 \tilde{\chi}^0_4$ & 92 & 104 & 82 & 89
& \raisebox{1.5ex}[-1.5ex]{9.9$^\star$} &
\raisebox{1.5ex}[-1.5ex]{10.5$^\star$} &
\raisebox{1.5ex}[-1.5ex]{6.2$^\star$} &
\raisebox{1.5ex}[-1.5ex]{6.2$^\star$} & 
\raisebox{1.5ex}[-1.5ex]{21.5} & \raisebox{1.5ex}[-1.5ex]{23.8} &
\raisebox{1.5ex}[-1.5ex]{21.1} & \raisebox{1.5ex}[-1.5ex]{23.2} \\ \hline
\rule[-6pt]{0pt}{18pt}$\tilde{\chi}^0_2 \tilde{\chi}^0_2$ & 74 & 90 & 56 & 66
& 11 & 8.2 & 5.2 & 5.2 & 17 & 18 & 18 & 19 \\ \hline
\rule[-6pt]{0pt}{18pt}$\tilde{\chi}^0_2 \tilde{\chi}^0_3$ & 16 & 37 &
7.0 & 2.9 & & & & & & & & \\ \cline{1-5}
\rule[-6pt]{0pt}{18pt}$\tilde{\chi}^0_2 \tilde{\chi}^0_4$ & 20 & 14 &
5.5 & 5.6 & \raisebox{1.5ex}[-1.5ex]{6.0$^\star$} &
\raisebox{1.5ex}[-1.5ex]{6.2$^\star$} &
\raisebox{1.5ex}[-1.5ex]{2.9$^\star$} &
\raisebox{1.5ex}[-1.5ex]{2.8$^\star$} & 
\raisebox{1.5ex}[-1.5ex]{1.9$^\star$} & \raisebox{1.5ex}[-1.5ex]{1.1$^\star$} &
\raisebox{1.5ex}[-1.5ex]{3.1$^\star$} &
\raisebox{1.5ex}[-1.5ex]{2.0$^\star$} \\ \hline 
\rule[-6pt]{0pt}{18pt}$\tilde{\chi}^0_3 \tilde{\chi}^0_3$ & 6.3 & 5.4
& 8.4 & 9.3 &  & & & & & & & \\ \cline{1-5}
\rule[-6pt]{0pt}{18pt}$\tilde{\chi}^0_3 \tilde{\chi}^0_4$ & 9.3 & 11 &
9.3 & 10 & -- & -- & -- & -- & 2.4$^\star$ & 3.1$^\star$ & 2.6$^\star$
& 3.4 $^\star$\\ \cline{1-5} \rule[-6pt]{0pt}{18pt}$\tilde{\chi}^0_4
\tilde{\chi}^0_4$ & -- & -- & -- & -- & 
& & & & & & & \\ \hline
\end{tabular}
\end{center}
\caption{The maximal significances $\cS$ of eq.(\ref{sdef})
and $\bar\cS$ of eq.(\ref{sbardef}) that can be found for choices of
phases which are compatible with all low--energy constraints. The
scenarios B1, B2 and B3 have been defined in
Table~\ref{scenarios}. Notation and calculational procedures are as in
Table~\ref{xsectable}.}
\label{signitable}
\end{table}
\end{center}

\subsection{Polarizations}

As emphasized earlier, the significances $\cS$ and $\bar\cS$ strictly
speaking only measure deviations from the CP--conserving MSSM; they do
not directly measure CP violation. Direct evidence for CP violation
could come from the measurement of the T--odd normal component of
$\tilde \chi$ polarization vectors introduced in Sec.~5.6. The maximal
possible absolute values of these ``polarization asymmetries'' for
scattering angle $\theta = \pi/2$ are summarized in
Table~\ref{pntable}. Recall that a nonzero asymmetry can emerge only
in the production of two {\em different} $\tilde \chi$ states, and
that the asymmetry will be larger for the {\em lighter} of the two
final--state particles. However, the polarization can only be measured
through the $\tilde \chi$ decay products; we therefore do not consider
the polarization of $\tilde \chi_1^0$, which is probably the LSP. 

\begin{center}
\begin{table}[ht!]
\begin{center}
\begin{tabular}{|c||c||c|c||c|c||c|c|}\hline
& $i$ & \multicolumn{2}{|c||}{B1} & \multicolumn{2}{|c||}{B2} &
\multicolumn{2}{|c|}{B3} \\ \hline 
$\tan\beta$ & & 3 & 12 & 3 & 12 & 10 &20 \\ \hline\hline
\rule[-6pt]{0pt}{18pt} $\tilde{\chi}^-_1 \tilde{\chi}^+_2$ &
$\tilde{\chi}_1^-$ & 1.4 & 0.2 & 57$^\star$ & 5.2$^\star$ &
1.6$^\star$&0.9$^\star$ \\ \hline \hline
\rule[-6pt]{0pt}{18pt}$\tilde{\chi}^0_1\tilde{\chi}^0_2$ &
$\tilde{\chi}^0_2$ & 6.4 & 7.8 & 34 & 33 & 31 & 31 \\ \hline
\rule[-6pt]{0pt}{19pt}$\tilde{\chi}^0_1 \tilde{\chi}^0_3$ &
$\tilde{\chi}^0_3\;(\tilde h)$ & 22 & 27 & & & & \\ \cline{1-4}
\rule[-6pt]{0pt}{19pt}$\tilde{\chi}^0_1 \tilde{\chi}^0_4$ &
$\tilde{\chi}^0_4\;(\tilde h)$ & 5.5 & 6.6 &
\raisebox{1.5ex}[-1.5ex]{ 7.2$^\star$}&\raisebox{1.5ex}[-1.5ex]{
2.4$^\star$} & \raisebox{1.5ex}[-1.5ex]{ 6.3} &
\raisebox{1.5ex}[-1.5ex]{6.8 } \\ \hline
\rule[-6pt]{0pt}{18pt}$\tilde{\chi}^0_2 \tilde{\chi}^0_3$ &
$\tilde{\chi}^0_2$ & 5.5 & 6.4 & & & & \\ \cline{1-4}
\rule[-6pt]{0pt}{18pt}$\tilde{\chi}^0_2 \tilde{\chi}^0_4$ &
$\tilde{\chi}^0_2$ & 45 & 30 &\raisebox{1.5ex}[-1.5ex]{23$^\star$ } &
\raisebox{1.5ex}[-1.5ex]{ 7.8$^\star$} &
\raisebox{1.5ex}[-1.5ex]{9.7$^\star$ } & \raisebox{1.5ex}[-1.5ex]{
9.9$^\star$} \\ \hline
\rule[-6pt]{0pt}{19pt}$\tilde{\chi}^0_3 \tilde{\chi}^0_4$ &
$\tilde{\chi}^0_3 \;(\tilde h)$ & 4.9 & 6.8 & -- & -- &
1.9$^\star$ & 1.8$^\star$ \\ \hline
\end{tabular}
\end{center}
\caption{Maximal absolute values of $P_N^{i,ij}$ in percent. The
scattering angle $\theta$ is set to $\frac{\pi}{2}$. Notations and
conventions are as in Table~\ref{xsectable}.}
\label{pntable}
\end{table}
\end{center}

We see that the chargino polarization is likely too small to be
useful, except in scenario B2 with large $|\mu|$ and small
$\tan\beta$. Recall from the discussion at the end of Sec.~5.6 that
this asymmetry (for the lighter chargino) scales like $|\mu| \,
\sin2\beta \, \sin\phi_\mu$; we saw in Sec.~4.3 that the upper bound
on $|\sin\phi_\mu|$ scales like $|\mu|^2$. Altogether the maximal
value of $P_N$ of the lighter chargino therefore scales like
$|\mu|^3$. The very rapid decrease of this polarization with
increasing $\tan\beta$ is partly due to the explicit $\sin 2\beta$
dependence, and partly due to the disappearance of the band around
$\phi_\mu \simeq \pi$, see Fig.~\ref{phimuphi1}b.

In scenarios with large $|\mu|$ (B2, B3) the $\tilde \chi_1^0 \tilde
\chi_2^0$ mode again proves most sensitive to CP--violating
phases. Eq.(\ref{pn12app}) shows that in this case a nonzero $P_N$ already
emerges at ${\cal O}(M_Z^0)$, and remains finite both for large
$|\mu|$ and large $\tan\beta$. This describes well the behavior seen
in cases where the perturbative diagonalization of the neutralino mass
matrix is reliable. Moreover, recall from Table~\ref{xsectable} that
this mode has a fairly high cross section. This is important, since
even for perfect (100\%) analyzing power one needs nearly 1,000 events
to detect a 10\% asymmetry at the $3\sigma$ level.

As expected from our earlier discussion of eqs.(\ref{pnhapp}), the
mixed gaugino--Higgsino modes also have sizable asymmetries even for
large $|\mu|$, the heavier $(2 \tilde h)$ mode being more
promising. However, the relatively small cross sections of these modes
imply that one would need a very large luminosity for a meaningful
measurement of polarization asymmetries in these modes, except in
scenario B1 with strong wino--Higgsino mixing. Indeed, in this last
case the $(13)$ and $(24)$ modes are far more promising than the
$(12)$ mode.

As noted earlier, the spin of the produced $\tilde \chi$ particles can
only be determined on a statistical basis by (partly) reconstructing
their decays. We find it encouraging that recent dedicated studies
demonstrated sensitivity to phases in the neutralino mass matrix using
T--odd variables constructed in $e^+ e^- \rightarrow \tilde \chi_1^0
\tilde \chi_i^0$ with $\tilde \chi_i^0 \rightarrow \tilde \chi_1^0
\ell^+ \ell^-$ \cite{bartl1}, $\tilde \chi_i^0 \rightarrow \tilde
\tau_1^\pm \tau^\mp \rightarrow \tau^+ \tau^- \tilde \chi_1^0$
\cite{cdgs}, and $\tilde \chi_i^0 \rightarrow \tilde \chi_1^0 Z$
\cite{bartl2}.

\subsection{Correlations between observables}

In addition to their absolute sizes, the correlations between various
phase--sensitive quantities are also of interest. Such correlations
can provide stringent tests of the MSSM, since they are a consequence
of the limited number of parameters affecting these leptonic
observables in the MSSM. Recall that all our ``high--energy''
variables (cross sections and polarizations) depend on the phase
$\phi_\mu$; most of them also depend on $\phi_1$, the exception being
observables related to chargino pair production. We saw in Sec.~4.3
that $\phi_\mu$ is tightly constrained by the ``low--energy''
observables $a_\mu$ and (especially) $d_e$, while $\phi_1$ in most
scenarios can take any value (for some combination of the other
phases). Moreover, the $d_e$ constraint enforces a tight correlation
between $\phi_\mu$ and $\phi_1$, see Fig.~\ref{phimuphi1}.

\begin{figure}[ht!]
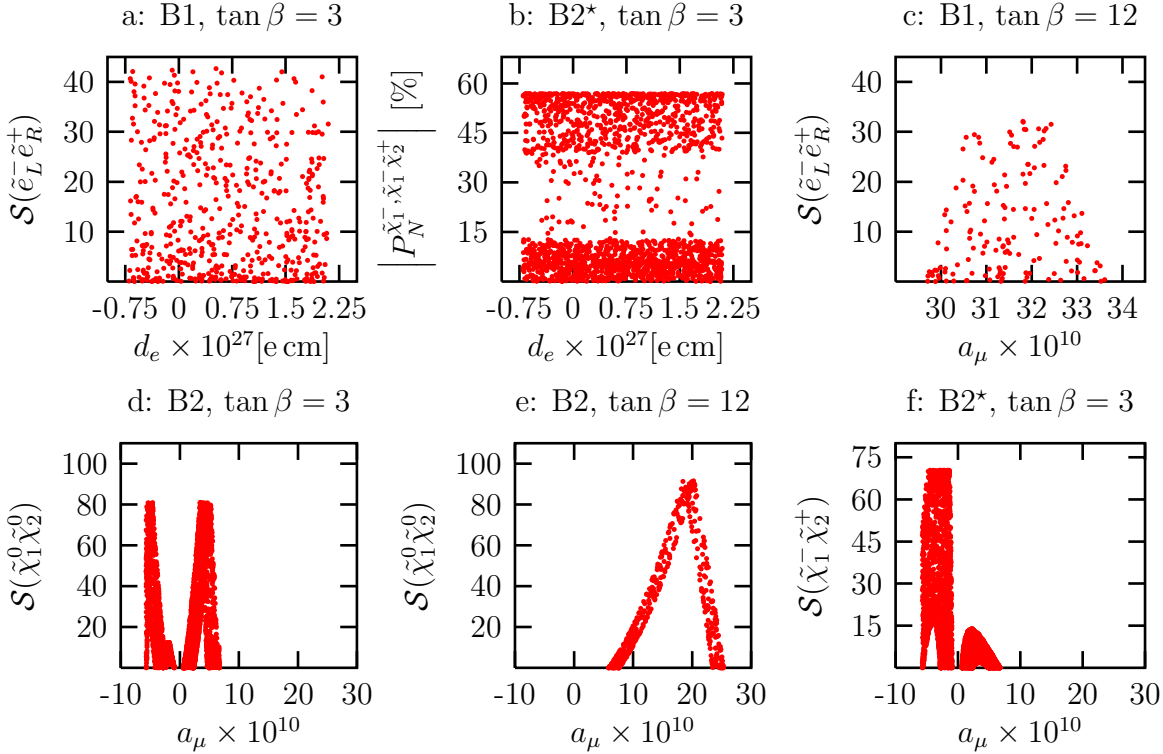

\begin{minipage}[ht!]{0.3\textwidth}
\input{desemleprt3_tex.tex}
\end{minipage}
\begin{minipage}[ht!]{0.3\textwidth}
\input{depnchar12t3_tex.tex}
\end{minipage}
\begin{minipage}[ht!]{0.3\textwidth}
\input{amusemleprt12_tex.tex}
\end{minipage}\hfill\\
\begin{minipage}[ht!]{0.3\textwidth}
\input{amusneut12t3_tex.tex}
\end{minipage}
\begin{minipage}[ht!]{0.3\textwidth}
\input{amusneut12t12_tex.tex}
\end{minipage}
\begin{minipage}[ht!]{0.3\textwidth}
\input{amuschar12t3_tex.tex}
\end{minipage}\hfill
\caption{Correlations between low-- and high--energy quantities. Most
high--energy observables have been computed at $\sqrt{s} = 500$ GeV,
except for panels b) and f), which are for $\sqrt{s} = 800$ GeV. The
parameter sets B1 and B2 have been defined in Table~1, and the
significance $\cS$ is defined via eq.(\ref{sdef}).}
\label{hilocor}
\end{figure}

In Fig.~\ref{hilocor} we compare high-- and low--energy quantities. We
see that the phase--sensitive high--energy quantities are {\em not}
correlated at all with $d_e$. This is true both for T--even variables
(Fig.~\ref{hilocor}a) and T--odd ones (b); in scenarios with strong
gaugino--Higgsino mixing (a) and in scenarios where this mixing is
suppressed (b); and for quantities that depend on both $\phi_1$ and
$\phi_\mu$ (a) as well as those that depend only on $\phi_\mu$
(b). This can be explained from the observation made at the end of
Sec.~4.3 that $d_e$ itself is not correlated with any of the phases
after scanning over the other two phases; recall that the low--energy
observables also depend on $\phi_A$. For example, except at the very
edges of the allowed range of $\phi_\mu$, $d_e$ can still take any
value within its experimentally allowed range even after $\phi_\mu$ is
fixed; this is due to the variation of $\phi_1$ and $\phi_A$.

On the other hand, in some cases we do observe significant
correlations between high--energy observables and $a_\mu$. We saw in
Fig.~\ref{amuphi1} that in scenarios B2 and B3 $a_\mu$ shows a
$\cos-$like dependence on $\phi_1$; in some cases (e.g. B2 at small
$\tan\beta$) two separate bands of $a_\mu$ values exist, corresponding
to $\cos\phi_\mu \simeq \pm 1$. However, in scenario B1 $a_\mu$ shows
very little correlation with $\phi_1$, see Fig.~\ref{amuphi1}a,d.
Correspondingly, Fig.~\ref{hilocor}c shows no correlation for scenario
B1, while Figs.~\ref{hilocor}d--f show significant correlations for
scenario B2. Comparison of panels d and e shows that this correlation
becomes stronger at larger $\tan\beta$. This is due to the diminished
role of $\phi_A$ and the reduced width of the allowed band in the
$(\phi_\mu, \phi_1)$ plane; the overall size of $|a_\mu|$ also
increases with increasing $\tan\beta$, see eq.(\ref{amuana}). Finally,
Fig.~\ref{hilocor}f shows that high--energy quantities whose only
phase sensitivity is through $\phi_\mu$ also correlate with $a_\mu$.
Note in particular that $\cS(\tilde \chi_1^+ \tilde \chi_2^-)$ is much
bigger for $a_\mu < 0$, which corresponds to $\phi_\mu \simeq \pi$,
than for $a_\mu > 0$, which corresponds to $|\phi_\mu| \ll 1$. This
confirms the explanation we gave in the discussion of
Table~\ref{signitable} for the very strong $\tan\beta$ dependence of
this quantity. For this class of observables the correlation with
$a_\mu$ also becomes stronger with increasing $\tan\beta$; however, as
remarked in Sec.~5.7, the sensitivity to $\phi_\mu$ disappears
$\propto \sin2\beta$ at least.

\begin{figure}[ht!]
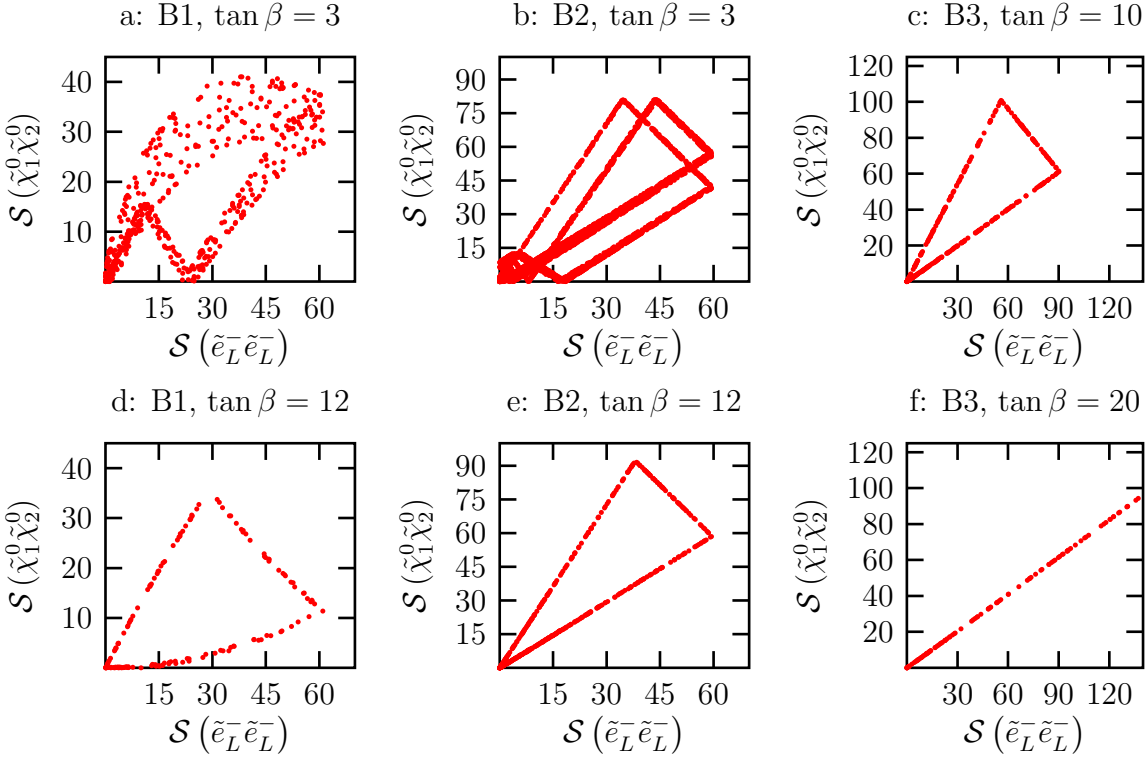

\begin{minipage}[ht!]{0.3\textwidth}
\input{sllsneut12b1t3_tex.tex}
\end{minipage}
\begin{minipage}[ht!]{0.3\textwidth}
\input{sllsneut12b2t3_tex.tex}
\end{minipage}
\begin{minipage}[ht!]{0.3\textwidth}
\input{sllsneut12b3t10_tex.tex}
\end{minipage}\hfill\\
\begin{minipage}[ht!]{0.3\textwidth}
\input{sllsneut12b1t12_tex.tex}
\end{minipage}
\begin{minipage}[ht!]{0.3\textwidth}
\input{sllsneut12b2t12_tex.tex}
\end{minipage}
\begin{minipage}[ht!]{0.3\textwidth}
\input{sllsneut12b3t20_tex.tex}
\end{minipage}\hfill
\caption{Correlations between the significances, defined as in
eq.(\ref{sdef}), for the processes $e^- e^- \rightarrow \tilde e^-_L
\tilde e^-_L$ and $e^+ e^- \rightarrow \tilde \chi_1^0 \tilde
\chi_2^0$, both taken at $\sqrt{s} = 500$ GeV.}
\label{sscor1}
\end{figure}

In most cases different phase sensitive high--energy observables are
strongly correlated with each other. This is illustrated by
Fig.~\ref{sscor1}, where we plot the two usually most promising
significances, for the $\tilde e^-_L \tilde e^-_L$ and $\tilde
\chi_1^0 \tilde \chi_2^0$ final states, against each other. The
simplest correlation obtains for scenario B3 for $\tan\beta = 20$,
shown in panel f. In this case the $a_\mu$ constraint excludes values
of $\phi_1$ near 0 as well as $\phi_\mu$ near $\pi$, see
Fig.~\ref{phimuphi1}f. Hence the minimization in the definition
(\ref{sdef}) of $\cS$ only goes over the single CPC point $\phi_\mu =
0, \, \phi_1 = \pi$. The strong correlation observed in
Fig.~\ref{sscor1}f then follows from the fact that both significances
shown here are essentially $\propto \cos\phi_1$ to leading order in
$M_Z$, as explained in Sec.~5.7. 

The next simplest situation obtains if both $\phi_1 = 0$ and $\phi_1 =
\pi$ are allowed, but $\phi_\mu = \pi$ is still forbidden, and
$\tan\beta$ is not small [panels c), d) and e)]. Now the minimization
in eq.(\ref{sdef}) goes over two CPC points. Recall that this
minimization is performed {\em independently} for the two
significances shown in Fig.~\ref{sscor1}. The upper (lower) branch
connected to the origin is populated by combinations of phases where
both minimizations pick the CPC point $\phi_1 = 0$ $(\phi_1 =
\pi)$. These two bands are connected by sets of points where our
algorithm picks the CPC point $\phi_1 = 0$ for $\cS(\tilde e^-_L
\tilde e^-_L)$, but chooses the point $\phi_1 = \pi$ for $\cS(\tilde
\chi_1^0 \tilde \chi_2^0)$.

Fig.~\ref{sscor1}a shows that in scenario B1 the correlations get
weaker at smaller $\tan\beta$. To understand this, recall that
scenario B1 has strong wino--Higgsino mixing, and hence a relatively
strong dependence on $\phi_\mu$ through the combination $\cos( \phi_1
+ \phi_\mu)$, which depends linearly on $\phi_\mu$ when $|\phi_1|$ and
$|\phi_1 - \pi|$ are sizable. In contrast, $\cos\phi_\mu$ depends only
quadratically on $\phi_\mu$ for small $|\phi_\mu|$, and can therefore
to good approximation be set to 1 in scenario B1, see
Fig.~\ref{phimuphi1}a,d. This dependence on $\phi_\mu$ will be
numerically different for the two modes present, loosening the
correlation. This effect is important only at small $\tan\beta$ for
two reasons. First, all contributions to our cross sections that are
sensitive to $\phi_\mu$ are suppressed by a factor $\sin2\beta$ at
large $\tan\beta$. Secondly, we saw that in scenario B1 the upper
bound on $|\phi_\mu|$ decreases with $\tan\beta$.

Fig.\ref{sscor1}a shows another new effect on the lower branch, where
both significances are evaluated with the CPC point $\phi_1 = \pi, \,
\phi_\mu = 0$. The cross section for $\tilde \chi_1^0 \tilde \chi_2^0$
production in this case shows a non--monotonous dependence on
$\cos\phi_1$. As expected from the expansion of the result
(\ref{dsigmaneutr}) in powers of $M_Z$ using
eqs.(\ref{neutdiagapp})--(\ref{deltami}), this cross section reaches
its absolute minimum at $\cos\phi_1 = +1$, where the $S-$wave
contribution vanishes. However, $\cos\phi_1 = -1$ is also a (local)
minimum, the maximum being reached at $\cos\phi_1 \simeq -0.8$; recall
that the expansion in powers of $M_Z$ is not reliable in this case,
since $M_2 = |\mu|$. As a result of this non--monotonous behavior, the
cross section at $\cos\phi_1 \simeq -0.6$ becomes identical to that at
$\cos\phi_1 = -1$. Since $\sigma(\tilde e^-_L \tilde e^-_L)$ does
decrease monotonically with $\cos\phi_1$, values of $\cos\phi_1 \simeq
-0.6$ give rise to scenarios with very small $\cS(\tilde \chi_1^0
\tilde \chi_2^0)$ but sizable $\cS(\tilde e^-_L \tilde e^-_L)$.

The comparison of Figs.~\ref{sscor1}b and e shows that the correlation
becomes weaker for smaller $\tan\beta$ also in scenario B2. This is
partly because the width of the allowed band in the
$(\phi_\mu,\phi_1)$ plane decreases with increasing $\tan\beta$, see
Fig.~\ref{phimuphi1}. In addition, in scenario B2 with $\tan\beta =
3$ the low--energy constraints also allow values of $\phi_\mu$ near
$\pi$. One can then find values of $\phi_1$ not far from $\pi$ where
$\sigma(\tilde \chi_1^0 \tilde \chi_2^0)$ for CPV points with
$|\phi_\mu| \ll 1$ is very close to this cross section at the CPC
point $\phi_\mu = \phi_1 = \pi$. This again leads to scenarios where
$\cS(\tilde \chi_1^0 \tilde \chi_2^0)$ is very small, but $\cS(\tilde
e^-_L \tilde e^-_L)$ is sizable. The existence of four different
allowed CPC points also explains the occurrence of additional bands in
Fig.~\ref{sscor1}b.

\begin{figure}[ht!]
\begin{minipage}[ht!]{0.99\textwidth}
\input{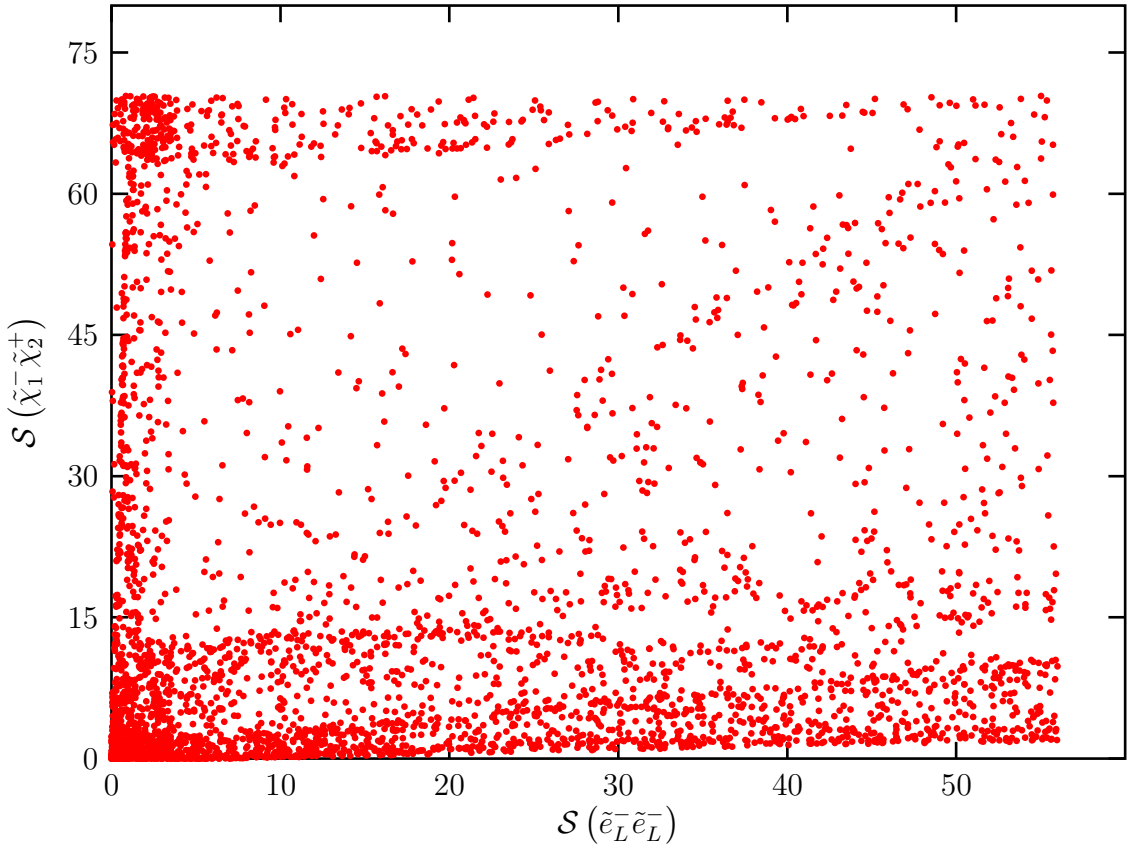}
\end{minipage}\hfill
\caption{Correlation between the significances for the processes $e^-
e^- \rightarrow \tilde e^-_L \tilde e^-_L$ and $e^+ e^- \rightarrow
\tilde \chi_1^- \tilde \chi_2^+$, both taken at $\sqrt{s} = 800$ GeV,
for scenario B2 with $\tan\beta = 3$.}
\label{sscor2}
\end{figure}

In some cases the correlations between different significances are
quite weak. The most extreme case we found is shown in
Fig.~\ref{sscor2}, and occurs for scenario B2 at $\tan\beta = 3$. We
saw in Table~\ref{signitable} that here (and only here) $\sigma(\tilde
\chi_1^- \tilde \chi_2^+)$ allows a significant probe of the phase
$\phi_\mu$, whereas $\cS(\tilde e^-_L \tilde e^-_L)$ is always mostly
determined by $\phi_1$. Moreover, Fig.~\ref{phimuphi1}b shows that in
the allowed band with $\phi_\mu \simeq \pi$, the deviation $|\phi_\mu
- \pi|$ becomes maximal for $\phi_1$ quite close to $\pm \pi$. This
leads to scenarios with large $\cS(\tilde \chi_1^- \tilde \chi_2^+)$,
but very small $\cS(\tilde e^-_L \tilde e^-_L)$. Conversely, $\big|
|\cos \phi_1| - 1 \big|$ can be quite large for small $|\phi_\mu|$,
leading to scenarios with $\cS(\tilde e^-_L \tilde e^-_L) \gg
\cS(\tilde \chi_1^- \tilde \chi_2^+)$, although the latter cannot be
strictly zero if the former is bigger than 10. However, we saw earlier
that other combinations of parameters do not allow meaningful probes
of $\phi_\mu$ using high--energy quantities. We therefore conclude
that in most cases, significances that can be large are also fairly
strongly correlated.

\begin{figure}[ht!]
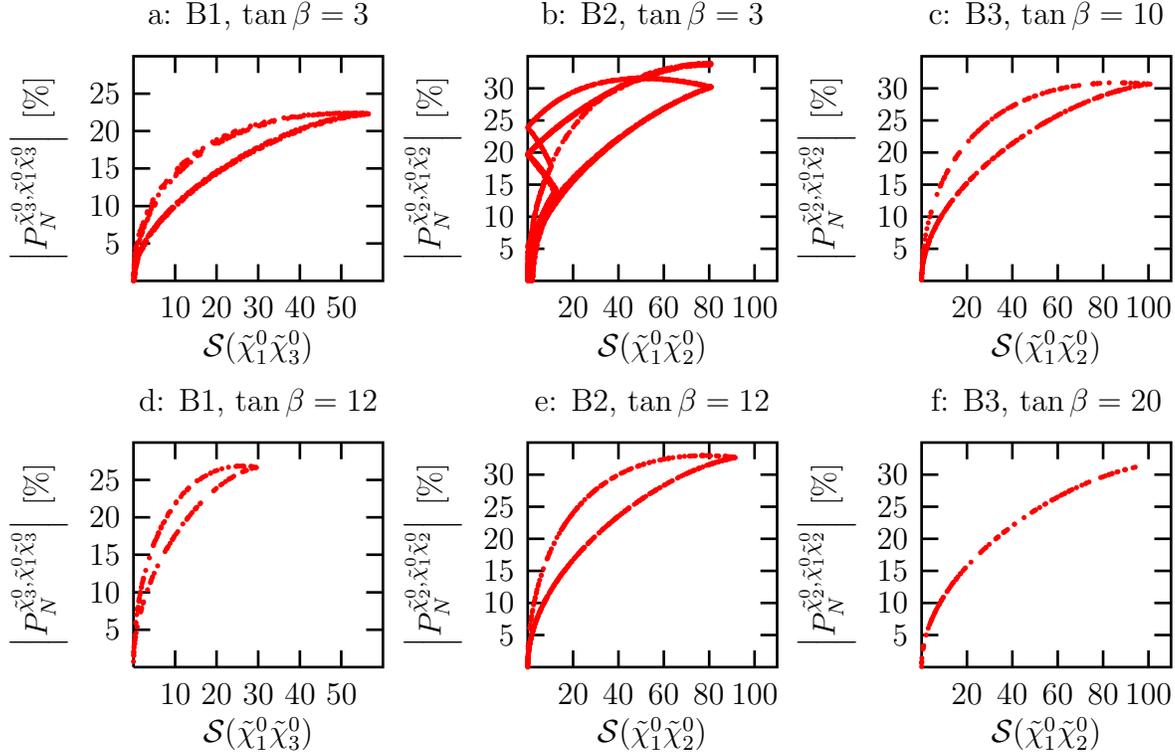

\begin{minipage}[ht!]{0.3\textwidth}
\input{sneu13p3b1t3_tex.tex}
\end{minipage}
\begin{minipage}[ht!]{0.3\textwidth}
\input{sneu12p2b2t3_tex.tex}
\end{minipage}
\begin{minipage}[ht!]{0.3\textwidth}
\input{sneu12p2b3t10_tex.tex}
\end{minipage}\hfill\\
\begin{minipage}[ht!]{0.3\textwidth}
\input{sneu13p3b1t12_tex.tex}
\end{minipage}
\begin{minipage}[ht!]{0.3\textwidth}
\input{sneu12p2b2t12_tex.tex}
\end{minipage}
\begin{minipage}[ht!]{0.3\textwidth}
\input{sneu12p2b3t20_tex.tex}
\end{minipage}\hfill
\caption{Correlation between the significance $\cS$ and the absolute
value of normal polarization $P_N$, measured at scattering angle
$\theta = \pi/2$, for mixed neutralino pair production at $\sqrt{s} =
500$ GeV. We consider $\tilde \chi_1^0 \tilde \chi_2^0$ production for
scenarios B2 and B3, but switch to the $\tilde \chi_1^0 \tilde
\chi_3^0$ final state for scenario B1.}
\label{spcor}
\end{figure}

Finally, in Fig.~\ref{spcor} we compare the normal component of the
polarization vector of the heavier neutralino in mixed neutralino pair
production with the significance of the same mode. We consider $\tilde
\chi_1^0 \tilde \chi_2^0$ production in scenarios B2 and B3, but
switch to $\tilde \chi_1^0 \tilde \chi_3^0$ production in scenario B1,
where this final state is far more promising, see
Tables~\ref{signitable} and \ref{pntable}. These figures look somewhat
simpler than those in Fig.~\ref{sscor1}, since now the existence of
two allowed CPC points only leads to two bands, as compared to three
in Fig.~\ref{sscor1}. Of course, scenarios with a single allowed CPC
point (Fig.~\ref{spcor}f) again only yield a single band. In panel b)
we again find scenarios with sizable phases, hence sizable
$|P_N^{\tilde \chi_2^0, \tilde \chi_1^0 \tilde \chi_2^0}|$, and yet
vanishing $\cS(\tilde \chi_1^0 \tilde \chi_2^0)$; we saw analogous
behavior in Fig.~\ref{sscor1}b. 

More importantly, Fig.~\ref{spcor} shows that the polarization $|P_N|$
increases much more quickly as the (relevant) phase $\phi_1$ is moved
away from 0 or $\pi$ than the significance $\cS$ does. The reason is
that $|P_N|$, being T--odd, has a sin--like dependence on $\phi_1$,
i.e. grows linearly with $|\phi_1|$ or $|\phi_1 - \pi|$. In contrast,
the T-- and CP--even quantity $\cS$ has cos--like dependence on all
phases, and thus only grows $\propto |\phi_1|^2$ or $|\phi_1 - \pi|^2$
as $\phi_1$ is moved away from a CPC point. T--odd observables like
$P_N$ are therefore in principle better suited to probe small
phases.

\section{Summary and Conclusions}
\label{sec:conclusion}

In this article we have discussed to what extent the phases of
dimensionful parameters in the SUSY Lagrangian can be determined from
leptonic observables. Since we assumed universal soft breaking
parameters for the first two generations of sleptons and did not
discuss processes involving third generation (s)particles, we only
have to deal with three phases: those of the Higgsino mass parameter
$\mu$, of the $U(1)_Y$ gaugino mass $M_1$, and of the leptonic
trilinear soft breaking parameter $A_l$, in all cases measured relative
to $M_2$ which we took to be real and positive by convention.

Our main focus was on quantities that can be measured at future
high--energy $e^+e^-$ and $e^-e^-$ colliders, but we first analyzed
the constraints that follow from the present measurements of the
leptonic dipole moments $d_e$ and $a_\mu$. We worked in a scenario
with moderately heavy sparticles; as well known, in this case sizable
CP--odd phases are possible only if neutralino and chargino loop
contributions to $d_e$ cancel to good approximation. In agreement with
earlier work \cite{nath,kane}, we found that, unless $|\mu| \gg M_2,
\, m_{\tilde l}$, the phases of $M_1$ and $A_l$ can take any value
(for some combination of the other phases), whereas the phase of $\mu$
is tightly constrained, the maximal allowed deviation from 0 or $\pi$
scaling like $|\mu|^2$. Our analysis of Sec.~4 also gave the perhaps
surprising result that in this case improved measurements of $d_e$
will not significantly reduce the allowed range for any one of the
three relevant phases after scanning over the other two. This is true
independently of whether this measurement leads to improved upper
bounds on $|d_e|$ or finds a non--vanishing result. On the other hand,
improved measurements of $a_\mu$ do have the potential to further
restrict the allowed ranges of these phases; however, here improved
measurements have to be combined with improved SM predictions for the
hadronic contributions to $a_\mu$.

Turning to high--energy observables, we first analyzed in detail the
phase sensitivity of total cross sections of various final states. To
that end we introduced ``significances'' that determine the statistical
significance with which the presence of non--trivial phases could be
determined in a given production channel. As pointed out in
ref.~\cite{thomas}, the cross section for $\tilde e^-_L \tilde e^-_L$
production depends very strongly on the relative phase between $M_1$
and $M_2$; we found that a deviation of $\sim 60$ to 90 standard
deviations from the predictions of the CP--conserving MSSM is possible
in this channel. However, this does not necessarily argue in favor of
constructing an $e^-e^-$ collider, since certain neutralino production
channels -- in particular, $\tilde \chi_1^0 \tilde \chi_2^0$
production for $|\mu| > M_2$ -- have comparable or better sensitivity
to the same phase. We also found a somewhat lower, but still
promising, sensitivity in the $\tilde e^-_L \tilde e^+_R$ final
state. For our choice $m_{\tilde l} \sim 200$ GeV, chargino pair
production can show significant phase dependence over the
experimentally allowed parameter space only for $|\mu| \geq 2
M_2$. Since the $d_e$--constraint on $\phi_\mu$ becomes weaker for
larger slepton masses, the minimal ratio $|\mu|/M_2$ where chargino
production channels can become useful for probing CP--violating phases
should be smaller for larger $m_{\tilde l}$. However, these chargino
modes will be useful only if $\tan\beta$ is quite small, since the
relevant significances scale like $\sin 2\beta$.

A deviation of any of these cross sections from the prediction of the
CP--conserving MSSM could perhaps also be explained by some extension
of the model which does not introduce new CP--odd phases. We therefore
also studied a CP--odd quantity: the component of the polarization of
produced charginos and neutralinos that is normal to the production
plane. We found that it can reach values exceeding 30\% for the
production of two different neutralinos; in scenarios with large
$|\mu|$ and small $\tan\beta$ the polarization vector of the lighter
chargino, produced in association with the heavier one, could have an
even larger normal component. Recent studies \cite{bartl1, cdgs,
bartl2} indicate that such large CP--odd polarizations might indeed
lead to measurable CP--odd asymmetries in the phase space distribution
of the $\tilde \chi$ decay products.

Finally, we studied correlations between the various phase--sensitive
observables. We found that the high--energy observables are
essentially not correlated at all with $d_e$. This is due to the
required rather precise cancellation between different contributions
to $d_e$; it implies that better measurements of $d_e$ will not
further restrict the possible ranges of phase--sensitive high--energy
quantities. However, there is some correlation between these
high--energy observables and $a_\mu$. Moreover, most pairs of
high--energy observables are quite strongly correlated with each
other. This follows from the fact that most of them basically probe
the phase of $M_1$, given the tight constraint on the phase of
$\mu$. Within the CP--violating MSSM the measurement of one phase
sensitive high--energy observable therefore allows to greatly
constrain the allowed range of other such quantities, thereby allowing
stringent tests of the model. However, at large $|\mu|$ and small
$\tan\beta$ the phase of $\mu$ can play an important role, in
particular in chargino production. In that case phase--sensitive
observables in the chargino sector correlate poorly with those in the
selectron or neutralino sector. This underscores the importance of
measuring as many phase--sensitive quantities as possible.

Total cross sections and CP--odd asymmetries offer complementary
access to CP--odd phases, since they depend on these phases through
cosine--like and sine--like functions, respectively. The former are
rather insensitive to these phases if they are small (the perhaps most
likely case). Measurements of, or bounds on, CP--odd asymmetries
should then lead to better determinations or constraints on these
phases. On the other hand, if some phase is near $\pi/2$, CP--odd
asymmetries will be near maximal, which means that they are not well
suited to precisely pinning down the value of this phase; precision
measurements of some cross sections will then have the edge. Of
course, there is also complementarity between high-- and low--energy
observables, since only the latter are sensitive to the phase of
$A_l$. 

We conclude that measurements at high energy colliders will be
necessary to pin down the phases of dimensionful parameters in the
SUSY Lagrangian. Both precision measurements of CP--even quantities
like masses and cross sections, and searches for CP--violating
asymmetries, are promising in certain regions of parameter
space. Linear $e^+e^-$ colliders seem to be ideally suited for
performing these measurements.

%%%%%%%%%%%%%%%%%%%%%%%%%%%%%%
\setcounter{footnote}{1}
\setcounter{equation}{0}
\renewcommand{\theequation} {\thesection.\arabic{equation}}
%%%%%%%%%%%%%%%%%%%%%%%%%%%%%%%%%%%%%%%%%%%%%%%%%%%%%%%%%%%%%%%%%
%%%%%%%%%%%%%%%%%%%%%%%%%%%%%%%%%%%%%%%%%%%%%%%%%%%%%%%%%%%%%%%%%
\section*{Acknowledgments}
The work of MD and BG was supported in part by the Deutsche
Forschungsgemeinschaft, project number DR 263 as well via SFB 375.
They would like to thank the KIAS school of physics and its members
for hospitality.  SYC was supported in part by the Korea Research
Foundation (KRF-2002-070-C00022) and in part by KOSEF through CHEP at
Kyungpook National University.
%%%%%%%%%%%%%%%%%%%%%%%%%%%%%%%%%%%%%%%%%%%%%%%%%%%%%%%%%%%%%%%%%
\setcounter{section}{0}
\renewcommand{\thesection}{Appendix \Alph{section}}
\section{Kinematics}
\label{app:kine}
\setcounter{equation}{0} \renewcommand{\theequation}
{\Alph{section}.\arabic{equation}} Working in the CMS frame with total
energy $\sqrt{s}$ and neglecting the electron mass, the first electron
and positron (second electron) momenta can be written as 
\ben \beqq
p_1^\mu &= \frac {\sqrt{s}} {2} \left( 1, 0, 0, 1 \right); \\
p_2^\mu &= \frac {\sqrt{s}} {2} \left( 1, 0, 0, -1 \right).
\eeqq \een
The outgoing momenta of the produced superparticles $b$ and $c$ are:
\ben \beqq
k_1^\mu &= \frac {\sqrt{s}}{2} \left( 1 + \frac{m_b^2-m_c^2}{s},
\lambda_{bc}^\frac{1}{2} \sin\theta, 0, \lambda_{bc}^\frac{1}{2}
\cos\theta \right), \\
k_2^\mu &= \frac {\sqrt{s}}{2} \left( 1 - \frac{m_b^2-m_c^2}{s},
-\lambda_{bc}^\frac{1}{2} \sin\theta, 0, -\lambda_{bc}^\frac{1}{2}
\cos\theta \right),
\eeqq\een
where $\lambda_{bc}$ denotes the usual two--body final state
kinematical function:
\ben \beqq
\lambda_{bc} &= \lambda \left(1, \frac{m_b^2}{s}, \frac{m_c^2}{s}
\right); \\
\lambda(1, x, y) &= 1 + x^2 + y^2 - 2(x+y+xy).
\eeqq\een
Furthermore the kinematical invariants (Mandelstam variables) are 
\ben \beqq
s &= (p_1+p_2)^2; \\
t &= (p_1-k_1)^2; \\ 
u &= (p_1-k_2)^2.
\eeqq\een
%
%%%%%%%%%%%%%%%%%%%%%%%%%%%%%%%%%%%%%%%%%%%%%%%%%%%%%%%%%%%%%%%%%
\section{Helicity amplitudes}
\label{app:heli}
\setcounter{equation}{0} \setcounter{footnote}{1}
\renewcommand{\theequation} {\Alph{section}.\arabic{equation}} 

We calculate the relevant helicity amplitudes using the formalism
introduced in \cite{Hagiwara:1985yu}\footnote{Our convention for a
momentum--dependent Weyl spinor for fermions going in the $-z$
direction differs by an overall sign from that of
\cite{Hagiwara:1985yu}.}. Using our definition of the kinematical
situation, we find the following results for the scalar and vectorial
fermionic string associated with massless fermions:
\ben \label{strings} \beqq 
\bar{v}(p_2, \sigma_2) P_\alpha u(p_1, \sigma_1) &= -\alpha \sqrt{s}
\delta_{\alpha\sigma_1} \delta_{\sigma_1\sigma_2}, \\
\bar{v}(p_2, \sigma_2) \gamma^\mu P_\alpha u(p_1, \sigma_1) &=
\sqrt{s} \delta_{\alpha\sigma_1} \delta_{\sigma_2,-\sigma_1} (0, 1,
i\sigma_1, 0),
\eeqq\een
where the four choices in eq.(\ref{strings}b) correspond to $\mu = 0,
\, 1, \, 2, \, 3$. In the case of neutralinos or charginos with
non--negligible masses only the vectorial string is required. It can
be written as
\beqa
\bar{u}_i( k_1, \lambda_1) \gamma^\mu P_\beta v_j(k_2, \lambda_2) &=&
\frac {\sqrt{s}} {2} \left[ \sqrt{ 1 - \eta_{\beta\lambda_1}^2 }
\delta_{\lambda_1\lambda_2} (\beta, \lambda_1 \sin\theta, 0, \lambda_1
\cos\theta ) \right.\\
&+& \left. \sqrt{( 1 + \beta \lambda_1 \eta_{\beta\lambda_1} ) ( 1 +
\beta \lambda_1 \eta_{-\beta\lambda_1} ) }
\delta_{\lambda_1,-\lambda_2} (0, \cos\theta, -i \lambda_1,
-\sin\theta ) \right],\nonumber
\eeqa
where
\beq \label{etadef}
\eta_{\beta\lambda_1} = \lambda^{\frac{1}{2}}_{ij} + \beta \lambda_1
\Delta_{ij},
\eeq
and
\beq
\Delta_{ij} = \frac{m_i^2-m_j^2}{s}.
\eeq
%
%%%%%%%%%%%%%%%%%%%%%%%%%%%%%%%%%%%%%%%%%%%%%%%%%%%%%%%%%%%%%%%%%
\section{Neutralino functions}
\label{app:nf}
\setcounter{equation}{0}
\renewcommand{\theequation} {\Alph{section}.\arabic{equation}}

After introducing two effective neutralino mixing coefficients 
\ben \beqq
V_L^j &= \frac {N_{1j}} {2\cw} + \frac {N_{2j}} {2\sw}, \\
V_R^j &= \frac{N_{1j}}{\cw},
\eeqq\een
we define two dimensionless neutralino functions for t- or u-channel
exchanges:
\ben \label{neufun} \beqq
M_{\alpha\beta} (s,t/u) &= \sum_{k=1}^4 m_{\tilde{\chi}_k^0} \sqrt{s}
V_\alpha^k V_\beta^k D_{t,u} ^k, \\
N_{\alpha\beta} (s,t/u) &= \sum_{k=1}^4 s V_\alpha^k V_\beta^{k\star}
D_{t,u}^k,
\eeqq\een
the propagators $D_t^k$ and $D_u^k$ have been defined in
eq.(\ref{tuprop}).  Very similar neutralino functions were introduced
in \cite{slepprod}; we saw in Sec.~\ref{sec:highenergy} that they
allow to give very compact expressions for the slepton production
amplitudes.

\end{document}